\documentclass[a4paper,11pt]{article}

\usepackage{geometry}
\usepackage{textcomp}
\usepackage{amsmath}
\usepackage{amssymb}
\usepackage{graphicx}
\usepackage{color,xcolor}
\usepackage{mathrsfs}
\usepackage{caption}
\usepackage[colorlinks=true, allcolors=blue]{hyperref}
\usepackage{tcolorbox}
\usepackage{tikz}
\usepackage{tikz-cd}
\usepackage{physics}
\usepackage{diagbox}
\usepackage{listings}
\usepackage{array}
\usepackage{subfig}

\makeatletter

\usetikzlibrary{backgrounds}

\numberwithin{equation}{section}

\def \be {\begin{equation}}
\def \ee {\end{equation}}
\def \ba {\begin{array}}
\def \ea {\end{array}}
\def \bea{\begin{eqnarray}}
\def \eea{\end{eqnarray}}

\def \G {\Gamma}

\def \D {\Delta}

\def \k {\kappa}

\def \r {\rho}

\def \p {\partial}

\def \nn {\nonumber}

\def \cP{\mathcal P}
\def \cO{\mathcal O}
\def \cA{\mathcal A}
\def \cB{\mathcal B}
\def \cC{\mathcal C}
\def \cD{\mathcal D}
\def \cH{\mathcal H}
\def \cL{\mathcal L}
\def \cK{\mathcal K}
\def \cS{\mathcal S}

\def \hs {\hspace}

\def \inf {\infty}

\def \Re {{\textrm{Re}}}

\def \Tr {{\textrm{Tr}}}

\newcommand{\greenfunction}[1]{\langle #1\rangle}
\DeclareMathOperator{\sign}{sign}
\newcolumntype{L}{>{$}l<{$}}
\newcommand{\glie}{\mathfrak{g}}        
\newcommand{\sllie}{\mathfrak{sl}}      
\newcommand{\solie}{\mathfrak{so}}      
\newcommand{\nor}{ d}                   
\newcommand{\CoefficientChi}{A(2-\Delta)}

\makeatother
\begin{document}
\title{On Galilean conformal bootstrap}
\vspace{14mm}
\author{
	 Bin Chen$^{1,2,3}$, Peng-xiang Hao$^4$, Reiko Liu$^1$ and Zhe-fei Yu$^1$\footnote{bchen01@pku.edu.cn, pxhao@tsinghua.edu.cn, reiko\_liu@pku.edu.cn, yuzhefei@pku.edu.cn}
}
\date{}
\maketitle
\begin{center}
	{\it
		$^{1}$School of Physics and State Key Laboratory of Nuclear Physics and Technology,\\Peking University, No.5 Yiheyuan Rd, Beijing 100871, P.~R.~China\\
		\vspace{2mm}
		$^{2}$Collaborative Innovation Center of Quantum Matter, No.5 Yiheyuan Rd, Beijing 100871, P.~R.~China\\
		$^{3}$Center for High Energy Physics, Peking University, No.5 Yiheyuan Rd, Beijing 100871, P.~R.~China\\
		$^4$Yau Mathematical Sciences Center,Tsinghua University, Beijing, 100084, China
	}
	\vspace{10mm}
\end{center}

\begin{abstract}
In this work, we develop  conformal bootstrap for Galilean conformal field theory (GCFT). In a GCFT, the Hilbert space could be decomposed into  quasiprimary states and its global descendants. Different from the usual conformal field theory, the quasi-primary states in a GCFT constitute multiplets, which are block-diagonized under the Galilean boost operator. More importantly the multiplets include the states of negative norms, indicating the theory is not unitary. We compute global blocks of the multiplets, and discuss the expansion of  four-point functions in terms of the global blocks of the multiplets. Furthermore we do the harmonic analysis for the Galilean conformal symmetry and obtain an inversion formula. As the first step to apply the Galilean conformal bootstrap, we construct  generalized Galilean free theory (GGFT) explicitly. We read the data of GGFT by using Taylor series expansion of four-point function and  the inversion formula independently, and find exact agreement. We discuss some novel features in the Galilean conformal bootstrap, due to the non-semisimpleness of the Galilean conformal algebra and the non-unitarity of the GCFTs.

\end{abstract}

\baselineskip 18pt
\newpage

\tableofcontents{}

\newpage

\section{Introduction}

Conformal bootstrap is a nonperturbative program to constrain or even read the spectrum and operator product expansion (OPE) coefficients of a conformal field theory(CFT) by imposing the conformal symmetry, unitarity and the crossing symmetry. It was first proposed in 1970s \cite{Ferrara:1973yt,Polyakov:1974gs} and was applied to solve the two-dimensional (2d) minimal models successfully in \cite{Belavin:1984vu}. In the past decade, conformal bootstrap has been revived, starting from the seminal work of \cite{Rattazzi:2008pe}. In this work, a numerical method has been proposed to extract the rigorous predictions from the conformal bootstrap equations without fully solving them. The method has been applied to study many models in various dimensions, for instance yielding precise critical exponents of the critical 3d Ising model \cite{ElShowk:2012ht,El-Showk:2014dwa}. For a review on the conformal bootstrap, especially the numerical techniques, see \cite{Poland:2018epd}.

Besides the numerical method, analytic approaches have been developed in modern conformal bootstrap.  These analytic approaches include the large spin perturbation theory  \cite{Komargodski:2012ek,Fitzpatrick:2012yx,Kaviraj:2015cxa,Kaviraj:2015xsa,Alday:2016njk}, holography from CFT \cite{Heemskerk:2009pn,Penedones:2010ue,Fitzpatrick:2011ia,Alday:2016htq}, Lorentzian inversion formula \cite{Caron-Huot:2017vep,Simmons-Duffin:2017nub}, analytic functional method \cite{Mazac:2016qev,Mazac:2018mdx,Mazac:2018ycv,Mazac:2019shk,Paulos:2019gtx}, etc.. The analytic studies not only help us to improve the numerical method, but also shed light on the AdS/CFT correspondence and the S-matrix bootstrap.

The usual conformal bootstrap is based on the conformal invariance and unitarity. It would be interesting to extend the program to field theories with other conformal-like symmetry. For example, conformal bootstrap has been studied in theories with Schr\"odinger symmetry in \cite{Goldberger:2014hca} and in Logarithmic conformal field theories (LCFTs) \cite{Banerjee:2019uxo}. In the present work, we would like to study the conformal bootstrap on non-relativistic field theories with Galilean conformal invariance. The global part of the symmetry could be obtained by a non-relativistic contraction of the conformal algebra \cite{Negro:1997,Lukierski:2005xy}, and it contains the translations, the isotropic scaling, the analogues of special conformal transformations and the Galilean symmetries instead of the Lorentzian symmetries.  Quite remarkably, it was found in \cite{Bagchi:2009my} that the Galilean conformal symmetry in any dimension is actually much larger and is generated by an infinite-dimensional Galilean Conformal Algebra (GCA), which can be obtained by taking the non-relativistic limit of conformal Killing equations and is shown to be the maximal subset of non-relativistic conformal isometries \cite{Duval:2009vt,Martelli:2009uc}.  In two dimensions the generators of GCA obey the following commutation relations
\begin{equation}
\begin{split}
&[L_n,L_m]=(n-m)L_{n+m}+c_L\delta_{n+m,0}(n^3-n),\\
&[L_n,M_m]=(n-m)M_{n+m}+c_M\delta_{n+m,0}(n^3-n),\\
&[M_n,M_m]=0.
\end{split}
\end{equation}
The 2d Galilean conformal field theory(GCFT) is of particular interest, as 2d GCA is isomorphic to the Bondi-Metzner-Sachs (BMS) algebra in three dimensions, which generates the asymptotic symmetries of 3d
flat spacetimes \cite{Bagchi:2010zz}. This motivates a lot of works establishing holography theory in asymptotic flat spacetimes (the so called BMS/GCA correspondence), see  \cite{Bagchi:2012cy,Barnich:2012xq,Bagchi:2012xr,Bagchi:2014iea,Jiang:2017ecm,Hijano:2018nhq,Hijano:2019qmi,Apolo:2020bld,Apolo:2020qjm}. In this paper, we will focus on 2d GCFT.

One typical feature in GCFT is that there are descendant states with negative norms. This fact suggests that GCFTs are not unitary. Even though the usual (both numerical and analytical) conformal bootstrap relies heavily  on the unitarity, it does not mean the bootstrap program can not be carried on in theories without unitarity\footnote{There are a few numerical \cite{Poland:2018epd} and (analytical) Polyakov-Mellin \cite{Gopakumar:2016wkt} bootstrap results for non-unitary CFTs.}. The essential requirement is that the block coefficients must be positive. The existence of the negative-norm states may not be fatal.  At technical level, as noted in \cite{Bagchi:2009pe}, the Galilean boost operator $M_0$ is in general not diagonalizable when acts on the descendant states. This indicates that the theory is not  unitary. Just like in LCFTs \cite{Hogervorst:2016itc}, multiplets appear. When we try to expand four-point functions in terms of GCA global blocks, in addition to the blocks of singlets calculated in \cite{Bagchi:2017cpu}, we have to count the contributions  from the global blocks corresponding to these multiplets. We discuss the multiplets in GCFT and compute their global blocks.

Motivated by the Lorentzian inversion formula \cite{Caron-Huot:2017vep} for analytic conformal bootstrap, we do  harmonic analysis for the global part of Galilean conformal symmetry, thus obtain a GCA inversion formula, which can be used to compute the spectrum and OPE coefficients. It turns out that the harmonic analysis in GCA is quite similar to the one for CFT$_1$ \cite{Maldacena:2016hyu,Murugan:2017eto}. Our result of harmonic analysis is new mathematically because the algebra considered here is not semi-simple, while the conformal algebra is.

In order to check our study, we discuss the generalized free theory with Galilean conformal symmetry.  The generalized free theories (GFTs) play an important role in conformal bootstrap. They provide the simplest examples of crossing-symmetric, conformally-invariant four-point functions. Their spectrum and OPE coefficients can be read off from the inversion formula \cite{Fitzpatrick:2011dm, Karateev:2018oml}.   Moreover, if one test the crossing condition in some specific regions\footnote{These regions are always the ones where conformal blocks expansion of a four-point function is not convergent uniformly,  this is also true for GCFTs.}, for example, in the lightcone limit for CFT,  GFTs are the leading contribution to the correlators at large spin \cite{Fitzpatrick:2012yx}. Furthermore GFTs provides the leading contribution to the correlators in bulk perturbation theory, from holographic point of view \cite{Heemskerk:2009pn}.  In the Galilean case at hand, the study of generalized Galilean free theories (GGFTs) is the first step towards the analytic Galilean conformal bootstrap. We expect that GGFTs will play  similar roles  as GFTs have played in the usual conformal bootstrap. 



In fact, there have been some earlier efforts towards 2d Galilean conformal bootstrap. In \cite{Bagchi:2016geg, Bagchi:2017cpu}, the crossing equation for four-point functions and the global GCA blocks for the singlets have been worked out. It was found that these kinematic quantities could be reproduced by taking non-relativistic limit of the corresponding ones in 2d (non-unitary) relativistic CFTs. It is tempting to think that other kinematic quantities in Galilean conformal bootstrap could be obtained by taking non-relativistic limit of parent CFTs. Our study shows that this is not always true\footnote{Note that in \cite{Bagchi:2009pe}, the constraint from the GCA analysis on the fusion rules is weaker than that from the limiting procedure of 2d CFTs. }.  For example,  as we will show in section 3, the Galilean conformal partial waves (GCPWs) can not be obtained by the limiting procedure, even though the inner product measure and the Casimir operators can be read by taking the limit, as there are subtleties in defining the Hilbert space. On the contrast, the GGFT can be reproduced by taking the limit on a 2d generalized free theory(GFT).

The remaining parts of this paper are organized as follows. In section 2, we revisit the block expansion of a four-point function in GCFTs, paying special attention to the contribution from the  multiplets.  In section 3, we do the harmonic analysis for the global GCA. Following the analysis in $SL(2,\mathbb{R})$ \cite{Maldacena:2016hyu}, we specify the Hilbert space and determine the Galilean conformal partial waves (GCPWs) as its complete orthogonal basis. Then we get an inversion formula and find that the blocks of the multiplets should appear as multiple poles in the inversion function. We also find that the limiting method is invalid to get the GCPWs. In section 4, we analyze the GGFTs from several different angles. Firstly, we construct the rank-$2$ multiplet directly from the level-$1$ quasiprimary operators in GGFT with two fundamental fields, and calculate its global block. This method show explicitly the forms of the ``double-trace" \footnote{We borrow the terminology ``double-trace'' in CFT here. } operators in GGFT, but the construction becomes awkward at higher levels.  In order  to get the information of GGFT, it is more effective to use other ways. One way is to  expand the four-point function into a double Taylor series and read the coefficients of the block expansion directly. The other way is to apply the GCA inversion formula obtained in section 3 to GGFTs. The result of these two methods match perfectly, and expectedly both match with the level-$1$ result from the constructive method. Moreover It  turns out that GGFT could be obtained by taking the non-relativistic limit of a 2d GFT. Especially, the multiplets appear in a remarkable way as  all superficially divergent terms under the limit cancel with each other.   In section 5, we go beyond GGFTs and discuss the spectral density of a GCFT.  We use the Hardy-Littlewood tauberian theorem to estimate the spectral density, and we check explicitly that the GGFTs satisfy the requirement of using the theorem. In Section 6, we discuss the shadow formalism and alpha space approach \cite{Hogervorst:2017sfd}, and find that with appropriate boundary condition the ``CPWs" in alpha space turns out to be the one obtained by shadow integral.  We end with conclusions and some discussions in section 7. Some technical details are left to Appendix.

\section{GCA revisited: multiplets}

In this section, we revisit  Galilean conformal field theory in two dimensional spacetime(GCFT$_2$). After a brief review on the basic knowledge on GCFT, including the symmetry, the primary operators and their two-point and three-point functions \cite{Bagchi:2009ca,Bagchi:2009pe,Bagchi:2016geg,Bagchi:2017cpu}, we turn to the quasi-primary operators which are essential in the global Galilean conformal bootstrap. We find that the quasi-primary operators typically form multiplets \footnote{Since similar structure of multiplets appears in Logarithmic CFT, we review the multiplets in LCFT in appendix \ref{LCFT}. In fact, our analysis of multiplets in GCFT  is inspired from the one in LCFT.}, which include negative-norm states and cannot be diagonalized under Galilean boost charge operator. We discuss the multi-point functions of multiplets and their global blocks.

\subsection{Review on Galilean CFT}

\subsubsection{Galilean conformal symmetry}
GCFT$_2$ is a non-relativistic field theory in 2d  spacetime. It has the scaling symmetry and the boost symmetry as follows,
\bea
x\rightarrow \lambda x, && y\rightarrow \lambda y.\\
x\rightarrow x,& & y\rightarrow y+v x.
\eea
The local Galilean conformal algebra is generated by $L_n, M_m \,(n, m\in \mathbb{Z})$,
\begin{center}
    \begin{tabular}{|L|L|L|l|}
    \hline
    \text{symmetry}                     & \text{label}      & \text{generator}                              & \text{finite transform}   \\
    \hline
    \text{Diff}(\mathbb{R}^1)       & L_{n}             & -x^{n+1}\partial_x-(n+1)x^n y\partial_y       & \begin{tabular}{@{}l}$x'=f(x) $\\ $y'=f'(x)y$\end{tabular} \\
    \hline
    \text{affine translations}      & M_{n}             & x^{n+1}\partial_y                             & \begin{tabular}{@{}l}$x'=x $\\ $y'=y+g(x)$\end{tabular}  \\
    \hline
\end{tabular}
\end{center}
and after considering the central extension, the commutation relations are
\begin{align}
[L_n,L_m]&=(n-m)L_{n+m}+c_L n(n^2-1)\delta_{n+m,0} ,\nn\\
[L_n,M_m]&=(n-m)M_{n+m}+c_M n(n^2-1)\delta_{n+m,0} ,\nn\\
[M_n,M_m]&=0.\nn
\end{align}

Analogous to $\sllie(2,\mathbb{R})\times \sllie(2,\mathbb{R})$ in the full Virasoro algebra, there is a maximal finite dimensional subalgebra $\glie\cong \mathfrak{iso}(2,1)$ corresponding to global Galilean conformal symmetry. The subalgebra is generated by $L_i\in \solie(2,1)$ and $M_i\in \mathbb{R}^3$, with  $i=\pm1,0$. In Table \ref{GGCG}, we list the representations of the generators and the corresponding finite transformations.
\begin{table}
\begin{center}
    \begin{tabular}{|L|L|L|L|}
    \hline
    \text{name}                    & \text{label}     & \text{vector field}          & \text{finite trans}   \\
    \hline
    x\text{-translation}  & L_{-1}  & -\p_x                & \begin{aligned}x'&=x+a\\y'&=y\end{aligned}\\
    \hline
    \text{dilation}                & L_{0}   & -x\p_x-y\p_y         & \begin{aligned}x'&=\lambda\,x\\y'&=\lambda\,y\end{aligned} \\
    \hline
    x\text{-SCT}          & L_{1}   & -x^2\p_x-2x y\p_y    & \begin{aligned}x'&=x/(1+\mu x)\\y'&=y/(1+\mu x)^{2}\end{aligned}\\
    \hline
    y\text{-translation}  & M_{-1}  & \p_y                & \begin{aligned}x'&=x\\y'&=y+b\end{aligned}\\
    \hline
    \text{boost}            & M_{0}   & x\p_y               & \begin{aligned}x'&=x\\y'&=y+v\,x\end{aligned}\\
    \hline
    y\text{-SCT}          & M_{1}   & x^2\p_y             & \begin{aligned}x'&=x\\y'&=y+\nu x^{2}\end{aligned}\\
    \hline
\end{tabular}
\end{center}
\caption{The generators of global Galilean conformal group. ``SCT" denotes ``special conformal transformation"}\label{GGCG}
\end{table}

\subsubsection{Primary operators}
The primary operators\footnote{By the state-operator correspondence we can talk about local operators and states interchangeably.} at origin $\cO=\cO(0,0)$ can be labelled by the eigenvalues $(\Delta,\xi)$ of $(L_0,M_0)$
\begin{equation}
[L_0,\cO]=\Delta \cO,\qquad [M_0,\cO]=\xi \cO.
\end{equation}
$\Delta$ and $\xi$ are referred to as the conformal weight and the boost charge of the operator respectively. The highest weight conditions are
\begin{equation}
[L_n,\cO]=0,\qquad [M_n,\cO]=0, \hs{3ex}n>0.
\end{equation}
 Then the descendant operators can be generated by acting $L_{-n},M_{-n}$ with $n>0$ successively on the primary operators. And the primary operator together with its descendants form a highest weight module.

The operators at other positions can be got by the translation operator $U=e^{x L_{-1}-y M_{-1}}$,
\begin{equation}
\cO(x,y)=U\cO(0,0)U^{-1}.
\end{equation}
Using the Baker-Campbell-Hausdorff (BCH) formula, the transformation law for the primary operators are,
\begin{align}
    [L_n,\cO(x,y)]&=(x^{n+1}\partial_x+(n+1)x^ny\partial_y+(n+1)(x^n\Delta-nx^{n-1}y\xi))\cO(x,y),\label{Lntrans}\\
    [M_n,\cO(x,y)]&=(-x^{n+1}\partial_y+(n+1)x^n\xi)\cO(x,y),\label{Mntrans}
\end{align}
and they can be integrated to finite one,
\begin{equation}
    \cO'(x,y)=|f'|^{\D}\,e^{-\xi\frac{g'+yf''}{f'}}\,\cO(x',y').  \label{Finite}
\end{equation}

By requiring the vacuum is invariant under the global symmetry, the two-point function and three-point function of primary operators are respectively
\begin{align}
    G_2(x_1,x_2,y_1,y_2) &= \nor \,\delta_{\Delta_1,\Delta_2}\delta_{\xi_1,\xi_2}|x_{12}|^{-2\Delta_1}e^{2\xi_1\frac{ y_{12}}{x_{12}}},\\
G_3(x_1,x_2,x_3,y_1,y_2,y_3) &= c_{123}|x_{12}|^{-\Delta_{123}}|x_{23}|^{-\Delta_{231}}|x_{31}|^{-\Delta_{312}}e^{\xi_{123}\frac{y_{12}}{x_{12}}}e^{\xi_{312}\frac{y_{31}}{x_{31}}}e^{\xi_{231}\frac{y_{23}}{x_{23}}}, \label{3ptprimary}
\end{align}
where $\nor$ is the normalization factor of the two-point function, $c_{123}$ is the coefficient of three-point function which encodes dynamical information of the GCFT$_2$, and
\begin{equation}
x_{ij}\equiv x_i-x_j,\ \ y_{ij}\equiv y_i-y_j,\ \ \Delta_{ijk}\equiv\Delta_i+\Delta_j-\Delta_k,\ \ \xi_{ijk}\equiv\xi_i+\xi_j-\xi_k.
\end{equation}

The four-point functions of primary operators can be determined up to an arbitrary function of cross ratios,
\begin{equation}
G_4=\langle \prod_{i=1}^4\cO_i(x_i,y_i)\rangle =\prod_{i,j}|x_{ij}|^{\sum_{k=1}^4 -\Delta_{ijk}/3}e^{\frac{y_{ij}}{x_{ij}}\sum_{k=1}^{4}\xi_{ijk}/3}\mathcal{G}(x,y)
\end{equation}
where the indices $i=1,2,3,4$ label the external operators $\cO_i$, $\mathcal{G}(x,y)$ is called the stripped four-point function and $x$ and $y$ are the cross ratios,
\begin{equation}
x\equiv\frac{x_{12}x_{34}}{x_{13}x_{24}}\qquad \frac{y}{x}\equiv\frac{y_{12}}{x_{12}}+\frac{y_{34}}{x_{34}}-\frac{y_{13}}{x_{13}}-\frac{y_{24}}{x_{24}}.  \label{cross ratio}
\end{equation}

\subsubsection{Global blocks of primary operator}

For simplicity, in the following we mainly focus on the case of four identical external operators with $(\D,\xi)$.  In principle, the stripped four-point function $\mathcal{G}(x,y)$ can be expressed in terms of  the OPE coefficients of primary operators and local Galilean conformal block. The local Galilean conformal block encodes all the contribution from a primary module. Unfortunately its form and properties has not been well-studied, as far as we know. In this work, we try to study the bootstrap  based on global Galilean conformal symmetry.

The contribution of a primary operator and its global descendant operators (which can be got by acting $L_{-1}$ and $M_{-1}$) to the stripped four-point function $\mathcal{G}(x,y)$ can be written as
\begin{equation}
\frac{1}{\nor}\,c_{12p}c_{34p}g_p(x,y)
\end{equation}
where  the index $p$ labels the propagating primary operator $\cO_p$. The function $g_p(x,y)$ is related to the global block $g_{\Delta_p,\xi_p}(x,y)$ by
\begin{equation}\label{gb0}
g_p(x,y)=x^{2\Delta}e^{-2\xi\frac{y}{x}}g_{\Delta_p,\xi_p}(x,y)
\end{equation}
The global block is the solution of the conformal Casimir equations with the OPE boundary conditions \cite{Bagchi:2016geg,Bagchi:2017cpu}
\begin{equation}
C_i g_{\Delta_p,\xi_p}(x,y)=\lambda_i g_{\Delta_p,\xi_p}(x,y),\ \ i=1,2
\end{equation}
where
\begin{align}
C_1&=M^2_0-M_1M_{-1}, \nn\\
C_2&=4L_0M_0-L_{-1}M_1-L_{1}M_{-1}-M_1L_{-1}-M_{-1}L_1. \label{quadraticC}
\end{align}
The eigenfunction  $g_{\Delta_p,\xi_p}(x,y)$ giving the global block of primary operators takes the form
\begin{equation}
g_{\Delta_p,\xi_p}(x,y)=2^{2\Delta_p-2}x^{\Delta_p-2\Delta}(1+\sqrt{1-x})^{2-2\Delta_p}e^{\frac{-\xi_p y}{x\sqrt{1-x}}+2\xi\frac{y}{x}}(1-x)^{-1/2},
\end{equation}
and the corresponding eigenvalues $\lambda_i$ are
\be
\lambda_1=\xi_p^2, \hs{3ex}
\lambda_2=2\xi_p(\Delta_p-1).
\ee

In this paper, we only concern about Galilean conformal field theories that do not have $\xi=0$ operators in their spectrum. The special case of  $\xi=0$  is very different and subtle. Even in this case, there exists the  multiplets as well,  and when the propagating operators have $\xi=0$ the corresponding singlet blocks are just the usual $SL(2,\mathbb{R})$ conformal blocks, but the multiplets blocks are quite different, which have non-trivial $y$ dependence. The following discussion will not involve this subtlety, though it is certainly very important\footnote{For example, see the discussion of BMS free scalar \cite{Hao:2021a}.}. We leave a thorough discussion on this degenerate case for a future project\cite{Hao:2021b}.

\subsection{Multiplets}

The  Hilbert space of a GCFT can be decomposed  into the highest weight modules of local GCA
\be
\cH=\sum_{\D, \xi}\cH_{\D,\xi}. \nn
\ee
However it is hard to find all the contribution of a module to the stripped four-point function due to its complicated structure. Alternatively we can examine the usage of global GCA in Galilean CFT in the spirit of modern conformal bootstrap. With respect to the global GCA,  the Hilbert space is composed of the quasi-primary states and their global descendants. This way is more under control and is related closely to the harmonic analysis and the inversion formula. However, the price we pay in this way is that $M_0$ usually acts non-diagonally on these quasi-primary operators, though $L_0,M_0$ act diagonally on the primary operators.

As an illustration, consider the following level-$2$ descendant operators of a primary operator $\cO_{\D,\xi}$,
\begin{equation}\label{example1}
\cA=L_{-2}\cO,\ \ \cB=M_{-2}\cO.
\end{equation}
on which $M_0$ acts as
\begin{equation}
[M_0,\cA]=\xi \cA+2\cB,\ \ [M_0,\cB]=\xi \cB.
\end{equation}
This gives rise to a rank-$2$ Jordan block. Generically, the action of $M_0$ on quasi-primary operators  can be written in the Jordan canonical form,
\begin{equation}
[M_0,\cO]=\tilde{\xi}\cO
\end{equation}
where $\cO$ are quasi-primary operators in the theory, and $\tilde{\xi}$ is block-diagonalized,
\begin{equation}
\tilde{\xi}=
\begin{pmatrix}
\ddots & & & \\
 & \tilde{\xi_i}& & \\
  & & \tilde{\xi_j}& \\
   & & & \ddots\\
\end{pmatrix}
\end{equation}
in which $\tilde{\xi_i}$ are Jordan blocks,
\begin{equation}
\tilde{\xi_i}=
\begin{pmatrix}
 \xi_i& & & \\
  1& \xi_i& & \\
  & \ddots&\ddots &\\
   & & 1& \xi_i\\
\end{pmatrix}_{r\times r}.
\end{equation}
The quasi-primary operators in the same Jordan block form a  multiplet\footnote{The situation is somehow similar to what happens in a logarithmic CFTs. Our analysis actually bases on the techniques developed in \cite{Gurarie:1993xq,Creutzig:2013hma,Hogervorst:2016itc}.}. The quasi-primary operators in a multiplet, together with their global descendants, compose  a reducible but indecomposable module of  global GCA.  If there are $r$ operators related to each other in a Jordan block, the multiplet they form will be referred to as of rank $r$, the same as the rank of the Jordan block. The above two operators $\cA,\cB$ form a multiplet of rank 2, and the primary operators introduced in previous subsections will be referred to as singlets or rank-$1$ multiplets.

In the following, we will discuss correlation functions of multiplets. Though there are some differences, our calculation of correlation functions follows the one in LCFT. To make this part more readable, we will not show all the details of the calculation in the subsection here. Instead, we set the details in appendix \ref{MutiGCFT}. Since these details closely follow the analysis of LCFT, we also give a review on multiplets in LCFT in appendix \ref{LCFT}.

In GCFT the correlation functions of multiplets differ from the ones of singlet, since the action of $M_0$ on the operators changes. Now the transformation under $M_n$ is
\begin{align}
[M_n,\cO_a(x,y)]&=(-\delta_{a,b}x^{n+1}\partial_y+(n+1)x^n\tilde{\xi})\cO_b(x,y),\quad \text{for } n\geq 0\\
[M_{-1},\cO_a(x,y)]&=-\partial_y\cO_a(x,y).
\end{align}
The two-point functions $\greenfunction{\cO_a \cO_b}$, where $\cO_a$ and $\cO_b$ belong to the same rank-$r$ multiplet\footnote{When two operators belong to different multiplets, their two-point function is vanishing. } with $(\D,\xi)$, can be determined by the Ward identities with respect to global symmetries,
\begin{equation}
\greenfunction{\cO_a(x_1,y_1)\cO_b(x_2,y_2)}=|x_{12}|^{-2\Delta}e^{2\xi y_{12}/x_{12}} A_{ab}(x_{12},y_{12})
\end{equation}
where $A_{ab}(x_{12},y_{12})$ is the following matrix,
\begin{equation}
A_{ab}(x_{12},y_{12})=\sum_{k=0}^{r-a-b}f_{k+a+b}\frac{1}{k!}(\frac{2y_{12}}{x_{12}})^k
\end{equation}
with $f_{k+a+b}$ being undetermined coefficients. One can always set $A_{ab}$ to a triangular matrix, by re-defining the operators in the multiplet, which eliminates $r$ degrees of freedom. This  simplifies the two-point functions into the following canonical form,
\begin{equation}\label{2pt}
\langle \cO_{k_1}(x_1,y_1)\cO_{k_2}(x_2,y_2)\rangle =\left\{\ba{ll}
0& \mbox{for $q<0$}\\
\nor_r\, |x_{12}|^{-2\Delta_1} e^{2\xi_1\frac{y_{12}}{x_{12}}}\frac{1}{q!}\left(\frac{2y_{12}}{x_{12}}\right)^q,& \mbox{otherwise}\ea \right.
\end{equation}
where
\begin{equation}
q=k_1+k_2+1-r,
\end{equation}
and $\nor_r$ is the overall normalization of this rank-$r$ multiplet. Here we denote $\cO_{k_i}$ ($k_i=0,\cdots,r-1$) as the $(k_i+1)$-th operator in the multiplet.\footnote{The two-point functions of the rank-$2$ case has also been discussed in \cite{Henkel:2013fja}.}

The three-point functions involving multiplets can also be determined by the Ward identities. Their  general forms are given by,
\begin{equation}
\langle \cO_i\cO_j\cO_k\rangle =A_{ijk}B_{ijk}C_{ijk}
\end{equation}
where
\bea
A_{ijk}&=&\exp(\xi_{123}\frac{y_{12}}{x_{12}}+\xi_{312}\frac{y_{31}}{x_{31}}+\xi_{231}\frac{y_{23}}{x_{23}}),\label{Acoef}\\
B_{ijk}&=&|x_{12}|^{-\Delta_{123}}|x_{23}|^{-\Delta_{231}}|x_{31}|^{-\Delta_{312}},\label{Bcoef}\\
C_{ijk}&=&\sum_{a=0}^{r_1-1}\sum_{b=0}^{r_2-1}\sum_{c=0}^{r_3-1}c_{ijk}^{(abc)}\frac{(q_i)^a(q_j)^b(q_k)^c}{a!b!c!},\label{Ccoef}
\eea
with
\begin{equation}
q_i=\partial_{\xi_i}\ln A_{ijk}.
\end{equation}
Note that $\cO_i,\cO_j,\cO_k$ can belong to different multiplets of rank $r_1,r_2,r_3$ respectively. The coefficient $c_{ijk}^{(abc)}$ encodes the dynamical information of the theory.  For the case $r_1=r_2=r_3=1$, the three-point function reduces to \eqref{3ptprimary}. Another simple example is when $r_1=r_2=1,r_3=2$, then
\begin{align}
\langle \cO\cO\cO_0\rangle &=ABc_0, \nn\\
\langle \cO\cO\cO_1\rangle &=AB(c_1+c_0q_{3})
\end{align}
where $\cO$ is a primary operator, $\cO_0,\cO_1$ belong to a rank-2 multiplet, and $q_3=\p_{\xi_3}A$ with $\xi_3$ being the charge of $\cO_0$. Here are two independent dynamical three-point coefficients. In general, for the three-point function of \textit{singlet-singlet-multiplet}, we have
\begin{align}\label{3pt}
\langle \cO\cO\cO_0\rangle &=ABc_0,\nn \\
\langle \cO\cO\cO_{k_i}\rangle &=\partial_{\xi_3}\langle \cO\cO\cO_{k_i-1}\rangle +ABc_{k_i}, \hs{3ex}k_i=1,\cdots r-1,
\end{align}
where $\xi_3$ is the charge of $\cO_0$. For a rank-$r$ multiplet, there are $r$ coefficients $c_0,\cdots,c_{r-1}$.

Now let us turn to the four-point functions. In the following discussion, we focus on the case where the four external operators are identical singlet. One can insert the identity operator
\begin{equation}
1=\Tr_{\mbox{\tiny all operators}}\frac{|\cO_p\rangle \langle \cO_p|}{\langle \cO_p|\cO_p\rangle }
\end{equation}
into the four-point functions
\begin{equation}
G_4=\Tr_{\mbox{\tiny all operators}}\left(\frac{\langle \cO\cO\cO_p\rangle \langle \cO_p\cO\cO\rangle }{\langle \cO_p|\cO_p\rangle }\right)
\end{equation}

We can get the global block expansion by collecting the contribution of each multiplet of quasi-primary operators and their global descendants.
\begin{equation}
G_4=\sum_{p}\sum_{i,j} T_{ij}\greenfunction{\cO\cO\cO_{p,i}}\greenfunction{\cO_{p,j}\cO\cO}
\end{equation}
where $\cO_{p,i}$ is the $(i+1)$-th operator in the rank-$r$ multiplet labelled by $p$ and $T_{ij}$ is the inverse of the Gram matrix $\bra{\cO_p}\ket{\cO_p}$ (which is a right-lower triangular matrix).

Different from the case of a singlet,  the global block of a multiplet is not the eigenfunction of the Casimir operators. Instead, the Casimir operators act on the multiplet as follows,
\begin{equation}
(C_i-\lambda_i)^r|\cO_{r,k_i}\rangle=0.
\end{equation}
This is also true for its global descendant operators. The stripped four-point functions can be expanded into
\begin{equation}
\mathcal{G}(x,y)=\sum_{\cO_r}\frac{1}{\nor_r}f[\cO_r],
\end{equation}
where the propagating quasi-primary operator $\cO_r$ is a rank-$r$ multiplet with an overall normalization $\nor_r$, and $f[\cO_r]$ satisfy the following Casimir equations
\begin{equation}\label{qq1}
(C_i-\lambda_i)^{r}f[\cO_r]=0,\ \ \mbox{for $i=1,2$}.
\end{equation}
The general solution is
\begin{equation}
f[\cO_r]=\sum_{s=0}^{r-1}A_sg^{(s)}_{\Delta_r,\xi_r}. \label{globalblock}
\end{equation}
Here $g^{(s)}_{\Delta_r,\xi_r}, s=0,\cdots r-1$ make up the global block (up to the square of the two-point function of the external operators) for the multiplet,
\begin{equation}
g^{(s)}_{\Delta_r,\xi_r}=\partial_{\xi_r}^sg^{(0)}_{\Delta_r,\xi_r}
\end{equation}
where $g^{(0)}_{\D_r,\xi_r}$ takes the form as $g_p(x,y)$ in \eqref{gb0}. To get the coefficients $A_s$ in \eqref{globalblock}, one need to consider the OPE limit of the global blocks by expanding them around $x=0,y=0$. Using the three-point functions in \eqref{3pt}, one gets the following coefficients $A_s$,
\begin{equation}\label{qq3}
A_s=\frac{1}{s!}\sum_{a,b|a+b+s+1=r}c_ac_b,
\end{equation}
where $c_a$'s are the three-point coefficients in \eqref{3pt}.
The global block expansion of the stripped four-point function in GCFT is
\begin{equation}
\mathcal{G}(x,y)=\sum_{\cO_r,\xi_r}\frac{1}{\nor_r}\sum_{s=0}^{r-1}\frac{1}{s!}\sum_{a,b|a+b+s+1=r}c_ac_b\partial_{\xi_r}^sg^{(0)}_{\Delta_r,\xi_r}.
\end{equation}

\subsection{Fermionic operators}

The quasiprimary operators we have introduced obey bosonic statistics as the ones in 1d CFT, and the correlation functions contains no sign functions. As in CFT$_1$, one may define fermionic operators. For fermionic operators we need to add sign functions into the finite transforms.

To be motivated,  recall that in 1d CFT the conformal group is $SO(2,1)\cong PSL(2,\mathbb{R})=SL(2,\mathbb{R})/ \{\pm 1\}$, and to get fermionic representations we lift it to the double covering group $SL(2,\mathbb{R})$. Then the finite transformation of fermionic operators is,
\begin{equation}
    \cO'(x)=\sign^{r}(cx+d)|cx+d|^{-2\D}\cO(x')
\end{equation}
where $r=0$ for bosons and $r=1$ for fermions, and $x$  transforms as $PSL(2,\mathbb{R})$
\be
x'=\frac{ax+b}{cx+d}. \nn
\ee
One can check that the fermionic one is indeed a representation by using
\begin{equation}
    \sign(1+(c+c')x)=\sign(1+c'x)\sign(1+c\frac{x}{1+c'x}).\nn
\end{equation}
Then the two-point functions of fermionic primaries are
\begin{equation}
    G_2(x_1,x_2)=\nor\,\delta_{\Delta_1,\Delta_2}\sign(x_{12})|x_{12}|^{-2\Delta_1},
\end{equation}
as those appearing in SYK model \cite{Murugan:2017eto} and 1d analytic bootstrap \cite{Mazac:2016qev}. Effectively the operators are anti-commutative.

Turning back to GCFT$_{2}$, the conformal group is $ISO(2,1)$ and the related spin group is $SL(2,\mathbb{R})\ltimes \mathbb{R}^3$. Following the procedure of 1d CFT, the transformation rule should be modified by adding $\sign^{r}(cx+d)$ when $L_i$ is involved, and the two-point functions of fermionic primaries are
\begin{equation}
    G_2(x_1,x_2,y_1,y_2) = \nor\,\delta_{\Delta_1,\Delta_2}\delta_{\xi_1,\xi_2}\sign(x_{12})|x_{12}|^{-2\Delta_1}e^{2\xi_1\frac{ y_{12}}{x_{12}}}.
\end{equation}

For external fermionic operators, the propagating operator should be bosonic due to parity conservation. And the global block expansion gets no modification. For the generalized free theory in section 4, the four-point functions of generalized free fermions equal $(s+t-u)$ channels, in contrast to  $(s+t+u)$ channels for generalized free bosons.

\section{Harmonic analysis}

\subsection{Overview of harmonic analysis}

In this section, we study the harmonic analysis on the symmetry group generated by GCA. Let us first review briefly the harmonic analysis in the conformal group.

One essential step in applying the inversion formula is to decompose the four-point function by using a set of complete basis of conformal group in the Euclidean space. The conformal group is now $SO(d+1,1)$ in $d$ dimensions. The complete basis consists of the normalizable eigenfunctions of the Hermitian Casimir operators. What one needs to do is specifying the Hilbert space which makes Casimirs Hermitian. This requires:
\begin{enumerate}
\item specifying the inner product;
\item specifying the boundary conditions.
\end{enumerate}
Then using the boundary conditions, we can obtain the eigenfunctions of Casimirs, which include the principal series representations and possible discrete ones. For example, in CFT$_1$, the Hermitian condition is,
\begin{equation}
(Cf,g)-(f,Cg) =0\label{eq:1}
\end{equation}
where $C$ is a Casimir operator, and $(f,g)=\int dx\mu f^{*}g$ is the inner product. Specifying the inner product means that we need
to specify the measure $\mu$ which makes the left side of \eqref{eq:1} reduce to some boundary terms, then by choosing appropriate boundary conditions (including the normalizable condition) we can set these boundary terms vanish. So the functions satisfying
the above boundary conditions, together with the above inner product defined on them, constitute the Hilbert space which makes the Casimir Hermitian.

The complete basis refers to a complete basis of the Hilbert space defined above. Note that the normalizable condition is considered as one requirement to define this Hilbert space, so only the normalizable part of the four-point function is in this Hilbert space, which means only this part can be decomposed into our complete basis. For the non-normalizable part of four-point function, the subtleties were explained in \cite{Simmons-Duffin:2017nub}.

 The  so-called conformal partial waves (CPWs) corresponding to principal series representations and possible discrete ones are the expected complete basis. The orthogonality is guaranteed by the Hermitian condition. For physical blocks, which have real dimensions and satisfy the unitarity bound, there exists no inner product making them orthogonal. The CPWs are necessary when we
try to invert the OPE.


Besides the usual CFT case, the above procedure has been  applied to the study of other models with conformal symmetry, including the SYK model \cite{Maldacena:2016hyu} and its supersymmetric version \cite{Murugan:2017eto}, different boundary condition for CFT$_1$ \cite{Hogervorst:2017sfd}, the defect CFT \cite{Lemos:2017vnx} and the  CFT at a finite temperature \cite{Iliesiu:2018fao}. In the following subsection, we will apply this procedure to the 2d field theories with Galilean conformal symmetry.

\subsection{Harmonic analysis on GCA}

As the group generated by GCA is not semi-simple, we cannot apply the formal harmonic analysis for conformal symmetry group. Here we just follow the discussion on the SYK model.

There are two independent Casimir operators $C_1$ and $C_2$ for 2d GCA. However, one can not define a Hilbert space which makes $C_{1}$ and $C_{2}$ Hermitian simultaneously. From the point of view of taking the non-relativistic limit, as we will discuss in section 3.4, ${C}_{2}$ comes from the difference between the holomorphic and the antiholomorphic quadratic Casimirs, and is not suppose to be Hermitian. To evade this obstacle, we introduce a quartic Casimir $C_{3}=C^2_2$, which is Hermitian together with $C_1$. In the following, we use $C_1$ and $C_3$ to do harmonic analysis.

Let us first write down the action of these two Casimirs. For the quadratic Casimir, its action is
\begin{equation}
C_{1}f_{\Delta,\xi}(x,y)=x^{2}(1-x)\partial_{y}^{2}f_{\Delta,\xi}(x,y)=\xi^{2}f_{\Delta,\xi}(x,y) \label{C1}
\end{equation}
and for the quartic Casimir, it acts like
\begin{equation}
\begin{split}C_{3}f_{\Delta,\xi}(x,y) & ={C}_{2}^{2}f_{\Delta,\xi}(x,y)\\
 & =[(3x-2)x y\partial_{y}^{2}+2x^{2}(x-1)\partial_{x}\partial_{y}+2x^{2}\partial_{y}]^{2}f_{\Delta,\xi}(x,y)\\
 & =4\xi^{2}(\Delta-1)^{2}f_{\Delta,\xi}(x,y)\label{C4}
\end{split}
\end{equation}
where $C_1,{C}_2$ are the quadratic Casimirs defined in \eqref{quadraticC} and could be obtained by taking the non-relativistic
limit on the quadratic Casimirs of CFT$_{2}$.

Now we can specify the Hilbert space. We define the inner product to be,
\begin{equation}
(f,g) =\int dxdy\mu(x,y)f^{*}g\label{eq:4}
\end{equation}
The integral domain of \eqref{eq:4} is restricted by the symmetry of the four-point function, which is the invariance under the exchange of $1\leftrightarrow2$ or $3\leftrightarrow4$, just as in the SYK model.\footnote{In the SYK model, the $1\leftrightarrow2$ or $3\leftrightarrow4$ invariance of the four-point functions leads to the symmetry $\chi\rightarrow\frac{\chi}{\chi-1}$ ($\chi$ is the cross ratio), from which one get the boundary condition $f'(2)=0$ and then determine the region to be $\chi\in[0,2]$, see \cite{Maldacena:2016hyu} for details.} In our GCA case, we can use the global coordinate transformation such that \cite{Bagchi:2017cpu}:
\begin{equation}
\{(x_i,y_i)\}\rightarrow\{(\infty,0),(1,0),(x,y),(0,0)\}
\end{equation}
Under the exchange of $1\leftrightarrow2$ or $3\leftrightarrow4$, using the above configuration and the expressing of the cross ratios \eqref{cross ratio}, one can easily obtain the symmetry as follows:
\begin{equation}
x\rightarrow\frac{x}{x-1},\qquad y\rightarrow\frac{-y}{(1-x)^{2}}.\label{symmetry}
\end{equation}
Using this symmetry, it is easy to see that the required region is a strip,
\begin{equation}
x\in[0,2],\qquad y\in(-\infty,+\infty).
\end{equation}
Now we use the Hermitian condition $(Cf,g)-(f,Cg) =0$ to determine the measure. Since we have two Casimirs, the required
Hilbert space should make both of the them Hermitian. Let us analyze them one by one to reduce the Hilbert space. For the quadratic Casimir $C_1$, it is easy to see that it becomes a Strum-Liouville problem with respect to the variable $y$, so the measure is found to be independent of $y$: $\mu(x,y)=\mu(x)$. Then as expected, the Hermitian condition reduces to the boundary terms,
\begin{equation}
\begin{split}(C_{1}f,g) -( f,C_{1}g)  & =\int_{0}^{2}dx\mu(x)\int_{-\infty}^{+\infty}dy\partial_{y}(g\partial_{y}f-f\partial_{y}g)\nn\\
 & =\int_{0}^{2}dx\mu(x)(g\partial_{y}f-f\partial_{y}g)\Big|_{-\infty}^{+\infty}.
\end{split}
\end{equation}
The normalizable condition for the function $f$ in $y$ direction is:
\begin{equation}
f\rightarrow 0 \text{ faster than }|y|^{-1/2},\hs{2ex}\text{ as }|y|\rightarrow\infty, \label{ybc}
\end{equation}
which ensures the boundary terms to be vanishing. The eigenfunctions of $C_{1}$ is of the form
\begin{equation}
\psi_{\xi}(x,y)=\exp\frac{\xi y}{x\sqrt{|1-x|}}. \label{eigenC1}
\end{equation}
The normalizable condition \eqref{ybc} requires that the quantum number $\xi$ to be imaginary $\xi=ir$, where $r$ is a real number.
Note here that \eqref{eigenC1} are the eigenfunctions of $C_1$, but the eigenvalues for $x\in(0,1)$ and $x\in(1,2)$ are different:  $\xi^{2}$ in $x\in(0,1)$, while $-\xi^{2}$ in $x\in(1,2)$.

For the quartic Casimir $C_3$, it is difficult to work with it directly. The idea is using the result for $C_{1}$ to decompose the Hilbert space into smaller ones. That is, using the above basis $\psi_{\xi}$ to write,
\begin{equation}
f(x,y)=\int_{-i\infty}^{+i\infty}d\xi\hat{f}_{\xi}(x)\psi_{\xi}(x,y).
\end{equation}
Now, we only need to decompose individual $\hat{f}_{\xi}(x)$. Substituting $\hat{f}_{\xi}(x)\psi_{\xi}(x,y)$ into the equation \eqref{C4}, we get a much simplified equation:
\begin{equation}
\hat{C_{3}}\hat{f}_{\xi}(x)=[x^{2}(x-1)\partial_{x}^{2}+\frac{1}{2}x(2+x)\partial_{x}-1]\hat{f}_{\xi}(x)=(1-\Delta)^{2}\hat{f}_{\xi}(x)\label{hatC4}
\end{equation}
where $\hat{C}_3$ means the reduced action on the function $\hat{f}_\xi(x)$. The dependence on $y$ and $\xi$ disappear in the above equation, and the equation reduces to a second order ordinary differential equation. Actually, $\hat{C_{3}}=-\,\hat{C_{2}^{2}}$, where,
\begin{equation}
\hat{C_{2}}=\sqrt{1-x}x\partial_{x}+\frac{x-2}{2\sqrt{1-x}}. \label{hatC2}
\end{equation}
This $\hat{C_{2}}$ can also be obtained by substituting $\hat{f}_{\xi}(x)\psi_{\xi}(x,y)$ into the eigen-equation of ${C}_{2}$: once again the $y$ dependent terms cancel mutually.

From the second order differential equation \eqref{hatC4} we again have a Sturm-Liouville problem, so the measure can be worked out:
\begin{equation}
\widetilde{\mu}(x)=\frac{1}{x^2(x-1)}\exp\int\frac{2+x}{2x(x-1)}dx=\frac{\sqrt{|1-x|}}{x^{3}}.
\end{equation}
Note that here we use $\widetilde{\mu}$ because this is not the final measure. Also, we drop the integration constant because it is not important for a measure. Strictly speaking, there could be a difference between $x\in(0,1)$ and $x\in(1,2)$ up to a multiplicative constant, but this will reduce to the difference in the matching condition at $x = 1$, so it does not matter.

To find the total measure, we need to write down the Hermitian condition for $C_{3}$ explicitly:
\begin{equation}
\begin{split} & ( C_{3}f,g)-( f,C_{3}g) \\
 & =\int_{0}^{2}dx\mu(x)\int_{-\infty}^{+\infty}dy[g(x,y)C_{3}f^{*}(x,y)-f^{*}(x,y)C_{3}g(x,y)]\\
 & =\int_{0}^{2}dx\mu(x)\int_{-\infty}^{+\infty}dy\int_{-i\infty}^{+i\infty}d\xi'\int_{-i\infty}^{+i\infty}d\xi''\psi_{\xi'}^{*}(x,y)\psi_{\xi''}(x,y)[\hat{C_{3}}\hat{f}_{\xi'}^{*}(x)\hat{g}_{\xi''}(x)-\hat{C_{3}}\hat{g}_{\xi''}(x)\hat{f}_{\xi'}^{*}(x)]\nn
\end{split}
\end{equation}
Substituting \eqref{eigenC1} into the above relation, and write $\xi'=is$, $\xi''=ir$, we find that the right-hand side becomes:
\begin{equation}
\begin{split} & \int_{-\infty}^{+\infty}dy\exp{i(r-s)\frac{y}{x\sqrt{|1-x|}}}\int_{0}^{2}dx\mu(x)\int_{-\infty}^{+\infty}ds\int_{-\infty}^{+\infty}dr[\hat{C_{3}}\hat{f}_{is}^{*}(x)\hat{g}_{ir}(x)-\hat{C_{3}}\hat{g}_{ir}(x)\hat{f}_{is}^{*}(x)]\\
 & =\int_{0}^{2}dx\mu(x)x\sqrt{|1-x|}\int_{-\infty}^{+\infty}ds\int_{-\infty}^{+\infty}dr\delta(r-s)[\hat{C_{3}}\hat{f}_{is}^{*}(x)\hat{g}_{ir}(x)-\hat{C_{3}}\hat{g}_{ir}(x)\hat{f}_{is}^{*}(x)].\nn
\end{split}
\end{equation}
So we have:
\begin{equation}
\mu(x)x\sqrt{|1-x|}=\widetilde{\mu}(x)=\frac{\sqrt{|1-x|}}{x^{3}}.
\end{equation}
Our final result for the measure is simply:
\begin{equation}
\mu(x)=\frac{1}{x^{4}},
\end{equation}
which can be actually obtained from the one in CFT$_2$ by taking the non-relativistic limit.

Next, we want to determine the Galilean conformal partial waves (GCPWs). Similar to the CFT case, this requires to find all the solution with same eigenvalues of $C_1$ and $C_3$. For given eigenvalues of quadratic and quartic Casimirs, there are four independent solutions:
\begin{equation}
\chi_{\Delta,\xi},\qquad\chi_{2-\Delta,\xi},\qquad\chi_{\Delta,-\xi},\qquad\chi_{2-\Delta,-\xi}\label{4chi}
\end{equation}
where,
\begin{equation}
\chi_{\Delta,\xi}=\frac{x^{\Delta}(1+\sqrt{1-x})^{2-2\Delta}}{\sqrt{1-x}}e^{\frac{-\xi y}{x\sqrt{1-x}}}.\label{chi}
\end{equation}
These four solutions  are related by the symmetry of the eigenvalues: $\Delta\leftrightarrow2-\Delta$ and $\xi\leftrightarrow-\xi$. More precisely, due to the fact that $\hat{C_{3}}=-\hat{C_{2}^{2}}$, to obtain these four solutions we only need to solve the eigen-equation of $\hat{C}_{2}$, i.e. Eq. \eqref{hatC2}, together with the $y$ dependent part \eqref{eigenC1}.

The above solutions should be taken with care. First of all, because the $\sqrt{1-x}$ factor is double-valued, we need to specify one of them. In fact, the other choice correspond to another independent solution which is just $\chi_{2-\Delta,-\xi}$. This is similar to 1D conformal block, where there are two branch points at $0$ and $1$. 
Secondly, the above solutions are not valid in the entire region $x\in(0,2)$, because there is a singular point at $x=1$, which is a branch point as well. Nevertheless, the above expressions of solutions are valid in the interval $(0,1)$ and $(1,2)$ separately. The point here is, for example, if we have a solution \eqref{chi} for $x\in(0,1)$, then continue it to $x\in(1,2)$, which means the matching between $x=1^{+}$ and $x=1^{-}$, we can not find \eqref{chi} anymore.  Instead, we will find a solution which is  a linear combination of the four solutions in \eqref{4chi}. Just like in the SYK model, the analytic continuation form $x<1$ either below or above the real axis to $x>1$ will not give a solution we need in $1<x<2$, only one of their combination does the job.

To obtain the GCPWs of CGA, we need to consider the solutions in different regions and match them properly. Let us first analyze the region $x\in(1,2)$ case. From the above analysis, we know that \eqref{4chi} in $x\in(1,2)$ are different from the ones in $x\in(0,1)$,  so for $x\in(1,2)$ we label these solutions as
\begin{equation}
\chi'_{\Delta',\xi'},\qquad\chi'_{2-\Delta',\xi'},\qquad\chi'_{\Delta',-\xi'},\qquad\chi'_{2-\Delta',-\xi'}\label{4chip}
\end{equation}
where
\begin{equation}
\chi'_{\Delta',\xi'}=\frac{x^{\Delta'}(1+i\sqrt{x-1})^{2-2\Delta'}}{\sqrt{x-1}}e^{\frac{i\xi' y}{x\sqrt{x-1}}}. \label{chip}
\end{equation}
Then the GCPWs in $x\in(1,2)$ are:
\begin{equation}
\Phi_{\Delta',\xi'}=a_{1}\chi'_{\Delta',\xi'}+a_{2}\chi'_{2-\Delta',\xi'}+a_{3}\chi'_{\Delta',-\xi'}+a_{4}\chi'_{2-\Delta',-\xi'},\qquad1<x<2. \label{CPWx2}
\end{equation}
Because of the symmetry \eqref{symmetry}, we get the boundary condition at $x=2$:
\begin{equation}\label{bc1}
\Phi_{\Delta',\xi'}(2,y)=\Phi_{\Delta',\xi'}(2,-y),
\end{equation}
which leads to
\begin{equation}
a_{1}=a_{3},\qquad a_{2}=a_{4}.
\end{equation}
Notice that $x=2,\,y=0$ is a fixed point, so we have another boundary condition\footnote{This boundary condition is similar with $f'(2)=0$ in \cite{Maldacena:2016hyu}.}:
\begin{equation}\label{bc2}
\partial_{x}\Phi_{\Delta',\xi'}(2,0)=0,\nn
\end{equation}
which leads to
\begin{equation}
a_{1}(1+i)^{2-2\Delta'}=a_{2}(1-i)^{2-2\Delta'}.\nn
\end{equation}
After setting $a_{1}=1$, we finally obtain
\begin{equation}
a_{1}=a_{3}=1, \qquad a_{2}=a_{4}=e^{i\pi(1-\Delta')}.
\end{equation}
Note that we can not write $e^{i\pi(1-\Delta')}=(-1)^{i\pi(1-\Delta')}$ because of the multi-valuedness.

Next, we have to continue the GCPWs to $0<x<1$ with the matching condition at $x=1$
\begin{equation}
\Phi_{\Delta,\xi}(1^{-},y)=\Phi_{\Delta',\xi'}(1^{+},y).
\end{equation}
As $x\rightarrow1$, we set $|\sqrt{1-x}|=\epsilon\rightarrow0$ in the following. For $x\rightarrow 1^+$, we find
\begin{equation}
\chi'_{\Delta',\xi'}(1^{+},y)=\frac{(1+i\epsilon)^{2-2\Delta'}}{\epsilon}e^{\frac{i\xi' y}{\epsilon}},
\end{equation}
then the GCPWs as $\epsilon\rightarrow0$
\bea
\Phi_{\Delta,\xi}(1^+,y)
 & =&(a_{1}+a_{2})\frac{1}{\epsilon}e^{\frac{i\xi' y}{\epsilon}}+(a_{3}+a_{4})\frac{1}{\epsilon}e^{\frac{-i\xi' y}{\epsilon}}\nn\\
 && +(a_{2}-a_{1})(2-2\Delta')ie^{\frac{i\xi' y}{\epsilon}}+(a_{4}-a_{3})(2-2\Delta')ie^{\frac{-i\xi' y}{\epsilon}}+\cdots.
\eea
Here we have neglected the  higher order terms, which vanish in the $\epsilon\rightarrow0$ limit.\footnote{Note that the exponential part is oscillatory and bounded, due to the normalizable condition in the $y$ direction \eqref{ybc}. This is true for both $x\in(1,2)$ and $x\in(0,1)$.} On the other hand, we write the GCPWs in $0<x<1$ as:
\begin{equation}
\Phi_{\Delta,\xi}=b_{1}\chi_{\Delta,\xi}+b_{2}\chi_{2-\Delta,\xi}+b_{3}\chi_{\Delta,-\xi}+b_{4}\chi_{2-\Delta,-\xi},\qquad0<x<1.
\end{equation}
When $x\rightarrow 1^-$, we have the GCPWs
\bea
\Phi_{\Delta,\xi}(1^-,y)
 & =&(b_{1}+b_{2})\frac{1}{\epsilon}e^{\frac{-\xi y}{\epsilon}}+(b_{3}+b_{4})\frac{1}{\epsilon}e^{\frac{\xi y}{\epsilon}}\nn\\
 &&+(b_{2}-b_{1})(2-2\Delta)e^{\frac{-\xi y}{\epsilon}}+(b_{4}-b_{3})(2-2\Delta)e^{\frac{\xi y}{\epsilon}}+\cdots.
\eea
The matching of the GCPWs at $x=1$ leads to the identification of the exponential factor
\begin{equation}
\xi'=i\xi\quad \mbox{or}\quad-i\xi,\qquad\Delta'=\Delta\quad \mbox{or}\quad2-\Delta.
\end{equation}
In fact, this relation just reflects the normalizable condition in $y$ direction, where $\xi\in \mathbb{R}$ for $x\in(0,1)$ and $\xi\in i\mathbb{R}$ for $x\in(1,2)$. The identification of the coefficients before the exponentials gives
\begin{equation}
\begin{split} & a_{1}+a_{2}=b_{1}+b_{2},\quad \quad i(a_{1}-a_{2})=b_{1}-b_{2},\\
 & a_{3}+a_{4}=b_{3}+b_{4},\quad \quad i(a_{3}-a_{4})=b_{3}-b_{4}.\nn
\end{split}
\end{equation}
Finally we get the coefficients
\begin{equation}
\begin{split} & b_{1}=b_{3}=\frac{1+i}{2}+\frac{1-i}{2}e^{i\pi(1-\Delta)},\\
 & b_{2}=b_{4}=\frac{1-i}{2}+\frac{1+i}{2}e^{i\pi(1-\Delta)},
\end{split}
\end{equation}
and the GCPWs
\begin{equation}
\Phi_{\Delta,\xi}(x,y)=
\begin{cases}
\chi'_{\Delta,\xi}+\chi'_{\Delta,-\xi}+e^{i\pi(1-\Delta)}(\chi'_{2-\Delta,\xi}+\chi'_{2-\Delta,-\xi}),\hs{3ex}\mbox{for $1<x<2$,}\\ \\
b_1(\chi_{\Delta,\xi}+\chi_{\Delta,-\xi})+b_2(\chi_{2-\Delta,\xi}+\chi_{2-\Delta,-\xi}),\hs{3ex}\mbox{for $0<x<1$}.
\end{cases}
\end{equation}
where $\chi'_{\Delta,\xi}$ and $\chi_{\Delta,\xi}$ are given in \eqref{chip} and \eqref{chi} respectively. As before, now using the normalizable condition to determine $\Delta$, we find that the solutions have the power law behaviour near $x=0$:
\begin{equation}
\chi_{\Delta,\xi}\sim\chi_{\Delta,-\xi}\sim x^{\Delta},\qquad\chi_{2-\Delta,\xi}\sim\chi_{2-\Delta,-\xi}\sim x^{2-\Delta}.
\end{equation}
Because $\widetilde{\mu}(x)\sim x^{-3}$, so to make GCPWs normalizable, one way is to allow all four terms. This leads to
\begin{equation}
\begin{cases}
2\Re\Delta-3\geq-1,\\
2(2-\Re\Delta)-3\geq-1.
\end{cases}
\end{equation}
The only solution to the above two constraints is $\Re\Delta=1$, that is:
\begin{equation}
\Delta=1+is, \hs{3ex}s\in \mathbb{R}.
\end{equation}
This is the usual quantum numbers of the principal series representation. Just like the SYK model, each term is marginally allowable. Another possible way to have normalizable CPWs is to let two of four terms vanish, namely, let $b_{1}=b_{3}=0$ or $b_{2}=b_{4}=0$. This leads to the discrete series:
\begin{equation}
\Delta=\frac{5}{2}+2n\qquad\text{or}\qquad\Delta=-\frac{1}{2}-2n\qquad n=0,1,2,\cdots.
\end{equation}
Note that the $(-\frac{1}{2})$-series is linearly dependent on the $\frac{5}{2}$-series, due to the symmetry $\D \leftrightarrow 2-\D$. This $\mathbb{Z}_{2}$ symmetry also appear in the principal series: $\Psi_{\Delta,\xi}$ and $\Psi_{2-\Delta,\xi}$ are linearly dependent. Besides, there is  another symmetry $\xi\leftrightarrow-\xi$ for $\Psi_{\Delta,\xi}$ in both case. Moreover, it is easy to see that  for both $\xi=ir$, $\Delta=1+is$ and $\xi=ir$, $\Delta=-1/2-2n$, the eigenvalues of the Casimirs are real.

To compare with CFT, the GCPWs can be written in a more symmetric form by multiplying a factor $e^{\frac{i\pi}{2}(\Delta-1)}$.
\begin{equation}
\Psi_{\Delta,\xi}(x,y)=
\begin{cases}
e^{\frac{i\pi}{2}(\Delta-1)}(\chi'_{\Delta,\xi}+\chi'_{\Delta,-\xi})+e^{\frac{i\pi}{2}(1-\Delta)}(\chi'_{2-\Delta,\xi}+\chi'_{2-\Delta,-\xi}),\hs{3ex}\mbox{for $1<x<2$,}\\ \\
A(\Delta)(\chi_{\Delta,\xi}+\chi_{\Delta,-\xi})+A(2-\Delta)(\chi_{2-\Delta,\xi}+\chi_{2-\Delta,-\xi}),\hs{3ex}\mbox{for $0<x<1$}.
\end{cases}
\end{equation}
where
\begin{equation}
A(\Delta)=e^{\frac{i\pi}{2}(\Delta-1)}b_1=\sin\frac{\pi\Delta}{2}+\cos\frac{\pi\Delta}{2},
\end{equation}
\begin{equation}
A(2-\Delta)=e^{\frac{i\pi}{2}(\Delta-1)}b_2=\sin\frac{\pi\Delta}{2}-\cos\frac{\pi\Delta}{2}.
\end{equation}
From now on, we use this symmetric form of the GCPWs for discussion.

Note that in the CFT case as well as the SYK case, for the principal series we have an alternative expression called the shadow integral representations. In GCA case, it will be interesting to ask whether there exists an analogue of shadow representation for our principal series. This will be the subject of section 6.

Now, let us work out the orthogonality and the completeness of our GCPWs. Firstly, because of the $\mathbb{Z}_{2}$ symmetry:
\begin{equation}
\Delta\leftrightarrow2-\Delta,\qquad\xi\leftrightarrow-\xi,
\end{equation}
the quantum number of our complete basis can be chosen to be: $\xi=ir,\,\Delta=1+is$ or $\D=-1/2-2n$ with $r,s\in \mathbb{R},\,n\in \mathbb{N}$. Now we determine the normalization factor in the inner product. For the principal series, we expect the inner product $(\chi_{\Delta,\xi},\chi_{\Delta',\xi'}) $ to be proportional to $\delta(r-r')\delta(s-s')$, so we only need to consider the singular part of the inner product, which comes from the integral over the small $x$ region, as the $y$ part  just gives $2\pi x\sqrt{|1-x|}\delta(r-r')$. Then it turns out to be
\begin{equation}
(\Psi_{1+is,ir},\Psi_{1+is',ir'}) =4\pi^2 N\delta(r-r')\delta(s-s'),
\end{equation}
where
\begin{equation}
N=-\cos\pi\Delta=A(\Delta)A(2-\Delta).
\end{equation}
For the discrete series, we can calculate the normalization factor directly:
\begin{equation}
(\Psi_{\Delta,\xi},\Psi_{\Delta',\xi'}) =N'\delta(r-r')\delta_{nn'},
\end{equation}
where
\begin{equation}
N'=\pi=A(\Delta)\partial_\Delta A(2-\Delta).
\end{equation}

Of course, the principal and the discrete series are orthogonal to each other:
\begin{equation}
(\Psi_{1+is,ir},\Psi_{-\frac{1}{2}-2n,ir'}) =0.
\end{equation}
For the orthogonality, here we cannot simply follow the fact that for the Hermitian operators the eigenfunctions with different eigenvalues are orthogonal to each other, because our GCPWs are not the eigenfunctions of the Casimirs in the whole region. As mentioned before, the GCPWs are the eigenfunctions only for $x\in(0,1)$ and $x\in(1,2)$ separately, but with different eigenvalues in these two regions. Nevertheless, based on this fact and just follow our analysis step by step one can easily find that our GCPWs are indeed orthogonal with each other.  The completeness relation is then:
\bea
\frac{1}{4\pi^2}\lefteqn{ \int_{0}^{+\infty}ds\int_{0}^{+\infty}dr\frac{1}{N}\Psi_{1+is,ir}(x,y)\Psi_{1+is,ir}(x',y')}\nn\\
 & &+\sum_{n=1}^{\infty}\int_{0}^{+\infty}dr\frac{1}{N'}\Psi_{-\frac{1}{2}-2n,ir}(x,y)\Psi_{-\frac{1}{2}-2n',ir}(x',y') =x^{4}\delta(x-x')\delta(y-y').
\eea

\subsection{Read Data from Inversion}

A four-point function admits global block expansion in which the expansion coefficients contain the data of the theory. It admits  the GCPW expansion as well, where the expansion coefficients can be obtained by using the inversion formula. The two expansions are related by the contour deformation. In this section, we want to explain how to get the date from the inversion formula.

The standard (Euclidean) inversion formula method takes the inner product of the four-point function and the CPWs to obtain the inversion function. After analytic continuation and contour deformation, one can read the spectrum and the OPEs from the poles and the residues of this inversion function. 

The contour deformation analysis is still valid in the GCFT case. Here we would like to point out some novel features  in the GCFT case. Firstly, we note that though our harmonic analysis include two regions, $0<x<1$ and $1<x<2$, we  can actually use any one of them to find the block expansion. In practice, we will work in the region $0<x<1$ in the following. Now we write the GCPWs expansion of a four-point function:
\begin{equation}
\mathcal{G}(x,y)=\int_{0}^{\infty} dr\frac{1}{4\pi^2 }\int_{0}^{\infty} ds  \frac{1}{N}\Psi_{\Delta,\xi}(x,y)(\Psi_{\Delta,\xi},\mathcal{G})+\int_{0}^{\infty} dr\sum_{n}\frac{1}{N'}\Psi_{\Delta,\xi}(x,y)(\Psi_{\Delta,\xi},\mathcal{G}).
\end{equation}
As we will show shortly, one can double the integral region and then try to use a contour deformation and find  that the contributions from  the arc at infinity (to the right hand side) can be dropped so that the above integral become a sum of residues which located at physical poles. 

Secondly, the $\xi=0$ part  should be treated separately as we mentioned in section 2.2. In this paper, we only deal with theories with no $\xi=0$ spectrum, for example, the Generalized Galilean Free Field Theories in section 4. For the treatment of the $\xi=0$ case, see \cite{Hao:2021b}.

Thirdly, the multiplets appear as the multiple poles in the inversion function. As we have already  shown, the multiplets are essential existence in GCFTs.  But only the singlet blocks are the eigenfunctions of the Casimirs, and the singlets appear as simple poles in the inversion function just as in the CFT case. So how to see the multiplets  in the inversion formula? The answer is that they appear as multiple poles. This can be understood by considering Fourier analysis. The analog of a conformal block is $e^{st}$ with $s$ real, which is generally not in the Hilbert space of the Fourier analysis\footnote{With $s$ pure imaginary, $e^{st}$ is a complete basis, so is the analog of CPW or GCPW. }. In fact, our GCA global blocks as well as  physical four-point functions are in the same case: they are generally not in the Hilbert space of the GCA harmonic analysis. To discuss these more general functions, we need to extend Fourier transform to  Laplace transform.  Here is a toy model for the inversion method: for a single ``conformal block'' $f_{s_0}(t)\equiv e^{s_0t}$, after the unilateral Laplace transform
\begin{equation}
F(s)=\int_0^\infty e^{-st}f_{s_0}(t)dt,
\end{equation}
 one get $F(s)=\frac{1}{s-s_0}$. So it appears as a simple pole at $s=s_0$, which means that we can read the spectrum $s=s_0$ directly (of course here the residue, which is related to the OPE, is 1).  For the GCA case, the multiplets involve additionally ``multi-blocks'':
 \begin{equation}
\partial_s^ne^{s_0t}=t^ne^{s_0t},
\end{equation}
  which corresponds to a multiple pole after the Laplace transform
\begin{equation}
\frac{n!}{(s-s_0)^{n+1}}.\nn
\end{equation}
 In fact, this is exactly what happens in the GCA inversion because the  block of multiplets is related to the global block by
 \be
 g_{\D,\xi}^{(n)}=\p^n_\xi g^{(0)}_{\D,\xi}, \hs{3ex}g^{(0)}_{\D,\xi}\propto \exp\left(-\xi\frac{y}{x\sqrt{(1-x)}}\right), \nn
 \ee
  such that $\xi$ plays the role of $s$ and $\frac{y}{x\sqrt{(1-x)}}$ plays the role of $t$ in the Laplace transform. Moreover, we observe that in order to get the block expansion, we only do the unilateral Laplace transform, which means we only need to use a half of the integral region. The full integral (bilateral Laplace transform) is actually divergent for a single block $e^{s_0t}$, but since we only concern its analytic structure, the unilateral Laplace transform is enough.  This is also the case for an GCA four-point function. In the next section when we discuss GGFT, we will use this logic and show explicitly the independence of the  integral region.

\par  Now let us make the above analysis more explicitly. For a stripped four point function $\mathcal{G}(x,y)$, we first use the GCA inversion formula to get a GCA inversion function:
\begin{equation}
I(\Delta,\xi)=(\Psi_{\Delta,\xi},\mathcal{G})
\end{equation}
Using the above GCA inversion formula, one gets the expansion with respect to the GCPWs.
the global block expansion is
\begin{equation}\mathcal{G}(x,y)=\sum_{\Delta,\xi,k}P_{\Delta,\xi,k}\partial_{\xi}^kg_{\Delta,\xi}^{(0)} ,\ \ \ x<1.
\end{equation}
Now we would like to compare the  block expansion and the GCPW decomposition for $\mathcal{G}(x,y)$ in the region $x<1$ to get the relation between $I(\Delta,\xi)$ and $P_{\Delta,\xi,k}$. For the GCPW expansion, we have
\begin{equation}
\mathcal{G}(x,y)=\int_{0}^{\infty} dr\frac{1}{4\pi^2 }\int_{0}^{\infty} ds  \frac{1}{N}\Psi_{\Delta,\xi}(x,y)(\Psi_{\Delta,\xi},\mathcal{G})+\int_{0}^{\infty} dr\sum_{n}\frac{1}{N'}\Psi_{\Delta,\xi}(x,y)(\Psi_{\Delta,\xi},\mathcal{G}),
\end{equation}
which contains the contribution to the poles in the inversion function 
\begin{figure}\centering
 \subfloat[$\D$-plane]{\includegraphics[width=6cm]{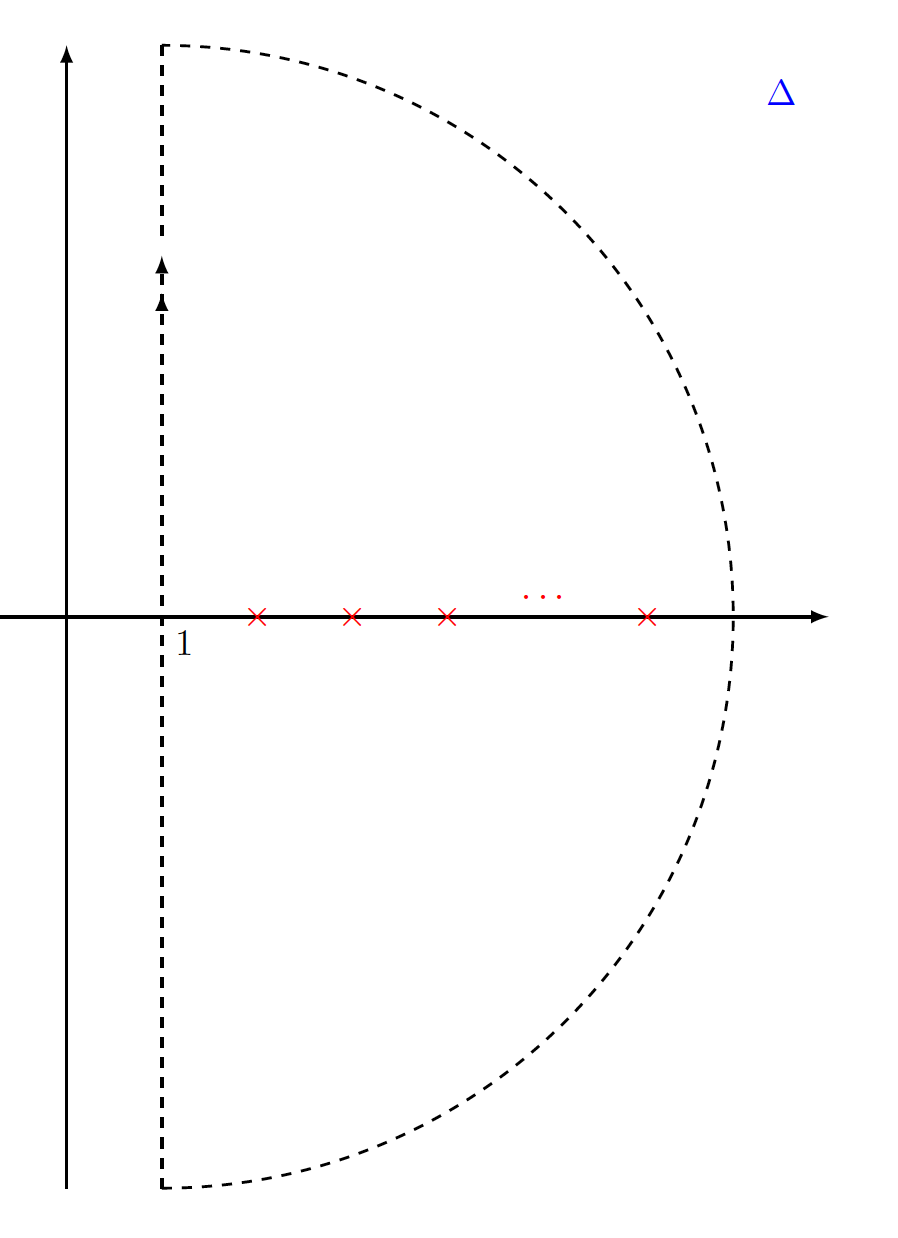}\label{Dcontour}}
\subfloat[$\xi$-plane]{\includegraphics[width=6.5cm]{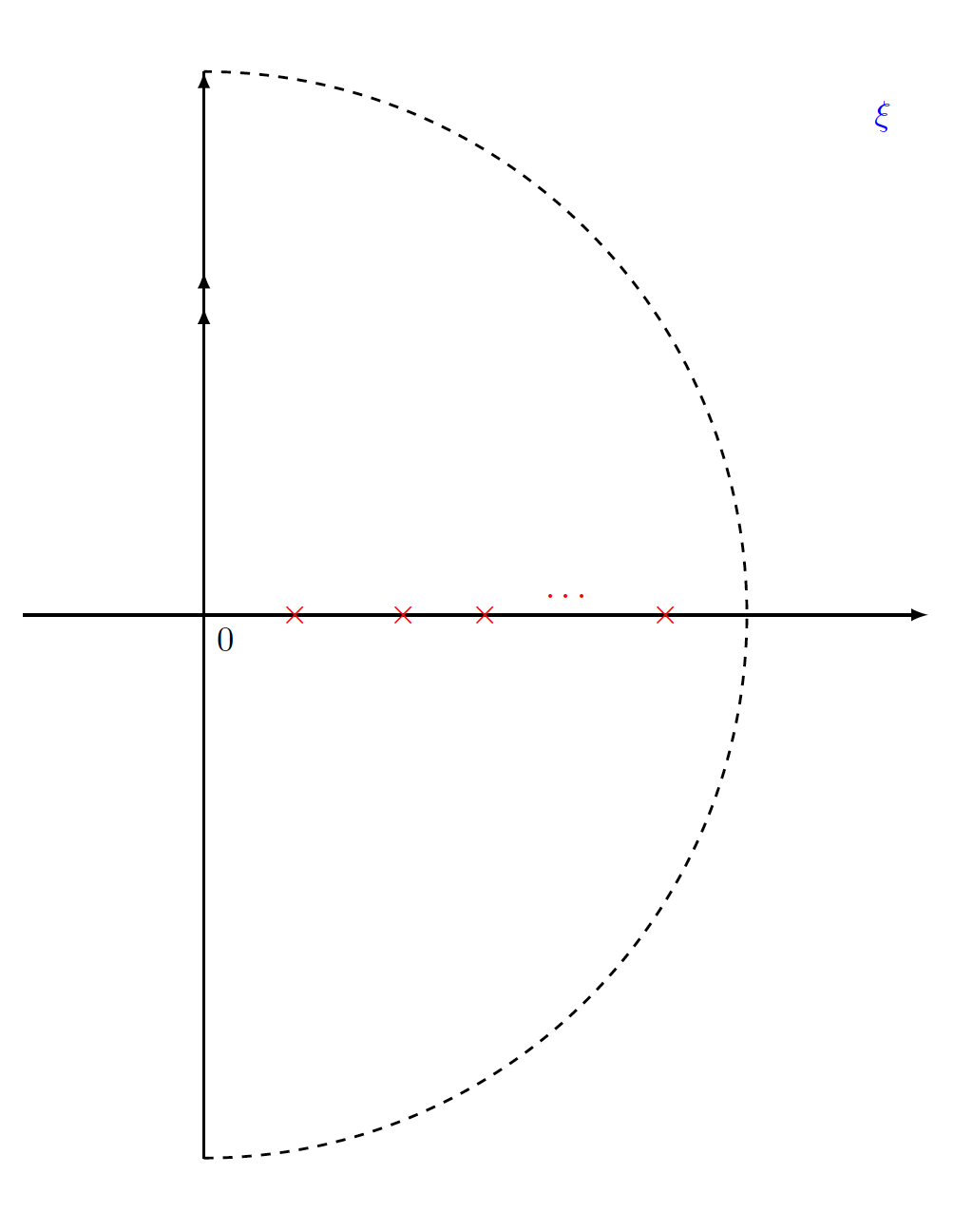}\label{Xicontour}}
\caption{The contours in the $\D$-plane and $\xi$-plane. }
\end{figure}
We want to move the integral contour on the complex $\Delta$-plane and complex $\xi$-plane.
We need  fix $\xi$ firstly to get the complex $\D$-plane, and choose the contour as in Fig.\ref{Dcontour} : the contour goes clockwise along the line $\Re (\D)=1$ and the arc at infinity. This will help us to read the potential poles corresponding to the physical spectrum. Then we may fix $\D$ to work on complex $\xi$-plane, and choose the contour as in Fig. \ref{Xicontour}: the contour goes clockwise along the line $\Re (\xi)=0$ and the arc at infinity. This will pick up the physical poles corresponding to $\xi$ spectrum.
 However, it is not obvious that the contribution from the arc at infinity is always  under well control, considering  all  four terms in the GCPWs. To make sure these contour integral are always well-defined, we need to use the symmetry of the GCPWs under $\Delta\rightarrow 2-\Delta$, $\xi\rightarrow -\xi$ to re-express the above expansion. By expressing it from the $\chi_{\Delta,\xi}$ term, the contribution from the principal series part in $\mathcal{G}(x,y)$ is
\begin{equation}\label{ww2}
\frac{1}{(2\pi i)^2}\int_{C_\xi} d\xi\int_{C_\D} d\Delta \frac{1}{N} I(\Delta,\xi)A(\Delta) \chi_{\Delta,\xi}=\frac{1}{(2\pi i)^2}\int_{C_\xi} d\xi\int_{C_\D} d\Delta I(\Delta,\xi) \chi_{\Delta,\xi}\frac{1}{A(2-\Delta)}\nn
\end{equation}
where each contribution from $\chi_{\Delta,-\xi}$ and $\chi_{2-\Delta,\pm\xi}$ is the same as the one from $\chi_{\Delta,\xi}$. This allows us to double the integral region twice. At this moment, the integral is on the complex $\Delta$ plane and complex $\xi$ plane. The relation $\Delta=1+is,\xi=ir$ gives additional $\frac{1}{i^2}$ as the pre-factor of the integrand.

Note that the factor $A(2-\Delta)$ has simple zeros at $\Delta=\Delta_n$, which gives the discrete spectrum. Deforming the integral contour, one also picks up the residues at these simple poles of $\frac{1}{A(2-\Delta)}$. The residue reads,
\begin{equation}
\sum_n \Res|_{\Delta=\Delta_n}\frac{1}{A(2-\Delta)}=\sum_n\frac{1}{\partial_\Delta A(2-\Delta)}|_{\Delta=\Delta_n}.\nn
\end{equation}
Considering the direction of the contour, this cancels the contribution from the discrete spectrum which is
\begin{equation}
\frac{1}{2\pi i}\int_{C_\xi} d\xi\sum_{n}\frac{1}{\partial_\Delta A(2-\Delta)}\chi_{\Delta,\xi}(x,y)(\Psi_{\Delta,\xi},\mathcal{G}).\nn
\end{equation}
Combining them, one gets the contribution from the physical spectrum as follows,
\begin{equation}
\mathcal{G}(x,y)=-\sum_{l}\Res|_{\xi_l}\sum_{m} \Res|_{\Delta_m}\frac{1}{A(2-\Delta)} I(\Delta,\xi) \chi_{\Delta,\xi},\ \ x<1,
\end{equation}
where $I(\Delta,\xi)$ has poles at $\xi=\xi_l$ and $\Delta=\Delta_m$, which constitute the physical spectrum. On the other hand,
the global blocks of multiplets read
\begin{equation}
g^{(k)}_{\Delta,\xi}=2^{2\Delta-2}\partial_{\xi}^k\chi_{\Delta,\xi},
\end{equation}
so the singular part in $I(\Delta,\xi)$ reads
\begin{equation}
I(\Delta,\xi)\sim-\sum_{\Delta_m,\xi_l,k}A(2-\Delta) \Gamma(k+1)\frac{2^{2\Delta_m-2}}{(\xi-\xi_l)^{k+1}}\frac{P_{\Delta_m,\xi_l,k+1}}{\Delta-\Delta_m}+\text{shadow poles}.
\end{equation}
We see that generically, there are multiple poles in $\xi$ and single poles in $\Delta$ in the inversion function. What's more, there are shadow poles in the inversion function. Consider
\bea
I(\Delta,\xi)&=&(\Psi_{\Delta,\xi},\mathcal{G})\nn\\
&=&A(2-\Delta)(\chi_{\Delta,\xi},\mathcal{G})+A(2-\Delta)(\chi_{\Delta,-\xi},\mathcal{G})+A(\Delta)(\chi_{2-\Delta,\xi},\mathcal{G})+A(\Delta)(\chi_{2-\Delta,-\xi},\mathcal{G}),\nn
\eea
one do not know {\it a priori} which of the four terms contributes to the physical poles. If $(\mathcal{G},\chi_{\Delta,\xi})$ contributes to the physical poles, it should have the following singular behaviour,
\begin{equation}
(\Psi_{\Delta,\xi},\mathcal{G})\sim-\sum_{\Delta_m,\xi_l,k} \Gamma(k+1)\frac{2^{2\Delta_m-2}}{(\xi-\xi_l)^{k+1}}\frac{P_{\Delta_m,\xi_l,k+1}}{\Delta-\Delta_m}
\end{equation}
We can see in the next section that this is the case for the GGFTs.

\subsection{Limiting Analysis}
The GCFT$_2$ can be seen as the non-relativistic limit of the CFT$_2$, which has the space time coordinates $z=x+y,\bar{z}=x-y$ \cite{Bagchi:2009pe,Bagchi:2017cpu,Bagchi:2019unf},
\begin{equation}
x\rightarrow x,\ \ \ y\rightarrow \epsilon y,\ \ \ \epsilon\rightarrow 0.
\end{equation}
This is to take the speed of light $c\rightarrow \infty$\footnote{There are actually two distinct contractions both leading to GCA from 2d conformal algebra. However, as explained in \cite{Bagchi:2017cpu},  only the non-relativistic limit  is relevant to the highest weight representations. We will focus on the non-relativistic contraction.}. In this sense, it is called the non-relativistic limit. Under this limit, the generators behave as,
\begin{equation}
L_n=\cL_n+\bar{\cL}_n,\ \ \ M_n=\epsilon(\cL_n-\bar{\cL}_n)
\end{equation}
where $\cL_n, \bar{\cL}_n$ are the generators of the two copies of Virasoro algebra of CFT$_2$,  and the Virasoro central charges $c,\bar{c}$ become,
\begin{equation}
c_L=\frac{c+\bar{c}}{12}, \ \ c_M=\frac{\epsilon(c-\bar{c})}{12}.
\end{equation}
Starting with the highest weight representation in CFT$_2$, one gets the highest weight representation in GCFT$_2$ after taking the non-relativistic limit. The weights and charges of operators are related as follows,
\begin{equation}
\Delta=h+\bar{h},\ \ \xi=\lim_{\epsilon\rightarrow 0}\epsilon(h-\bar{h})=\lim_{\epsilon\rightarrow 0}\epsilon (\pm J), \label{Jlimit}
\end{equation}
where $h,\bar{h}$ are (anti-)holomorphic weight of the operators in CFT$_2$.

Consequently, many kinematic quantities can be obtained by this limiting method, including the Casimirs, the two-point functions, three-point functions, the null vectors and the global conformal blocks of singlets, etc.  Interestingly, we will show in the next section that this limiting method can even be valid in the dynamical level, namely, it can be used to obtain the results on GGFT. Particularly, one can easily see that how the multiplets appear after taking the limit. On the other hand, when we try to revisit the harmonic analysis by using the limiting method, surprising things happen: we find that it is impossible to reach the result  of GCA harmonic analysis, though the symmetry is preserved well in the limiting procedure.

Now, we turn to discuss the harmonic analysis of GCA from the point of view of taking the non-relativistic limit. We will show that it is in principle impossible to obtain the GCA harmonic analysis result by taking the limit. The reason is, in a word, that the Hilbert space in the GCA harmonic analysis can not be obtained from the one in 2d conformal harmonic analysis. We will discuss two different aspects of this claim, both related to the definition of the Hilbert space. The first one is the normalizable condition, which determine the quantum number of GCPWs.  The other one is the boundary condition.\footnote{Note that all the previous results (for example, the ones in \cite{Bagchi:2009pe,Bagchi:2017cpu,Bagchi:2019unf}) that can be obtained by the limiting method do not involve the boundaries.} 

\par Before explaining the two unsuccessful aspects of the limiting method, let us note that there is actually something still valid.  After taking the limit, the measure becomes
\begin{equation}
\mu(x,y)=\lim_{\epsilon\to 0}\frac{1}{z^2\bar{z}^2}=\lim_{\epsilon\to 0}\frac{1}{(x+\epsilon y)^2(x-\epsilon y)^2}=\frac{1}{x^4},
\end{equation}
which perfectly match with the one in GCA harmonic analysis. Notice that the measure is independent of the normalizable condition as well as the boundary condition.

It is most obvious to see the ineffectiveness of the limiting method by considering the quantum numbers.
For the quantum number of the GCPWs, the principal values of $\Delta$ are actually the same as the ones in 2d conformal harmonic analysis.   However,  it is impossible to obtain the discrete values of $\D$ or the quantum number of $\xi$. Let us explain this point in detail. Firstly, a function $\mathcal{G}(z,\bar{z})$ in the Hilbert space of the 2d conformal harmonic analysis has the following CPW decomposition:
\begin{equation}
\mathcal{G}(z,\bar{z})=\sum_{J=0}^{\infty}\int_{1-i\infty}^{1+i\infty}\frac{d\Delta}{2\pi i}c(\Delta,J)F_{\Delta,J}(z,\bar{z}), \label{HA}
\end{equation}
Taking the limit on both sides, the question is whether or not that the right-hand side (RHS) will become the GCPW decomposition of the left hand side. We want to look at the quantum number on the RHS. From \eqref{Jlimit}, a finite $J$  always leads to $\xi=0$ after taking the limit. This is a trivial case because both sides will be $y$ independent under the limit. In order to obtain a non-trivial result, the spin appeared in the CPW
decomposition of $\mathcal{G}(z,\bar{z})$  must be of order $\frac{1}{\epsilon}$.
 Then after the limit, the function on the left-hand side (LHS) still has $y$ dependence and the quantum number $\xi$ on the RHS is nonvanishing and finite  (in fact, GGFTs in the next section are in this case).
There are several possible problems. The first one is: do CPWs have reasonable (finite) limiting results? We will show this is true in the next paragraph. Another subtle point is on the commutativity between doing the $J$ summation and doing the $\Delta$ integral in the $\epsilon\to 0$ limit. Even if we ignore the above subtle issues, it seems impossible to obtain a continuous spectrum of $\xi$ by taking the limit. No matter what Hilbert space we use, the spectrum of $\xi$ should be  continuous because the $y$ direction is not compact.  As the spin is discrete, it seems impossible to translate the $J$ summation in the RHS of \eqref{HA} into an integral of $\xi$ after taking the limit. In fact, the true principal value of $\xi$ should be pure imaginary  which originates  from the normalizable condition of the $y$ direction. Moreover, the possible discrete series of $\Delta$, which does not appear in a 2d CFT at all, cannot appear  from taking the limit on the RHS of \eqref{HA} either. This discreteness also originates from the normalizable condition for the $x$ direction.

\par Next, let us turn to the boundary conditions for GCPWs, which will be more explicit. It has  already been known that a single conformal block can be obtained by taking the limit  \cite{Bagchi:2017cpu}. In fact, it turns out that the full CPWs is well
defined (finite) in the $\epsilon\to0$ limit. Recall that the CPW in a $d$ dimensional CFT takes the following form:
\begin{equation}
F_{\Delta,J}(z,\bar{z})=K_{\Delta,J}G_{\Delta,J}(z,\bar{z})+K_{d-\Delta,J}G_{d-\Delta,J}(z,\bar{z})
\end{equation}
where $G_{\Delta,J}(z,\bar{z})$ is the conformal block, and
\begin{equation}
K_{\Delta,J}=\frac{\Gamma(\Delta-1)}{\Gamma(\Delta-\frac{d}{2})}\kappa_{\Delta+J}, \qquad \kappa_\beta=\frac{\Gamma(\frac{\beta}{2}-a)\Gamma(\frac{\beta}{2}+a)\Gamma(\frac{\beta}{2}-b)\Gamma(\frac{\beta}{2}+b)}{2\pi^2\Gamma(\beta-1)
\Gamma(\beta)}.
\end{equation}
In the $d=2$ case, if we take the limit on the  CPWs, divergences appear in both the two terms. The point here is that only the relative ratio between the coefficients $\frac{K_{d-\Delta,J}}{K_{\Delta,J}}$ is meaningful, which reaches a finite value under the limit. We observe that all gamma functions in $\frac{K_{d-\Delta,J}}{K_{\Delta,J}}$ could be combined into  Beta functions:
\begin{equation}
B(b,c)\equiv\frac{\Gamma(b)\Gamma(c)}{\Gamma(b+c)}=\int_0^1y^{b-1}(1-y)^{c-1}dy.
\end{equation}
Then in the limit $\epsilon\to0$, we can use the saddle point approximation for these Beta functions just as in \cite{Bagchi:2017cpu}, and find  that the relative ratio between the coefficients of  the block and its shadow is $1$. This is very different from the GCPWs in our intrinsic harmonic analysis.

 Now, we  explain what goes wrong in this story. In short, the origin of the mismatch stems from the boundary conditions. More precisely, unlike the measure, the computation of the GCPWs are closely related to the boundary conditions. In fact,  the differential equation \eqref{C1} and \eqref{C4} can be obtained from taking the limit \cite{Bagchi:2017cpu}, but we are not sure whether the boundary conditions can also be preserved in the limiting procedure. Let us look more carefully. Notice that unlike the ones in usual CFT,  the GCPWs in Galilean conformal theory are linear combinations of four blocks. This is not strange because $J=|h-\bar{h}|$, so the combination of $\xi$ block and $-\xi$ block correspond to a single block in CFT. It gives naturally the
$\xi\leftrightarrow-\xi$ symmetry of GCPWs in GCA. Moreover, in CFT$_2$ there is a symmetry
\begin{equation}
z\rightarrow\frac{z}{z-1} \qquad \bar{z}\rightarrow\frac{\bar{z}}{\bar{z}-1},  \label{ssss}
\end{equation}
which transforms into
\begin{equation}
\begin{split}
&2x=z+\bar{z}\rightarrow\frac{z}{z-1}+\frac{\bar{z}}{\bar{z}-1}=\frac{2x}{x-1}+O(\epsilon)\\
&2\epsilon y=z-\bar{z}\rightarrow\frac{z}{z-1}-\frac{\bar{z}}{\bar{z}-1}=\frac{-2\epsilon y}{(x-1)^2}+O(\epsilon^2).
\end{split}
\end{equation}
After taking the $\epsilon\to0$ limit, this is just the symmetry \eqref{symmetry} for GCA . It seems that the boundary conditions in harmonic analysis could also be obtained by taking the limit because they come from the symmetry. However, this $\epsilon\to 0$ limit for the boundary conditions is illegal because the symmetry property for the CPWs can not be preserved in the limiting procedure.
The point here is that, when we do the limiting analysis, because $\xi=\epsilon J$, we must analytically  continue the spin $J=h-\bar{h}$\footnote{As noted in \cite{Murugan:2017eto}, 2d CPW $F_{h,\bar{h}}$ has a meromorphic continuation in $h$, but need to keep the spin $h-\bar{h}$ fixed to be an integer. So in the following we will see that multi-valuedness appear when we analytically  continue the spin.}, then the CPWs for 2d conformal group will generally do not have the symmetry \eqref{ssss}. In fact, with respect to \eqref{ssss} the CPWs of 2d conformal group transforms as \cite{Murugan:2017eto}
\begin{equation}
F_{h,\bar{h}}(\frac{z}{z-1},\frac{\bar{z}}{\bar{z}-1})=(-1)^{h-\bar{h}}F_{h,\bar{h}}(z,\bar{z}) ,\label{trans}
\end{equation}
which holds only for integer spin so $(-1)^{h-\bar{h}}$ make sense. More precisely, the above symmetry \eqref{trans} only holds for CPWs  with even values of $J=h-\bar{h}$. Because \eqref{ssss} arises from the invariance of the four-point function under $z_1\leftrightarrow z_2$, this means a four-point function
that possesses the $z_1\leftrightarrow z_2$ symmetry can only receive contributions from $F_{h,\bar{h}}$ with even spin. However, under the  $\epsilon\to0$ limit,  this even spin condition gets lost and the relation  \eqref{trans} does not hold any more. More concretely, because $z=1$, $\bar{z}=1$ become branch points when $J$ is analytically continued,  the symmetry \eqref{ssss} can be rewritten as:
\begin{equation}
1-z\rightarrow\frac{1}{1-z}, \qquad 1-\bar{z}\rightarrow\frac{1}{1-\bar{z}}.
\end{equation}
Consequently in the arguments of $1-z\equiv re^{i\theta}$ and $1-\bar{z}\equiv re^{-i\theta}$, the transformation \eqref{ssss} results in the transformation
\begin{equation}
\theta\rightarrow -\theta,
\end{equation}
which means that the RHS and LHS of \eqref{trans} may live on different sheets with respect to the branch point $z=1$, $\bar{z}=1$. Thus, when $J$ is not restrict to be an integer, \eqref{trans} is not correct. Only a single conformal block have a definite transformation law. For $J$ being an integer, the transformation law of a conformal block coincides with its shadow, so \eqref{trans} holds.   If one consider  the fixed point of \eqref{ssss}:
\begin{equation}
z=\bar{z}=2, \qquad \text{or} \qquad  1-z=1-\bar{z}=-1,
\end{equation}
then $\theta=(2n+1)\pi$, after the transformation \eqref{ssss},
\begin{equation}
\Delta\theta\equiv\theta-(-\theta)= (2n+1)2\pi.
\end{equation}
Thus, when we consider the CPWs, the fixed point $(z,\bar{z})=(2,2)$ is at a different sheet after the transformation \eqref{ssss}. Only when $J$ is an even integer, the CPWs are single valued and \eqref{trans} holds and the CPWs have the symmetry \eqref{ssss}, given the Euclidean condition $\bar{z}=z^*$. In this case at $(z,\bar{z})=(2,2)$ we have the boundary conditions satisfied by CPWs just like in the 1d case\footnote{For the 1d case \cite{Maldacena:2016hyu}, a single conformal block indeed has similar transformation law for $\chi\rightarrow\frac{\chi}{\chi-1}$ ($\chi$ is the 1d cross ratio), so only when the quantum number of $\Delta$ is an even integer can it be a CPW. This fact is reflected  on the appearance of the discrete series. }.
However, when $J$ is analytically continued,  taking the limit does not lead to any boundary condition.
 In short,  after taking the limit, for CPWs we do not have the symmetry \eqref{symmetry} as well as the boundary conditions \eqref{bc2}. Note that there is another symmetry $z\leftrightarrow \bar{z}$ in 2d CPWs, which becomes the symmetry $y\leftrightarrow -y$ after taking the limit, so the boundary condition \eqref{bc1} still hold. In fact this is related to the $\xi\leftrightarrow-\xi$ symmetry.

Finally,  our harmonic analysis is like the one in CFT$_1$,  which relies on the space of real $x$ and $y$, while the $d\geq2$ conformal Harmonic analysis stand on the Euclidean space $\bar{z}=z^*$ with $z$ and $\bar{z}$ complex. In the $d\geq2$ conformal Harmonic analysis one needs a single valued boundary condition at $z=\bar{z}=1$ to ensure that the CPWs are Euclidean single valued. Our GCA case do not have such a boundary condition. Instead, near $x=1$, we have a matching condition similar to the 1d case, which makes our GCPWs have different expressions for $0<x<1$ and $1<x<2$. One can not see this piece-wise structure when take the the limit of a CPW, note that this matching condition is also related to the normalizable condition for the Hilbert space.

\section{Generalized Galilean Free Field Theories}

In this section, we study  generalized free theories (GFT) with Galilean conformal symmetries. The correlation functions in  GFTs can be read from the Wick contraction. We consider a generalized Galilean free field theory (GGFT) which contains two fundamental scalar type operators $\cO_1,\cO_2$ with the conformal weights $\D_1,\D_2$ and the charges $\xi_1,\xi_2$  respectively. The two-point functions read
\bea
\langle \cO_1\cO_1\rangle &=&x_{12}^{-2\Delta_1}e^{2\xi_1y_{12}/x_{12}}, \nn\\
\langle \cO_2\cO_2\rangle &=&x_{12}^{-2\Delta_2}e^{2\xi_2y_{12}/x_{12}},\nn\\
\langle \cO_1\cO_2\rangle &=&0.
\eea
The four-point function of $\cO_1\cO_1\cO_2\cO_2$ is simply the product of two-point functions.

\subsection{GGFTs from operator construction}
In this subsection, we consider the spectrum and three-point coefficients in such GGFT by constructing the quasi-primary operators explicitly. We want to study the operator product expansion (OPE) of $\cO_1\cO_2$. Due to the Wick theorem, only composite operators containing $\cO_1\cO_2$ can appear in the expansion. Moreover, $L_{-1},M_{-1}$ which are the derivatives with respect to $x,y$ can also appear in the construction. We give a detailed analysis of the number of the operators in appendix \ref{opernumber}. It turns out that the generating function of the total number of linear-independent operators at each level is
\be
  Z(q) =\frac{1}{(1-q)^4}
   = \sum_{N=0}^{\infty}P(N)q^N.
\ee
where $P(N)$ is the number of operators at level $N$ with a weight $\D_1+\D_2+N$. According to the appendix \ref{opernumber}, we list the information on the operators and the multiplets in the following table.
\begin{center}
    \begin{tabular}{|c|c|c|c|c|c|}
\hline
Level &  0 &  1 &  2 &  3 & 4 \\
\hline
 Number of operators &  1 &  4 &  10 &  20 &  35  \\
\hline
 Number of quasi-primaries &  1 &  2 &  3 &  4 &  5 \\
\hline
 Rank &  1 &  2 &  3 &  4 &  5 \\
\hline
 Number of multiplets &  1 &  1 &  1 &  1 &  1 \\
\hline
\end{tabular}
\end{center}
Note that there is one multiplet at each level.

Let us construct the level $1$ operators explicitly. At level $1$, there are four linearly independent operators denoted  by $\cA,\cB,\cC,\cD$, respectively,
\begin{equation}
\cA:\cO_1M_{-1}\cO_2,\hs{3ex} \cB:M_{-1}\cO_1\cO_2\nn
\end{equation}
\begin{equation}
\cC:\cO_1L_{-1}\cO_2,\hs{3ex} \cD:L_{-1}\cO_1\cO_2.\nn
\end{equation}
At level $1$, there are two
 quasi-primary operators, which are the linear combination of the above four operators
\begin{equation}
\cP_i=a_i\cA+b_i\cB+c_i\cC+d_i\cD, \hs{3ex}i=0,1.
\end{equation}
 They should satisfy
\begin{equation}
L_1\cP_i=0,\hs{5ex} M_1\cP_i=0,
\end{equation}
\begin{equation}
M_0\cP_0=\xi_p \cP_0=(\xi_1+\xi_2)\cP_0,\ \hs{3ex} M_0\cP_1=\xi_p \cP_1+\cP_0,
\end{equation}
\begin{equation}
L_0\cP_0=\Delta_p \cP_0=(\Delta_1+\Delta_2+1)\cP_0,\ \hs{3ex} L_0\cP_1=\Delta_p \cP_1.
\end{equation}
From  acting  the generators on $\cA,\cB,\cC,\cD$,
\begin{equation}
L_1\cC=2\Delta_2\cO_1\cO_2,\ \hs{3ex} L_1\cD=2\Delta_1\cO_1\cO_2,\nn
\end{equation}
\begin{equation}
M_1\cC=2\xi_2\cO_1\cO_2,\ \hs{3ex} M_1\cD=2\xi_1\cO_1\cO_2,\nn
\end{equation}
\begin{equation}
L_1\cA=2\xi_2\cO_1\cO_2,\ \hs{3ex} L_1\cB=2\xi_1\cO_1\cO_2,\nn
\end{equation}
\begin{equation}
M_1\cA=0,\ \hs{3ex} M_1\cB=0,\nn
\end{equation}
\begin{equation}
M_0\cA=(\xi_1+\xi_2)\cA,\ \hs{3ex} M_0\cB=(\xi_1+\xi_2)\cB,\nn
\end{equation}
\begin{equation}
M_0\cC=(\xi_1+\xi_2)\cC+\cA,\ \hs{3ex} M_0\cD=(\xi_1+\xi_2)\cD+\cB,\nn
\end{equation}
one can obtain that
\begin{equation}
\cP_0:a_0=\xi_1,\ \hs{3ex} b_0=-\xi_2,\ \hs{3ex} c_0=d_0=0
\end{equation}
\begin{equation}
\cP_1:a_1=a,\ \ b_1=-\frac{a\xi_2}{\xi_1}-\frac{\Delta_2\xi_1-\Delta_1\xi_2}{\xi_1},\ \ c_1=\xi_1,\ \ d_1=-\xi_2.
\end{equation}
Note that there is still one undetermined parameter $a$ in $\cP_1$.

On the other hand, in order to define the composite operators properly and calculate their correlation functions,  we may apply  the point-splitting regularization,
\begin{equation}
(\cO(x_1,y_1)\cO(x_2,y_2))_{\text{r}}=\lim_{\substack{x_2\rightarrow x_1\\y_2\rightarrow y_1}}\cO\cO-\langle \cO\cO\rangle
\end{equation}
such that there is no singularities as $x_2\rightarrow x_1$. Note that there is no singularities as $y_2\rightarrow y_1$ in the two-point function, so we can simply take the same $y$. The above composite operators should be taken as the regularized ones.  Now we can calculate the two-point functions of $\cP_0$ and $\cP_1$,
\bea
\langle \cP_0\cP_0\rangle&=&0, \\
\langle \cP_1\cP_0\rangle&=&d e^{2(\xi_1+\xi_2)y_{12}/x_{12}}x_{12}^{-2(\Delta_1+\Delta_2+1)},\\
\langle \cP_1\cP_1\rangle&=&\langle \cP_1\cP_0\rangle\frac{2y_{12}}{x_{12}}+\langle \cP_1\cP_0\rangle f(a),
\eea
where $d=2\xi_1\xi_2(\xi_1+\xi_2)$ is the overall normalization. To normalize the two-point functions as required, one should choose $a$ so that $f(a)=0$. This leads to
\begin{equation}
a=\frac{\Delta_1\xi_2^2-\Delta_2\xi_1(\xi_1+2\xi_2)}{2\xi_2(\xi_1+\xi_2)}.
\end{equation}

Let us turn to
 the multiple-point functions. There are two three-point functions. One  is
\begin{equation}
\langle \cO_1\cO_2\cP_0\rangle=ABC
\end{equation}
where $A$ and $B$ share the same structure with \eqref{Acoef} \eqref{Bcoef}, and
\begin{equation}
C=c_0=-2\xi_1\xi_2.
\end{equation}
The other one is
\begin{equation}
\langle \cO_1\cO_2\cP_1\rangle=ABc_1+c_0AB\partial_{\xi_p}\ln A
\end{equation}
where
\begin{equation}
c_1=-\frac{\Delta_2\xi_1^2+\Delta_1\xi_2^2}{\xi_1+\xi_2}.
\end{equation}
For the four-point function $\langle \cO_1\cO_1\cO_2\cO_2\rangle$,  the contribution of the global block from the quasiprimary operators to the stripped four-point functions is
\begin{equation}
\mathcal{G}(x,y)=\frac{1}{d}(A_0g^{(0)}_{\Delta_1+\Delta_2+1,\xi_1+\xi_2}+A_1g^{(1)}_{\Delta_1+\Delta_2+1,\xi_1+\xi_2})+\cdots
\end{equation}
where the ellipsis represents the contributions from other levels, and
\begin{equation}
A_0=2c_0c_1,\ \ \ A_1=c_0^2.
\end{equation}
To compare with the calculations using other methods where we consider the global block contribution to the $t$-channel of the stripped four point function with identical external operators $\cO$, we take $\Delta_1=\Delta_2\equiv\Delta_\cO$, $\xi_1=\xi_2\equiv\xi_\cO$ such that
\begin{equation}\label{order1}
\mathcal{G}(x,y)=\Delta_\cO g^{(0)}_{2\Delta_\cO+1,2\xi_\cO}+\xi_\cO g^{(1)}_{2\Delta_\cO+1,2\xi_\cO}+\cdots
\end{equation}
Note that it shows just the explicit contribution from the level-$1$ quasi-primary operators. For the quasi-primary operators at higher levels, the construction becomes more and more tedious. We will use the inversion formula and double series expansion method to read the relevant information on the quasi-primary operators at higher levels.

\subsection{Radial coordinates in Galilean CFT}

In many cases, it turns out to be more convenient to
work in radial coordinates in Galilean CFT. The configuration of four points is $p_1=(-\rho,-\tau),p_2=(\rho,\tau),p_3=(1,0),p_4=(-1,0)$, where $(\rho,\tau)\in D= (0,1)\times\mathbb{R}$. For simplicity,  the slope $\kappa=\frac{\tau}{\rho}\in \mathbb{R}$ will be used. In terms of $(\r, \k)$, the cross ratios $(x,y)\in(0,1)\times\mathbb{R}$ could be written as
\begin{equation}
\begin{split}
        x&=\frac{x_{12}x_{34}}{x_{13}x_{24}}=\frac{4\rho}{(\rho+1)^2},\\
    \frac{y}{x}&=\frac{y_{12}}{x_{12}}+\frac{y_{34}}{x_{34}}-\frac{y_{13}}{x_{13}}-\frac{y_{24}}{x_{24}}=\kappa \frac{1-\rho}{1+\rho},
\end{split}
\end{equation}
and inversely
\begin{equation}
\begin{split}
    \rho=\frac{1-\sqrt{1-x}}{1+\sqrt{1-x}},\hs{3ex}
    \kappa=\frac{y}{x\sqrt{1-x}}.
\end{split}
\end{equation}
The conformal block with correct $s$-channel OPE behavior is
\begin{equation}
\chi_{\Delta, \xi}=\frac{x^{\Delta}(1+\sqrt{1-x})^{2-2\Delta}}{\sqrt{1-x}}e^{\frac{-\xi y}{x\sqrt{1-x}}}.
\end{equation}
In terms of the radial coordinates it is of the form
\begin{equation}
    \chi_{\Delta,\xi}=\frac{4}{1-\rho^2}\rho^{\Delta}e^{-\xi \kappa}.
\end{equation}
Now the blocks for the multiplets are just
\begin{equation}
    \chi_{\Delta,\xi,k}=\frac{4}{1-\rho^2}\rho^{\Delta}(-\kappa)^k\,e^{-\xi \kappa},
\end{equation}
where $(-\kappa)$ is to keep the relation $\chi_k=\partial^k_{\xi} \chi$.

Notice that since $|\rho|<1$, the $s$-channel block expansion is a possibly convergent series. To get convergence we need  to control the coefficients further.

\subsection{Taylor series expansion of four-point functions}
In radial coordinates the block expansion has a simpler form, and for GGFT we can calculate the block coefficients of all orders by a Taylor series expansion. The $s$-channel conformal block expansion of four identical quasi-primary operators with the weight and charge $(\Delta_\cO,\xi_\cO)$ is
\begin{equation}
    \mathcal{G}=\frac{\greenfunction{\cO\cO\cO\cO}}{x^{-2\Delta_{\cO}}\exp(\frac{2\xi_{\cO}y}{x})}=\sum_{\Delta,\xi,k}P_{\Delta,\xi,k}\chi_{\Delta,\xi,k}.
\end{equation}
In GGFT, it turns out that there is no $\xi= 0$ part and more restrictively, the $\xi$-spectrum contains only $\xi=2\xi_\cO$. From the viewpoint of operator construction, the condition $\xi=2\xi_\cO$ is consistent with that only double trace operators, schematically $\cO\partial^a_{x}\partial^b_{y}\cO$, enter in the $\cO\cO$ OPE. Besides, the limiting procedure also suggests this spectrum, see section \ref{LimitGFT}. Hence now the summation is over $\Delta$ and the multiplet number $k$,
\begin{equation}
    \sum_{\Delta,k} P_{\Delta,k}\,\rho^{\Delta}(-\kappa)^{k}=\frac{1}{4}(1-\r^2)e^{\xi \kappa}\mathcal{G}. \label{matchingtwosides}
\end{equation}
For $t$-channel, we have
\begin{equation}
    \text{RHS}=2^{4\Delta_{\cO}-2}(1-\rho^{2})(1-\rho)^{-4\Delta_{\cO}}\rho^{2\Delta_{\cO}}\exp(-2\xi_{\cO}\kappa\frac{2\rho}{1-\rho}). \label{tcontribution}
\end{equation}
For $u$-channel, we have
\begin{equation}
    \text{RHS}=2^{4\Delta_{\cO}-2}(1-\rho^{2})(1+\rho)^{-4\Delta_{\cO}}\rho^{2\Delta_{\cO}}\exp(2\xi_{\cO}\kappa\frac{2\rho}{1+\rho}), \label{ucontribution}
\end{equation}
which is related to $t$-channel by  $\rho\to-\rho$, as in the case of CFT.

In the $t$-channel's contribution \eqref{tcontribution}, apart from the $\rho^{2\Delta_{\cO}}$ factor, the remaining part has a Taylor series expansion of $\rho$ (an expansion of $\rho^n$, where $n\in\mathbb{N}$). So matching the left hand side of \eqref{matchingtwosides} to its right hand side (which is \eqref{tcontribution} for the $t$-channel), we get $\Delta_n=2\Delta_O+n$. This is also true for the $u$ channel's contribution \eqref{ucontribution}. Now we calculate each channel separately, to compare it with the inversion formula and the operator construction method. The relation in the $t$-channel is,
\begin{equation}
\sum_{n,k} P^{t}_{n,k}\rho^{n}(-\kappa)^{k}=2^{4\Delta_{\cO}-2}(1-\rho^{2}) \,(1-\rho)^{-4\Delta_{\cO}}\exp(\frac{-4\xi_{\cO}\kappa\rho}{1-\rho})
\end{equation}
The key feature is that $\kappa$ only appears in exponential factor together with $\rho$ due to the condition $\xi=2\xi_\cO$. Also it is analytic at $\rho=0$. Hence the summation range is $n\geq0$ and $k=0,\cdots,n$.
Expanding the right hand side we get the coefficients,
\begin{equation}\label{coef1}
P^{t}_{n,k}=\frac{2^{4\Delta_\cO+2k-2}\xi_\cO^k}{k!}A_{n,k}
\end{equation}
where $A_{n,k}$ is a polynomial of $\Delta_\cO$ of degree $n-k$.
\begin{equation}A_{n,k}=\frac{4\Delta_\cO+2n-k-2}{4\Delta_\cO+n-2}\binom{4\Delta_\cO+n-2}{n-k}\end{equation}
We list explicitly $A_{n,k}$ of the first few levels in order to compare with other methods.
\begin{center}
    \begin{tabular}{|l|l|l|l|l|}
    \hline
    \diagbox{$n$}{$k$} & $0$&$1$&$2$&$3$  \\
    \hline
    $0$ & $1$ & & &\\
    \hline
    $1$ & $4 \Delta_\cO$ & $1$ & & \\
    \hline
    $2$ & $8 \Delta_\cO^2+2 \Delta_\cO-1$&$4 \Delta_\cO+1$&$1$&\\
    \hline
    $3$ & $\frac{32 \Delta _\cO^3}{3}+8 \Delta _\cO^2-\frac{8 \Delta _\cO}{3}$ & $8 \Delta _\cO^2+6 \Delta _\cO$ & $4 \Delta _\cO+2$&$1$\\
    \hline
\end{tabular}
\end{center}
At level $1$, we have,
\begin{equation}
    P^t_{1,0}=2^{4\D_{\cO}}\D_{\cO},\quad P^t_{1,1}=2^{4\D_{\cO}}\xi_{\cO},
\end{equation}
and the corresponding block expansion is,
\begin{align}
\mathcal{G}_t(x,y)&=2^{4\D_{\cO}}\D_{\cO}\,\chi_{2\Delta_\cO+1,2\xi_\cO,0}+2^{4\D_{\cO}}\xi_{\cO}\,\chi_{2\Delta_\cO+1,2\xi_\cO,1}+\cdots\nn\\
                &=\Delta_\cO g^{(0)}_{2\Delta_\cO+1,2\xi_\cO}+\xi_\cO g^{(1)}_{2\Delta_\cO+1,2\xi_\cO}+\cdots
\end{align}
where the different conventions of blocks are related by $g^{(k)}_{\Delta,\xi}=2^{2\Delta-2}\chi_{\Delta,\xi,k}$. It matches the operator construction result \eqref{order1}.

Now it is easy to sum over the $t$ and $u$ channels:
\begin{equation}P^{t+u}_{n,k}=\begin{cases}
\frac{2^{4\Delta_\cO+2k-1}\xi_\cO^k}{k!}\frac{4\Delta_\cO+2n-k-2}{4\Delta_\cO+n-2}\binom{4\Delta_\cO+n-2}{n-k},& n\text{ is even}\\
0, & n \text{ is odd.}
\end{cases}\end{equation}
This is the result for the generalized free boson, and for the generalized free fermions we only need to interchange even and odd in the above formula.

\subsection{Inversion function of GGFT}
It is interesting that we are able to extract the spectrum data of generalized free theory, although the inversion function $I=(\Psi_{\Delta,\xi},\mathcal{G})$ is divergent due to the pathological behaviors of two-point functions with non-vanishing $\Re\xi_\mathcal{O}$\footnote{As we will show in the next subsection, if we choose $\xi_\mathcal{O}$ to be pure imaginary such that the inversion function is convergent in the sense of distribution, however, the result has no $\Delta$ poles related to the physical operators, closely relating to the fact that there is a branch cut in the $\xi$-plane in the inversion function.}. Without lose of generality we assume $\Re\xi_\mathcal{O}>0$ in this subsection.

The way out is to introduce a regularization, and analyticity ensures the results are cut-off free. We regularize the two-point functions by defining
\begin{equation}\greenfunction{\cO \cO}_r\equiv \theta(r-\frac{y_{12}}{x_{12}})\greenfunction{\cO \cO}\end{equation}
which converges to $\greenfunction{\cO \cO}$ when $r\to+\infty$. And equivalently the integration region of inversion formula restricts to $D_r=(0,1)\times(-\infty,r)$. In the previous subsection we have known that the block expansion is analytic in $D$. The rigidity of analytic functions ensures that nothing will be changed if restricting to $D_r$, yet the inversion integral will be convergent.

To be concrete, the structure of regularized inversion function turns out to be
\begin{equation}
\label{Schwinger1}I_{r}(\Delta,\xi)=\sum^{\infty}_{n=0}\sum^n_{k=0}P^{\text{inversion}}_{n,k}\frac{\CoefficientChi}{(\Delta-\Delta_n)(\xi-2\xi_\cO)^{k+1}}+\text{shadow poles}+\text{regular w.r.t. }\xi.
\end{equation}
The shadow poles are from the three shadow blocks in $\Psi_{\Delta,\xi}$. The third term, contributed from $D_r\backslash D_0$ and $(1,2)\times\mathbb{R}$, though depends on $r$ and contains the poles of $\Delta$, is regular with $\xi$. If we deform the contour of $\Delta$ and $\xi$, only the first term $I_{1}$ gives non-vanishing contributions. The first term $I_{1,r}$ and also the shadow part are independent of $r$ and remaining finite in the limit $r\to+\infty$ and we can safely evaluate it at $r=0$, i.e.
\begin{equation}I_{1,r=\infty}=I_{1,r=0}\end{equation}
which means taking $D_{0}$ as the integral region. Equivalently it is a unilateral Laplace transform.

Now let us calculate $I_1$,
\be
I_1=A(2-\D)(\chi_{\Delta,\xi},\mathcal{G}_t)_{D_0}.
\ee
It is more convenient to transform the inversion integral into $\rho$ coordinates. The integration region $D_0$ in radial coordinates is $(\r,\kappa)\in (0,1)\times (-\inf,0)$,  and we denote it also by $D_0$. The inner product is,
\begin{equation}(f,g)=\int_{D_0} dx dy \frac{1}{x^4} f^* g=\int_{D_0} d\rho d\kappa \frac{\left(1-\rho^2\right)^2}{16 \rho^3}f^* g. \end{equation}
The input $t$-channel contribution is
\begin{equation}\mathcal{G}_t=\frac{(1-x)^{-2\Delta_\cO}\exp(\frac{-2\xi_\cO y}{1-x})}{x^{-2\Delta_\cO}\exp( \frac{2\xi_\cO y}{x})}=2^{4\Delta_\cO} \left(\frac{\rho}{(1-\rho)^2}\right)^{2 \Delta_\cO} e^{-2 \kappa \xi_\cO \frac{1+\rho}{1-\rho}}. \end{equation}
The conformal partial wave is a linear combination of four blocks, but as we have discussed in \eqref{Schwinger1}, to read out the physical spectrum of GGFT we only need one of them. The block is $\chi_{\Delta,\xi}=\chi^{*}_{2-\Delta,-\xi}$, and
\begin{equation}
(\chi_{\Delta,\xi},\mathcal{G}_t)_{D_0}=2^{4 \Delta_{\cO}-2} \int^1_0 d\rho \rho^{-\Delta+2 \Delta_{\cO}-1} f(\rho)\label{inversionint}\end{equation}
in which $\kappa$ has been integrated out: $\int^0_{-\infty} e^{\kappa a}=\frac{1}{a}$, and
\begin{equation}f(\rho)=\frac{ (1+\rho) (1-\rho)^{2-4 \Delta_{\cO}} }{(1-\rho)\xi-2\xi_\cO(1+\rho)}. \label{frho}\end{equation}
We will see that the independence of $\Delta$ in $f(\rho)$ simplifies the calculation drastically.

The integral \eqref{inversionint} can be represented as an Appell function $F_1$. However to figure out the analytic structure it is more convenient using the following method. Recall that $\rho^{a-1}$ is a distribution on $\rho\in(0,1)$ and is meromorphic with respect to $a$, with poles at $a=-n,\,n\in \mathbb{N}$ and the residues $\Res_{a=-n}=(-1)^n\frac{\delta^{(n)}(\rho)}{n!}$,
\begin{equation}\label{Gelfand}
\int^1_0d\rho \rho^{a-1}\sum_n c_n\rho^n=\sum_n \frac{c_n}{a+n}
\end{equation}
We use this to read the poles and the residues of $\Delta$, and the $\xi$-pole structure automatically appears in the residues of $\Delta$ - there are multiple poles at $\xi=2\xi_\cO$ corresponding to the multiplets at each level, whose order ranges from $1$ to $n+1$.
To extract the pole structures we need to calculate the $n$-th derivatives of $f(\rho)$ at $\rho=0$, and then expand them with respect to $\frac{1}{\xi-2\xi_\cO}$. The result is
\begin{equation}\label{DeltaFirst}
    (\chi_{\Delta,\xi},\mathcal{G}_t)_{D_0}=\sum^{\infty}_{n=0}\sum^n_{k=0}\frac{1}{-\Delta+2\Delta_{\cO}+n}\frac{1}{(\xi-2\xi_\cO)^{k+1}}P^{t,\text{inversion}}_{n,k}
\end{equation}
where,
\begin{equation}P^{t,\text{inversion}}_{n,k}=k!\,P^{t}_{n,k}.\end{equation}
It matches with \eqref{coef1} obtained by the Taylor series expansion. The factorial $k!$ is due to reading out the block expansion from higher order poles. And the clock-wise direction of contours with respect to $\xi$ and $\D$ gives rise to two minus sign but totally $(-1)^2=1$. When reading out the spectrum data, we firstly deform the $\D$-contour picking up the poles at $\D_n$, and then deform the $\xi$-contour giving rise to multiplets.

\subsection{Branch cut of inversion function}

The order of $\xi$-poles in \eqref{DeltaFirst} is unbounded above as $n\to\inf$. This suggests that   $\xi=2\xi_\cO$ could be a logarithmic type branch point in $\xi$-plane. This fact could be shown explicitly if we change the order in doing the integrals. In the above discussion, we took the contour integrals first in $\D$-plane  and then in $\xi$-plane. If we take the $\xi$-contour integral first, we will find a  discontinuity along the $\xi$-branch cut. With this discontinuity we can still recover the $t$-channel stripped four-point function, as we will show.  

\begin{figure}[h]
\centering
 \includegraphics[width=0.75\textwidth]{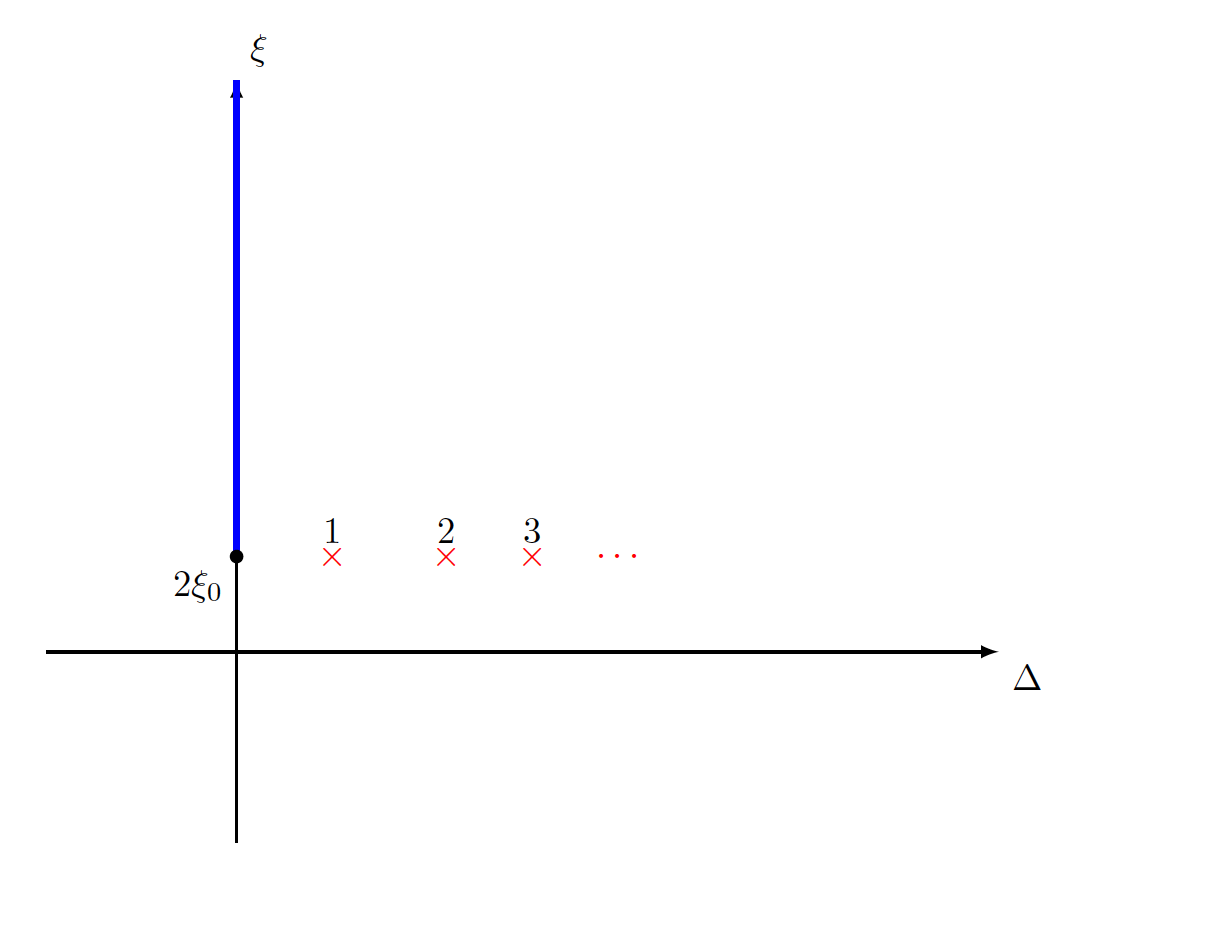}
\caption{The analytic structure of $\left(\chi_{\Delta,\xi},\mathcal{G}_{t}\right)_{D_0} $. From the $\xi$-viewpoint, it has a branch cut, while from $\D$-viewpoint it has double trace multiple poles.}
\label{fig:street lamp}
\end{figure}

To see the branch cut we discuss a toy model firstly. Consider an improper integral of $t$,
\begin{equation}\int^1_0 dt\,\frac{1}{t-a}=\log(1-\frac{1}{a}).\end{equation}
It has a branch cut with respect to $a$ at $(0,1)$. This is because when $a\in(0,1)$ the integrand should be understood as a distribution, and the ambiguity of definition comes into play,
\begin{equation}\label{dispersion}\frac{1}{t-a\pm i\epsilon}=\text{PV}(\frac{1}{t-a})\mp \pi i \delta(t-a).\end{equation}
This is  also the Fourier transform (integral with respect to $\kappa$) of Heaviside step function supporting on $\mathbb{R}^{>0}$ or $\mathbb{R}^{<0}$. And the discontinuity is just twice the integration of the imaginary part,
\begin{equation}\text{Disc}=\int_0^1 dt 2\pi i \delta{(t-a)}. \end{equation}

For the inversion function $I_1$, which is proportional to  \eqref{inversionint}, there is a singularity  in $f(\r)$, Eq. \eqref{frho}, lying at
\begin{equation}
    \r_{\text{sing}}=\frac{\xi-2\xi_{\cO}}{\xi+2\xi_{\cO}},
\end{equation}
and the resulting branch cut in $\xi$-plane is along $\xi\in (2\xi_{\cO},+\inf)$. Then the discontinuity is,
\begin{align}
\text{Disc}\left(\chi_{\Delta,\xi},\mathcal{G}_{t}\right)_{D_0}
    &=2\pi i \,2^{4 \Delta_{\cO}-2} \int^1_0 d\rho \rho^{-\Delta+2 \Delta_{\cO}-1}  \frac{(1+\rho) (1-\rho)^{2-4 \Delta_{\cO}}}{-(\xi+2\xi_\cO)}\delta\left(\r-\r_{\text{sing}}\right)
    \nn\\
    &=-2\pi i\,2^{-4\Delta_{\cO}+3}\,\xi \xi_{\cO}^{2-4\Delta_\cO}(\xi-2\xi_\cO)^{2\Delta_\cO-\Delta-1}(\xi+2\xi_\cO)^{\Delta+2\Delta_\cO-3}.
    \label{disc}
\end{align}
Now if  we reverse the order of doing $\D$ and $\xi$ integrals, opposed to the ones in the previous subsection, using \eqref{ww2} and enclosing the $\xi$-contour onto the branch cut, we get,
\begin{align}
    \mathcal{G}_t&=\frac{1}{2\pi i}\int_{C_{\D}}d\D\frac{1}{2\pi i}\int_{C_{\xi}}d\xi\,\left(\chi_{\Delta,\xi},\mathcal{G}_{t}\right)_{D_0} \chi_{\D,\xi}\nn\\
    &=\frac{1}{2\pi i}\int_{C_{\D}}d\D\frac{1}{2\pi i}\int^{\inf}_{2\xi_{\cO}}d\xi\,\text{Disc}\left(\chi_{\Delta,\xi},\mathcal{G}_{t}\right)_{D_0} \chi_{\D,\xi}\nn\\
    &=\frac{1}{2\pi i}\int_{C_{\D}} d\D\int^{\inf}_{2\xi_{\cO}}d\xi K(\xi)\frac{1}{1-\rho^2}\rho^{\Delta} e^{-2\xi_\cO \kappa}\label{annoyingfactor}
\end{align}
where the integrand $K(\xi)$ is
\begin{equation}
    K(\xi)=-2^{-4\Delta_{\cO}+5}\xi \xi_{\cO}^{2-4\Delta_\cO}(\xi-2\xi_\cO)^{2\Delta_\cO-\Delta-1}(\xi+2\xi_\cO)^{\Delta+2\Delta_\cO-3}
    e^{-(\xi-2\xi_\cO) \kappa}.
\end{equation}
Notice that the factor $(\xi-2\xi_\cO)^{2\Delta_\cO-\Delta-1}$ plays the role of $\rho^{-\Delta+2 \Delta_{\cO}-1}$ in \eqref{inversionint}, and provides the double-trace poles $\frac{1}{\D-\D_n}$. And there are no contributions of double-trace poles from $\xi-2\xi_\cO\to\inf$ due to the exponential decay.

To check \eqref{annoyingfactor} indeed recovers the $t$-channel contribution $\mathcal{G}_t$, we borrow the identity of confluent hypergeometric function $\,_1F_1$ and get,
\begin{equation}
    \int^{\inf}_{2\xi_{\cO}}d\xi K(\xi)=k_1(\D)\,\G(2\Delta_\cO-\D)+k_2
\end{equation}
where,
\begin{multline}
k_1(\D)=\frac{2^{4 \Delta_{\cO}+1} \Gamma\left(2-4 \Delta_{\cO}\right)}{\Gamma\left(-\Delta-2 \Delta_{\cO}+3\right)}
\Big(\left(2 \Delta_{\cO}-1\right)\,_{1}F_{1}\left(2 \Delta_{\cO}-\Delta , 4 \Delta_{\cO}-2 ; 4 \kappa \xi_{\cO}\right)\\
+\left(\Delta-2 \Delta_{\cO}\right)\,_{1}F_{1}\left(-\Delta+2 \Delta_{\cO}+1 , 4 \Delta_{\cO}-1 ; 4 \kappa \xi_{\cO}\right)\Big)
\end{multline}
and the $k_2$-term contains no poles of $\D$. The factor $\G(2\Delta_\cO-\D)$ of $k_1$-term gives double trace poles $\D_n$, with residue $\frac{(-1)^n}{n!}$. After doing the $\D$-contour integration, the clock-wise contour of $\D$ gives an extra minus sign $(-1)$, and the right hand side of \eqref{annoyingfactor} is,
\begin{align}
    \text{RHS}=\sum_{n=0}^{\inf} \frac{(-1)^{n+1}}{n!}k_1(\D_n)\frac{1}{1-\rho^2}\rho^{\Delta_n} e^{-2\xi_\cO \kappa}
\end{align}
where $k_1(\D_n)$ is a polynomial of $\kappa$ since $_{1}F_{1}(-n,a;z)$ truncates to the generalized Laguerre polynomials. Now we can re-expand with respect to $\kappa$ and get back to the Taylor series expansion of $\mathcal{G}_t$.

Finally we discuss the subtlety of inversion integral for purely imaginary\footnote{The physical operators lie on the unitary principal series, which also happens in celestial CFTs, see e.g.  \cite{Fotopoulos:2019tpe}.} $\xi_{\cO}$. In this case the inversion integral is convergent in $D=(0,1)\times\mathbb{R}$ since the part involving $y$ is only a phase factor, giving rise to a delta function of $x$, and the integral
\begin{equation}\label{cspec}
\left(\chi_{\Delta,\xi},\mathcal{G}_{t}\right)_{D}\sim\xi (\xi-2\xi_\cO)^{2\Delta_\cO-\Delta-1}(\xi+2\xi_\cO)^{\Delta+2\Delta_\cO-3}
\end{equation}
has no double-trace pole structures and coincides with \eqref{disc}. This is because for purely imaginary $\xi_{\cO}$ we can freely choose the integration region as $D_0$ or $D\backslash D_{0}=(0,1)\times (0,+\inf)$ since both are convergent. Adding these two up we get the result \eqref{cspec} and lose track of pole structures coming from $\text{PV}(\frac{1}{t-a})$ in \eqref{dispersion}.

\subsection{Limit for GGFT}\label{LimitGFT}
\par Now, we show how to reach the GGFT by taking the limit on a 2d GFT. Our starting point will be the block expansion of a four-point function of four identical fundamental operators (we use the same notation $\cO$ for this operator) with the same quantum numbers $\Delta_\cO$ and $J_\cO$ in a GFT.
In 2d, such a correlator can be easily analysed by its  factorization property, namely, it can be factorized into holomorphic and anti-holomorphic sectors.  In each sector, the theory is a generalized free theory. For simplicity, we will focus on the identity in
$t$ channel, which has the following form:
\begin{equation}
1^{(t)}=\left(\frac{z}{1-z}\right)^{2h_\cO}\left(\frac{\bar{z}}{1-\bar{z}}\right)^{2\bar{h}_\cO}
\end{equation}
where $h_0$ and $\bar{h}_0$ are the input quantum number and satisfy:
\begin{equation}
h_\cO+\bar{h}_\cO=\Delta_\cO, \qquad h_\cO-\bar{h}_\cO=J_\cO.
\end{equation}
The holomorphic part has a block expansion with respect to the $SL(2,\mathbb{R})$ symmetry:
\begin{equation}
\left(\frac{z}{1-z}\right)^{2h_\cO}=\sum_{n=0}^\infty a_nG_n(z)
\end{equation}
where $G_n(z)$ is the conformal block corresponding to a double trace operator with dimension $h_n=2h_\cO+n$ and
\begin{equation}
a_n=\frac{(2h_\cO)_n^2}{n!(4h_\cO+n-1)_n}.
\end{equation}
The same thing holds for the anti-holomorphic part. So we have:
\begin{equation}
\left(\frac{z}{1-z}\right)^{2h_\cO}\left(\frac{\bar{z}}{1-\bar{z}}\right)^{2\bar{h}_\cO}=\sum_{n,m=0}^\infty a_n\bar{a}_mG_n(z)G_m(\bar{z}). \label{tidentity}
\end{equation}

Now we take a limit on both sides of the  equation \eqref{tidentity}. The limit is defined by setting \begin{equation}
z=x+\epsilon y, \quad \bar{z}=x-\epsilon y, \quad \Delta=h+\bar{h}, \quad \xi=\epsilon(h-\bar{h})=\epsilon J   \label{limit}
\end{equation}
and taking the $\epsilon \to 0$ limit while keeping $\xi$ to be finite.
It is easy to work out the left-hand side of \eqref{tidentity} after taking $\epsilon\rightarrow0$:
\begin{equation}
\lim_{\epsilon \to 0}\left(\frac{z}{1-z}\right)^{2h_\cO}\left(\frac{\bar{z}}{1-\bar{z}}\right)^{2\bar{h}_\cO}=\left(\frac{x}{1-x}\right)^{2\Delta_\cO}e^{-2\xi_\cO\frac{y}{x(1-x)}},
\end{equation}
which is just what we want: the  identity part of the $t$ channel correlator  in GGFT.

Next we deal with the right-hand side of \eqref{tidentity}. First, we want to know which operators appear in the block expansion. The quantum numbers of double trace operators in 2d GFT are:
\begin{equation}
h_n=2h_\cO+n, \qquad \bar{h}_m=2\bar{h}_\cO+m, \hs{3ex}n,m\in N.
\end{equation}
After taking the limit, we have:
\begin{equation}
\Delta=2\Delta_\cO+n+m, \qquad \xi=2\xi_\cO+\epsilon(n-m)
\end{equation}
We see that the spectrum of $\Delta$ is $\epsilon$ independent, while the $\xi$ spectrum depends on $\epsilon$. The point here is when we take the $\epsilon\to 0$ limit, the spectrum of $\xi$ is localized to $\xi=2\xi_\cO$ and the
$\Delta$ spectrum always stay at the ``double trace" location, which is just the result of the previous sections.

However, here comes a question: how do multiplets appear in this procedure? The answer is that we need to
count the small shift of $\epsilon$, namely, we should expand the block at $\xi=2\xi_\cO+\epsilon(n-m)$ in terms of the block at $\xi=2\xi_\cO$. Then, one naturally obtain the terms of some derivatives with respect to $\xi$ of the
singlet blocks. The crucial point here is that even at the higher order of the $\epsilon$ expansion, these terms indeed give finite contributions, due to the combination of the $\frac{1}{\epsilon}$ factor in the coefficients $a_n\bar{a}_m$ under the limit. Furthermore, all the superficially divergent terms cancel with each other, and the coefficient function only has terms of finite order in $\frac{1}{\epsilon}$, which gives a truncation of the rank of the multiplets. The above analysis indeed has two kinds of $\epsilon$ corrections:  one comes from the order $O(\epsilon)$ shift of $\xi$ away from $2\xi_\mathcal{O}$, as mentioned above. While the other comes from higher order corrections in $\epsilon$ for a 2d conformal blocks:
\begin{equation}
G_n(z)G_m(\bar{z})=g_{2\Delta_\cO+n+m,2\xi_\cO+\epsilon(n-m)}+O(\epsilon) \label{second kind}
\end{equation}
where $g_{\Delta_{n+m},2\xi_\cO+\epsilon(n-m)}$ is the GCA singlet block. Both these two kinds of $\epsilon$ corrections are possible to give finite contributions when combining with higher $\frac{1}{\epsilon}$ order terms in the coefficients $a_n$ and $\bar{a}_m$. In the argument above, we have only concerned about the first one, as the second kind of correction is always at higher  order in $\epsilon$ in the calculation and does not contribute.
The final picture
is that the limit of the right-hand side of \eqref{tidentity} is well defined and match perfectly with our pervious calculation!

Let us illustrate the picture by showing several leading terms. Denoting $G_{h,\bar{h}}$ as the 2d conformal block,  we  relabel it as $G_{\Delta,\xi}$ according to \eqref{limit}.
The leading term corresponds to $n+m=0$ which requires  $n=m=0$, and has:
\begin{equation}
\Delta=2\Delta_\cO, \quad \xi=2\xi_\cO, \quad a_0\bar{a}_0=1.\nn
\end{equation}
This is just the singlet block in the previous sections. The next-to-leading terms correspond to $n+m=1$, which requires
\begin{equation}
n=1,  m=0, \qquad\mbox{or}\qquad n=0, m=1,\nn
\end{equation}
then the two terms contribute to:
\bea
\lefteqn{a_1\bar{a}_0(G_{\Delta_1,2\xi_\cO}+\epsilon\partial_{\xi}G_{\Delta_1,2\xi_\cO})+a_0\bar{a}_1(G_{\Delta_1,2\xi_\cO}-\epsilon\partial_{\xi}G_{\Delta_1,2\xi_\cO})}\nn\\
&=&\Delta_\cO G_{\Delta_1,2\xi_\cO}+\xi_\cO\partial_{\xi}G_{\Delta_1,2\xi_\cO}\nn\\
&=&\Delta_\cO [g_{\Delta_1,2\xi_\cO}+O(\epsilon)]+\xi_\cO\partial_{\xi}[g_{\Delta_1,2\xi_\cO}+O(\epsilon)]\nn\\
&\stackrel{\epsilon\to0}{=}&\Delta_\cO g_{\Delta_1,2\xi_\cO}+\xi_\cO\partial_{\xi}g_{\Delta_1,2\xi_\cO}.
\eea
This is the contribution from the rank $2$ multiplet in the previous section. From above analysis it is obvious that the second kind of corrections is at higher  order in $\epsilon$. When we go to the higher orders, some possible divergences appear. Interestingly, these terms cancel with each other. For example, at rank $3$, the three possible terms are:
\begin{equation}
\begin{split}
&a_2\bar{a}_0\left(G_{\Delta_1,2\xi_\cO}+2\epsilon\partial_{\xi}G_{\Delta_1,2\xi_\cO}+\frac{(2\epsilon)^2}{2!}\partial_{\xi}^2G_{\Delta_1,2\xi_\cO}\right)
+a_1\bar{a}_1G_{\Delta_1,2\xi_\cO}\\
&+a_0\bar{a}_2\left(G_{\Delta_1,2\xi_\cO}-2\epsilon\partial_{\xi}G_{\Delta_1,2\xi_\cO}+\frac{(2\epsilon)^2}{2!}\partial_{\xi}^2G_{\Delta_1,2\xi_\cO}\right)
+O(\epsilon)\\
&=A_0G_{\Delta_1,2\xi_\cO}+A_1\partial_{\xi}G_{\Delta_1,2\xi_\cO}+A_2\partial_{\xi}^2G_{\Delta_1,2\xi_\cO}+O(\epsilon)\\
&\stackrel{\epsilon\to0}{=}A_0g_{\Delta_1,2\xi_\cO}+A_1\partial_{\xi}g_{\Delta_1,2\xi_\cO}+A_2\partial_{\xi}^2g_{\Delta_1,2\xi_\cO}.
\end{split}
\end{equation}
It is easy to see that $A_2$ has no divergence. For $A_0$ and $A_1$, one can find the potential divergence disappear due to the cancellation. In general, the terms with $k=n$ are not divergent, while the terms with $k<n$ consist of superficially divergent terms which cancel with each other such that these terms are finite as well.
One can prove this cancelation precisely and find the final results are identical to the ones obtained before. The details of the proof can be found in the appendix \ref{Cancelation}.

\section{Tauberian theorem and the spectral density} \label{Tauberian}

The tauberian theory plays an important role in the modern conformal bootstrap program  \cite{Pappadopulo:2012jk, Fitzpatrick:2012yx, Qiao:2017xif, Mukhametzhanov:2018zja}( see \cite{Goldberger:2014hca} for the Schrodinger  case). In this section, we want to use it to estimate the spectral density. The $\rho$ and $\kappa$ coordinates will  simplify our discussion here as well. The key point is that  the stripped four-point function in $t$ channel is dominated by the identity operator in the $x\to 1$ (or $\rho\to 1$) limit. This allows us to use the Hardy-Littlewood tauberian theorem to estimate the spectral density.

 Because the $y$ (or $\kappa$) dependence is always exponential, we want to expand it and match the asymptotical behaviour near $x=1$ (or $\rho=1$) in each order of $y$ (or $\kappa$). We will see that tauberian theorem can be used in each order. The crossing equations are:
\begin{equation}
\frac{1}{x^{2\Delta_{\cO}}}e^{2 \xi_{\cO}\frac{y}{x}}\sum_{\Delta,\xi,k}b_{\Delta,\xi,k}g^{(k)}_{\Delta,\xi}(x,y)=
(x\to1-x,y\to-y)
\end{equation}
where we pack the dynamical data into the coefficient $b_{\Delta,\xi,n}$,
and the GCA block is:
\begin{equation}
g^{(k)}_{\Delta,\xi}(x,y)=2^{2\Delta-2}\chi_{\Delta,\xi,k}(x,y)=2^{2\Delta-2}\partial_{\xi}^k\chi_{\Delta,\xi}(x,y).
\end{equation}
Our strategy is to expand both sides in $y$ and match the asymptotical behaviour at $x\to1$ order by order. In the $t$ channel, the general block is:
\begin{equation}
g_{\Delta,\xi}^{(k,t)}(x,y)=g^{(k)}_{\Delta,\xi}(1-x,-y)=2^{2\Delta-2}\frac{(1-x)^{\Delta}(1+\sqrt{x})^{2-2\Delta}}{\sqrt{x}}e^{\xi\frac{y}{(1-x)\sqrt{x}}}(\frac{y}{(1-x)\sqrt{x}})^k.
\end{equation}
Moving the input factor on the left-hand side to the right, we have the prefactor (which is just the $t$ channel identity because $g_{\text{identity}}^{(t)}$=1)
\begin{equation}
\frac{x^{2\Delta_\cO}}{(1-x)^{2\Delta_\cO}}e^{-2\xi_\cO\frac{y}{x(1-x)}}.
\end{equation}
Now we look at the $x\to1$ behaviour. Expanding both sides in $y$, we can get the asymptotical behaviour near $x=1$ from the right-hand side. In fact, at each order of $y$, the $x\to1$ behaviour is controlled by
the $t$-channel identity only. For example, in the $y^0$ order, the identity gives the most singular power law behaviour so we have (we use $\mathcal{G}^{y^k}$ or $\mathcal{G}^{\kappa^k}$ denote the $k$-th order part of the four-point function in the following):
\begin{equation}
\mathcal{G}^{y^0}\sim(1-x)^{-2\Delta_\cO}.
\end{equation}
When we go to the higher order of $y$, in both the  prefactor and the GCA ($t$ channel) blocks, $y$ is always combined with a  power law term $\frac{1}{1-x}$,  so at each order of $y$, the identity always dominates near $x=1$. At the order $y^k$, the asymptotical behaviour near $x=1$ is
\begin{equation}
\mathcal{G}^{y^k}\sim(1-x)^{-2\Delta_\cO-k}.
\end{equation}
Then we can estimate the spectral density for each order. We  use the $\rho$ and $\kappa$ coordinates to make the argument more explicitly. In these coordinates, all the
argument above is still valid, except that now we expand in $\kappa$ and look at the $\rho\to0$ behaviour. For technical reason, we need to write every $\kappa^k$ part of the four-point function as the Laplace transform of a ``weighted spectral density'' . This allows us to use the Hardy-Littlewood tauberian theorem in its general form, otherwise we need to rewrite the theorem in a different form. It turns out that\footnote{In the following of this section, by saying a four point function or its $\kappa^k$ order part, we always multiply them an additional factor $\frac{1-\rho^2}{4}$ because we want the $\rho$ dependence of the expanding block become a single pure power law.}
\begin{equation}
\mathcal{G}^{\kappa^k}(\beta)=\int_0^{\infty}d\Delta f^{(k)}(\Delta)e^{-\beta\Delta}
\end{equation}
where $e^{-\beta}=\rho$ and $f^{(k)}(\Delta)$ is the spectral density function. Because the GCA block in $\rho$ coordinate has a simple power-law form, one can easily read the spectral density function $f^{(k)}(\Delta)$ from
the dynamical coefficient $b_{\Delta,\xi,k}$. For example, at order $\kappa^0$
the spectral density function $f^{(0)}(\Delta)$ for the GGFT is
\begin{equation}
f^{(0)}(\Delta)=\sum_nP^t_{n,0}\delta(\Delta-\Delta_n)
\end{equation}
where $P^t_{n,0}$ is the GGFT coefficient \eqref{coef1}. Note that for GGFT, $\xi$ is fixed to be $2\xi_0$. Generically, $f^{(k)}(\Delta)$ will include the contributions from all possible $\xi$ for the specific $\Delta$. For the higher order of $\kappa$, the contributions from the multiplets should be included in
$f^{(k)}(\Delta)$. At order $\kappa^0$,  from the $t$ channel, we know that as $\rho\to1$ (that is, $\beta\to0$)
\begin{equation}
\mathcal{G}^{\kappa^0}(\beta)\sim 2^{4\Delta_\cO-1}(1-\rho)^{1-4\Delta_\cO}\sim 2^{4\Delta_\cO-1}\beta^{1-4\Delta_\cO} \qquad (\beta\to0).  \label{asymptotical behaviour}
\end{equation}
We may define the integrated spectral density function
\begin{equation}
F(\Delta)\equiv\int_0^{\Delta}d\Delta'f^{(0)}(\Delta').
\end{equation}
Then by the Hardy-Littlewood tauberian theorem, we get:
\begin{equation}
F(\Delta)\sim 2^{4\Delta_\cO-1}\frac{\Delta^{4\Delta_\cO-1}}{\Gamma(4\Delta_\cO)},  \qquad (\Delta\to\infty).  \label{integrated}
\end{equation}

There remains an important question: does the condition of tauberian theorem hold in the GCA case? For the case of CFT, thanks to the unitarity, the spectral density function is always non-negative so the
tauberian theorem is valid. Our theory is generically not unitary, so we have no convincing argument to show the validness of the tauberian theorem at present. However, note that the condition for the
tauberian theorem is actually more relaxed than the non-negativity of spectral density function, it is (see Theorem I.15.3 in \cite{2007Tauberian}):
\begin{equation}
f^{(0)}(\Delta)\geq -C\Delta^{4\Delta_\cO-2}  \label{tauberian condition}
\end{equation}
where $C\geq0$ is some constant, and the above condition need only to hold for sufficiently large $\Delta$ ($\Delta'\leq\Delta\leq\infty$ for some $\Delta'\geq0$). This condition seems gentle, and our
result for the integrated spectral density is only valid for those theories who satisfy this condition. In fact, we have checked that GGFTs marginally satisfy this condition in the appendix \ref{check density}. As a result, we also have checked that
the integrated spectral density of the GGFT has the above asymptotical behaviour.

 Just as in the CFT case, we can actually discuss the error estimates in the Hardy-Littlewood asymptotics \cite{Pappadopulo:2012jk}. The error is controlled by $\Delta*$, the lowest dimension of operators in the $\cO\times\cO$ OPE.
\begin{equation}
\mathcal{G}^{\kappa^0}(\beta)=2^{4\Delta_\cO-1}\beta^{1-4\Delta_\cO}(1+O(\beta^{\Delta*})).
\end{equation}
As usual, this power-suppressed error will not translate into a power-suppressed error term for the integrated spectral density function.  Instead, the best possible error estimate is only logarithmic:
\begin{equation}
F(\Delta)=2^{4\Delta_\cO-1}\frac{\Delta^{4\Delta_\cO-1}}{\Gamma(4\Delta_\cO)}(1+O(1/\text{log}\Delta)).
\end{equation}

\section{Shadow formalism and alpha space approach}
In CFT we have several approaches to the inversion formula besides the harmonic analysis, one of which is the shadow formalism \cite{Ferrara:1972ay,Ferrara:1972uq,Ferrara:1972xe,Ferrara:1973vz,Osborn_2012,SimmonsDuffin:2012uy,Karateev:2018oml}. In this section we set the stage for the shadow formalism of Galilean CFT. Besides, in one and two dimensions there is another slightly different method of diagonalizing the Casimir equations named the alpha space method \cite{Hogervorst:2017sfd,Rutter:2020vpw}, where an unusual boundary condition at $z=1$ is selected and the Hilbert space is supported only on $z\in(0,1)$.  Unlike the situation in CFT, we find that in GCFT, the  alpha space approach could be related to the shadow formalism.

\subsection{Shadow formalism of GCFT}
The starting point of the shadow formalism is the shadow transform, an intertwining operator between two principal series representations of the conformal group. Another highlighting concept is the Plancherel measure, a canonical measure on the unitary dual of the conformal group, appearing in the decomposition of the regular representation as a direct integral of unitary irreducible representations. In the following we explore on similar objects of the Galilean conformal group $ISO(2,1)$.

After analytic continuing the dimensions of a bosonic primary operator $\cO$ onto the principal series $(\Delta=1+is,\xi=ir),\,r \in \mathbb{R}^{*},\,s\in\mathbb{R}$,\footnote{For $\xi=0$, the discussion is the same as in CFT$_1$, hence we focus on the $\xi\neq0$ sector.}, we can construct a shadow operator $\Tilde{\cO}$, non-local with respect to $\cO$,
\begin{align}
    \Tilde{\cO}(x,y)&=A(\Delta,r)\int_{\mathbb{R}^2}dx'dy'\greenfunction{\Tilde{\cO}(x,y)\Tilde{\cO}(x',y')}\cO(x',y')\nonumber\\
                &=A(\Delta,r)\int_{\mathbb{R}^2}dx'dy'\frac{1}{|x-x'|^{4-2\Delta}}e^{-2ir \frac{y-y'}{x-x'}}\cO(x',y')\nonumber
\end{align}
transforming as an primary operator of weight $(2-\Delta,-\xi)$. The undetermined pre-factor $A(\Delta,r)$ is to ensure the equivalence of two representations $\Tilde{\Tilde{\cO}}=\cO$, or equivalently
\begin{equation}\label{shadow1}
    \int dxdy\,\greenfunction{\Tilde{\cO}(x_1,y_1)\Tilde{\cO}(x,y)}\greenfunction{\cO(x,y)\cO(x_2,y_2)}=\frac{\delta(x_1-x_2)\delta(y_1-y_2)}{A(\Delta,r)A(2-\Delta,-r)}.
\end{equation}
This leads to the vertex-graph identity,
\begin{equation}
\greenfunction{\phi(x_1,y_1)\phi(x_2,y_2)\cO(x,y)}=\int_{\mathbb{R}^2}dx_0dy_0\greenfunction{\phi(x_1,y_1)\phi(x_2,y_2)\Tilde{\cO}(x_0,y_0)}\greenfunction{\cO(x_0,y_0)\cO(x,y)}.
\end{equation}
Actually, from
\eqref{shadow1}, we can read its left-hand side
\begin{align}
\text{LHS}&=\int_{\mathbb{R}^2}dxdy\,|x-x_1|^{-2\Delta}|x-x_2|^{2\Delta-4}e^{-2i r\frac{y(x_1-x_2)+x(y_1-y_2)+x_1y_2-x_2y_1}{(x-x_1)(x-x_2)}}\nonumber\\
    &=\frac{\pi}{| r|}\delta(x_1-x_2)\int_{\mathbb{R}}dx\,|x-x_1|^{-2}e^{-2i r\frac{y_1-y_2}{x-x_1}}\nonumber\\
    &=\frac{\pi^2}{ r^2}\delta(x_1-x_2)\delta(y_1-y_2),
\end{align}
where in the first line the integration of $y$ contributes to $\delta(x_1-x_2)$, in the second line the simplification is due to $x_1=x_2$ and in the last line we change the variable $\frac{1}{x-x_1}=t$. This determines $A(\D,r)=\frac{i|r|}{\pi}$.

There are several remarkable points:
\begin{itemize}
    \item All the terms are distributional and hence are taken as principal values or as analytic continuations if necessary. For $\xi=0$ the integral is divergent due to the reducibility of representations (or null states), and we need to adopt the shadow transform of $SL(2,\mathbb{R})$.
    \item For $4-2\Delta=-2n-1,\,n\in \mathbb{N}$, the distribution $|x|^{-2n-1}$ in the shadow transform seems to be ill-defined as in 1d CFT, where this phenomenon corresponds to the discrete series of $SL(2,\mathbb{R})$. Here the key point is that we must integrate out $y$ firstly, then the apparent divergence due to $|x|^{-2n-1}$ disappears and $\Gamma(\D)$-like factors do not appear in $A(\D,r)$.
    \item The condition of principal series $(\Delta=1+is,\xi=ir),\,r \in \mathbb{R}^{*},\,s\in\mathbb{R}$ means that the representation admits an inner product,
    \begin{equation}
        (\cO_1,\cO_2)=\int_{\mathbb{R}^2}dxdy\,(\cO_1(x,y))^*\cO_2(x,y)
    \end{equation}
    which is invariant under \eqref{Finite}.
    \item The factor $A(\Delta,r)A(2-\Delta,-r)\sim r^2$, compared with CFT, should be proportional to the Plancherel measure\cite{Simmons-Duffin:2017nub}. We find in page 114 of \cite{ASENS_1973_4_6_1_103_0} the Plancherel measure of Poincare group. The measure of massive and tachyonic representations have non-vanishing supports on the unitary dual, and our result matches the tachyonic ones up to numerical factors.
    \item The co-adjoint orbits (mass-shell) of tachyonic representations are one-sheeted hyperboloids, and imaginary mass $im\leftrightarrow -im$ label the same representation. Our shadow transform $\xi\leftrightarrow -\xi $ shares the similar feature.
\end{itemize}
As in CFT, the partial waves for four identical primary operators $\phi$ with $(\Delta_0, r_0),\, r_0\in \mathbb{R}^{\times}$ can be constructed as (we use the same notation $\Psi$ here),
\begin{equation}\label{shadow2}
        \Psi_{\Delta,i r}(x,y)=\frac{1}{\greenfunction{\phi_1\phi_2}\greenfunction{\phi_3\phi_4}}\int_{\mathbb{R}^2}dx_5dy_5\,\greenfunction{\phi_1\phi_2 \cO_{\Delta,i r}(x_5,y_5)}\greenfunction{\Tilde{\cO}(x_5,y_5)\phi_3\phi_4}
\end{equation}
where $ r\in \mathbb{R}^{\times}$, $\phi_i=\phi(x_i,y_i)$.
The detailed evaluation of \eqref{shadow2} is as follows:
\begin{align}
    \Psi_{\Delta,i r}(x,y) &=\frac{1}{\greenfunction{\phi_1\phi_2}\greenfunction{\phi_3\phi_4}}\int_{\mathbb{R}^2}dx_5dy_5\,\greenfunction{\phi_1\phi_2 O_{\Delta,i r}(x_5,y_5)}\greenfunction{\Tilde{O}(x_5,y_5)\phi_3\phi_4}\nonumber\\
    &=A(\Delta, r)\int_{\mathbb{R}^2}dx_5dy_5\,\left(\frac{|x_{12}|}{|x_{15}||x_{25}|}\right)^{\Delta}\left(\frac{|x_{34}|}{|x_{35}||x_{45}|}\right)^{2-\Delta}e^{-i r\left(\frac{y_{15}}{x_{15}}+\frac{y_{25}}{x_{25}}-\frac{y_{35}}{x_{45}}-\frac{y_{45}}{x_{45}}+\frac{y_{34}}{x_{34}}-\frac{y_{12}}{x_{12}}\right)}\nonumber\\
    &=A(\Delta, r)\int_{\mathbb{R}^2}dx_5dy_5\,\left(\frac{|x|}{|x_5||x-x_5|}\right)^{\Delta} |1-x_5|^{\Delta-2} e^{-i r\left( y_5\frac{x^2_5-2x_5+x}{x_5(x-x_5)(1-x_5)}+\frac{y}{x}\frac{x_5}{x-x_5}\right)}\nonumber
\end{align}
where $ r\in \mathbb{R}^{*}$, $\phi_i=\phi(x_i,y_i)$. In the third line we have fixed the gauge to the standard conformal frame by conformal covariance. Since the exponential factor is just a phase, the integration of $y_5$ gives a delta function,
\begin{equation}\label{shadow4}
    \delta\left( r\frac{x^2_5-2x_5+x}{x_5(x-x_5)(1-x_5)}\right)=\frac{1}{| r|}\frac{|x_5(x-x_5)(1-x_5)|}{|x_{+}-x_{-}|}\left(\delta(x_5-x_{+})+\delta(x_5-x_{-})\right)\nonumber
\end{equation}
where $x_{\pm}=1\pm\sqrt{1-x}$. Then the resulting partial waves are
\begin{align}
    \Psi_{\Delta,i r}(x,y)=\frac{i\,\text{sign}( r)}{2\pi}\left(\frac{x^{\Delta}(1+\sqrt{1-x})^{2-2\Delta}}{\sqrt{1-x}}e^{-i r\frac{y}{x\sqrt{1-x}}}+(\Delta\to2-\Delta,\, r\to- r)\right)\label{shadow3}
\end{align}

The conformal blocks can be split out by different monodromy around $x=0$ as in  \cite{SimmonsDuffin:2012uy}. However notice that the second delta function in \eqref{shadow4} is located in $x_5\in(1,\infty)$. If we restrict the integration region to $x_5\in (0,1)$, the physical block is automatically selected. And this also happens in 1d CFT. In higher dimensional CFT, we pick out the physical block by changing the integration region to Lorentzian spacetime $x_5\in \mathbb{R}^{d-1,1}$.

There are several different features from the harmonic analysis in section 3. The shadow integral \eqref{shadow2} vanishes in $x > 1$. The shadow integral automatically guarantees that the partial waves are eigenfunctions of the Casimir operators, restricting the ones in $(1,\infty)$ to zero. The discrete partial waves at $\Delta=\frac{5}{2}+n,\,n\in \mathbb{N}$ are absent and the boundary condition at $x=1$ is different.

\subsection{Alpha space approach to GCFT}
The alpha space approach in CFT is defined by solving the Sturm-Liouville problem of the Casimir operators restricted in the region $z\in (0,1)$, leading to an integral transform slightly different from the inversion formula. We seek a similar transform for Galilean CFT.

Now we restrict the region to $(x,y)\in D=(0,1)\times\mathbb{R}$ mimicking the alpha space method, and in radial coordinates\footnote{For the Harmonic analysis in section 3, the transformation to radial coordinates are piece-wise defined in the region $(x,y)\in \mathbb{R}^2$ and we get no simplification.} the region is $(\r,\k)\in(0,1)\times\mathbb{R}$ denoted also by $D$. The Casimir equations are
\begin{align}
    C_1:&\qquad\left(\p^2_{\k}-\xi^2\right)f_{\D,\xi}(\r,\k)=0,\nonumber \\
    C_2:&\qquad\left(\r\p_{\r}\p_{\k}- \frac{1+\r^2}{1-\r^2}\p_{\k}+\xi(\D-1)\right)f_{\D,\xi}(\r,\k)=0.
\end{align}
By changing of variables $f_{\D,\xi}(\r,\k)=\frac{1-\r^2}{\r}\,h_{\D,\xi}(\r,\k)$ and $t=-\log \r \in (0,+\inf)$ the new region is denoted by $(t,\k)\in D= (0,+\inf)\times \mathbb{R}$, and we get,
\begin{align}
    C_1:&\qquad\left(\p^2_{\k}-\xi^2\right)h_{\D,\xi}(t,\k)=0, \nonumber\\
    C_2:&\qquad\left(\p_{t}\p_{\k}-\xi(\D-1)\right)h_{\D,\xi}(t,\k)=0,
\end{align}
and the general solutions are linear combinations of,
\begin{align}
    h_{\D,\xi}(t,\k)&=e^{(\D-1)t}e^{\xi \k},\nonumber\\
    h_{2-\D,-\xi}(t,\k)&=e^{(1-\D)t}e^{-\xi \k}.
\end{align}
The measure and inner product are simplified to,
\begin{equation}
(f,g)=\int_{D} dx dy \frac{1}{x^4} f^* g=\int^{\inf}_0 dt \int_{\mathbb{R}} d\kappa f^* g.
\end{equation}
Now we need to specify the boundary conditions. Formally this is to determine possible self-adjoint extensions of symmetric operators, see eg. chapter 10 of \cite{reed1975ii}. The quadratic Casimir $C_1$ sets $\xi=ir,\,r\in\mathbb{R}$ as before. Then the Hermitian condition of $C_2$ is
\begin{align}\label{alphabc}
    (C_2 f,g)-(f,C_2 g)&=\int^{\inf}_0 dt \int_{\mathbb{R}}d\k\,\left[\p_t(\p_{\k}f^*\,g)-\p_{\k}(f^*\p_t g)\right]\nonumber\\
                        &=\int_{\mathbb{R}}d\k\,  (\p_{\k}f^*(0,\k))\, g(0,\k).
\end{align}
In the second line, other boundary terms have been dropped by the fall-off conditions at infinity, leading to $\Delta=1+is$.

To make the remaining boundary term in \eqref{alphabc} vanish, we need to choose a specific boundary condition at $t=0$, and then check if this condition leads to a self-adjoint extension of $C_2$. As we will show later there are infinitely many in-equivalent boundary conditions corresponding to different self-adjoint extensions. We pick one of them to match the result of shadow formalism,
\begin{equation}\label{shadowbc}
    f(0,\k)=f(0,-\k),\qquad g(0,\k)=g(0,-\k),
\end{equation}
namely, boundary values are even functions of $\k$. Then the basis are $(h_{\D,\xi}+h_{2-\D,-\xi})$, and back to the $(x,y)$ coordinates, are,
\begin{equation}
    \Psi_{\Delta, r}= \chi_{\Delta, \xi}+ \chi_{2-\Delta,-\xi},
\end{equation}
in which the normalization is fixed by,
\begin{equation}
    (\Psi_1,\Psi_2)=4\pi^2\delta( r_1- r_2)\delta(s_1-s_2).
\end{equation}
In fact, we need to use the theory of self-adjoint extension of unbounded operators \cite{reed1975ii}. Recall that the deficiency subspaces of a closed symmetric operator $A$ are defined by
\begin{align}
    D_{+}&=\ker(A^\dagger-i),\nonumber\\
    D_{-}&=\ker(A^\dagger+i),
\end{align}
and different self-adjoint extensions $A_{U}$ correspond to different unitary operators $U$ from $D_{+}$ to $D_{-}$. Then the domain of $A_U$ is spanned by
\begin{align}
    D(A_U)=\left\{\phi+\phi_{+}+U(\phi_+):\phi\in D(A),\,\phi_+\in D_{+}\right\},
\end{align}
and the action of $A_U$ are extended by
\begin{equation}
    A_U(\phi+\phi_{+}+U(\phi_+))=A(\phi)+i\phi_{+}-i U(\phi_+).
\end{equation}
Now for $C_2$, starting from $D(C_2)=\{f\in L^2(D):f(0,\k)=0\}$, the deficiency subspaces are
\begin{align}
    D_{+}&=\left\{h_{a,+}(t,\k)=e^{-t/a}e^{-ia\k}:a>0\right\},\nonumber\\
    D_{-}&=\left\{h_{a,-}(t,\k)=e^{-t/a}e^{ia\k}:a>0\right\}.
\end{align}
Any unitary operator from $D_+$ to $D_-$ leads to a different boundary condition. We choose $U$ such that $U(h_{a,+})=h_{a,-}$ then,
\begin{align}
    D(C_{2,U})=\left\{f+h_{a,+}+h_{a,-}:f\in D(C_2),\,h_{a,\pm}\in D_{\pm}\right\},
\end{align}
then for a generic function $f\in D(C_{2,U})$ we have $f(0,k)=f(0,-k)$, matching with \eqref{shadowbc}. As a result we can construct an alpha space method sharing similar features with the shadow formalism in Galilean CFT$_2$. This is different from the case in CFT.

\section{Conclusion and Discussions}
In this work, we tried to develop Galilean conformal bootstrap, paying special attention to its analytical aspects. A general Galilean conformal field theory is not unitary. Even though all the primary operators have positive conformal dimensions, the quasi-primary states have negative norms. Such a non-unitarity is mild in the sense that the primary states are always of positive norm\footnote{Similar phenomenon happens in holographic warped CFT, where the Kac level is negative leading to mild non-unitarity \cite{Apolo:2018eky}. It was shown that the modular bootstrap is still viable in this case.}. Our study suggests that the conformal bootstrap might still be viable. In particular we showed that analytic Galilean bootstrap is valid, at least for generalized Galilean free theory.

Our study on the Galilean conformal bootstrap is based on the global symmetry. One reason is that the local GCA block is unknown and the infinite dimensional symmetry cannot be applied. Nevertheless, there appear some novel features in our study. Due to the presence of multiplets, their contribution should be taken into account appropriately. Our first study was on the multiplets and their multi-point functions. Especially we computed the global blocks of the multiplets and showed how the four-point functions can be expanded in terms of  these global blocks. The appearance of multiplets is an essential feature in Galilean conformal bootstrap.

Our second study was on the harmonic analysis of GCA, which paves the way for further analytic study. Due to the fact that GCA is not semi-simple, the harmonic analysis is quite different from the usual conformal case and is more subtle. Due to the semidirect product structure of GCA, the technical treatment followed closely  the one in CFT$_1$,  but the Galilean conformal partial wave is very different. Especially, in order to define a {\it bona fide} Hilbert space, we had to use quartic Casimir and introduce proper measure and inner product.  We found that for GCA, there are principal series representation, as well as discrete series representation, similar to CFT$_1$.  With the GCPWs, we established an inversion formula which  allow us to read data of GCFT. We pointed out that a rank-$n$ multiplet appear as a multipole of order $n$ in the inversion function. By the way, we showed  that the GCPWs could not be reached by taking non-relativistic limit on CPWs of 2d conformal group, as the normalizable condition and boundary conditions should be analyzed in a way independent of the non-relativistic limit.

In order to test our formalism, we studied the generalized Galilean free theory in several different ways. Firstly we  constructed explicitly the level-$1$ double trace operators, which form a rank-$2$ multiplet. Secondly we studied the crossing equations by doing double series expansion and read the data of GGFT.  Thirdly we tried to do  inversion integral and found the same set of data successfully from the inversion function. Finally, we found  that GGFTs can be obtained by taking the non-relativistic limit of 2d GFTs. Our study of GGFT provided a nontrivial example of using analytic conformal bootstrap to a non-unitary theory.

In this work, we focused on the GCFT whose spectrum is made of the operators with nonvanishing $\xi$ charge. The $\xi=0$ case is quite different and subtle, but could be essential to have a complete bootstrap analysis. It is indispensable to have a thorough study of this case \cite{Hao:2021a,Hao:2021b}.

The study in the present work can be extended to several directions.   Considering the fact that the similar multiplets appear in LCFT, it would be interesting to discuss the conformal bootstrap for such CFT,   see \cite{Banerjee:2019uxo} for relevant study. Moreover, the viability of applying conformal bootstrap to the theories with conformal-like symmetry and mild non-unitarity suggests that other theories sharing similar features, like warped CFT \cite{Hofman:2011zj,Detournay:2012pc,Chen:2020juc} and anisotropic Galilean CFTs \cite{Chen:2019hbj}, should be investigated in more details.

Our study focused on the 2d case. It would be interesting to study higher dimensional Galilean CFT. In this case, one has to take   the angular momentum quantum number into account\cite{Chen:2020}. Another interesting direction is to investigate Galilean conformal bootstrap numerically. This is a field which has not been explored. There are some efforts to do numerical investigations on the theories without unitarity, see the review \cite{Poland:2018epd} for details. It is also interesting to further study the Galilean conformal bootstrap analytically. For example, it is possible to develop the analytic functional method \cite{Mazac:2016qev,Mazac:2018mdx,Mazac:2018ycv,Mazac:2019shk,Paulos:2019gtx} for the Galilean conformal bootstrap. One can also try to develop our GCA inversion formula further just like in the case of CFT ($d\geq2$) \cite{Caron-Huot:2017vep,Simmons-Duffin:2017nub} and CFT$_1$ \cite{Mazac:2018qmi}.

One essential obstacle in the Galilean conformal bootstrap beyond GGFT is the non-independence of the blocks of multiplets. For example, a rank-$1$ block can be rewritten as an infinite summation over the blocks of arbitrary rank up to infinity,
\begin{equation}
\chi_{\Delta_0,\xi_0+c,1}(\rho,\kappa)=\sum_{r=0}^{\infty}\frac{c^r}{r!}\chi_{\Delta_0,\xi_0,r+1}(\rho,\kappa).
\end{equation}
For the GGFT, this is not a problem as the spectrum of $\xi$ is fixed. However,
if we consider a generic Galilean CFT rather than generalized free theories, the block expansion of a four-point function is not unique. One can easily see this from the above observation that the blocks of multiplets are not independent. Another way to understand this is as follows. Recall that our GCA blocks have the following form:
\begin{equation}
\frac{1-\rho^2}{4}\chi_{\Delta,\xi,k}(\rho,\kappa)=\rho^{\Delta}(-\kappa)^ke^{-\xi \kappa}
\end{equation}
 We can multiply a 4-point function $\mathcal{G}(\rho,\kappa)$ with a factor $\frac{1-\rho^2}{4}$, then Taylor expand it in $\kappa$.  We try to match the 4-point function and its block expansion order by order in $\kappa$. At order $\kappa^m$, only blocks $\rho^{\Delta}(-\kappa)^ke^{-\xi \kappa}$ with $k\leq m$ contribute. So at every order $\kappa^m$, new blocks $\rho^{\Delta}(-\kappa)^me^{-\xi \kappa}$ are introduced. However, it is easy to see that there are no constraints on the spectrum of $\xi$ of these new introduced blocks. So at every order of $\kappa$, these $\xi$s will be free parameters. As a result, the block expansion is not unique. In the GGFT case, we have the condition $\xi=2\xi_\cO$ so we get a unique block expansion.

This phenomenon does not occur in the CFT case, since it relies on the existence of infinite rank multiplets. Even in LCFTs, e.g. Logarithmic generalized free theories \cite{Banerjee:2019uxo}, there are only finite rank multiplets in the quasiprimary $\cO\cO$ OPE.
Hence to make further progress on Galilean conformal bootstrap analytically and numerically, we need more physical inputs for GCFTs. One possible choice is to consider the local Galilean conformal symmetry, giving more tight constraints on the block expansions. Another way is to assume that the spectrum contains only finite rank multiplets as in the GGFT case that we expand the four-point functions at $\xi=2\xi_\cO$, then  figure out the implications in the crossing equations.

The third option is from the observation  that the sign of block coefficients\eqref{coef1} $P^t_{n,k}$ is positive if $\Delta_\cO\geq 1/4$ and $\xi_{\cO}\geq 0$, although there are states of negative norms from the descendants and multiplets. If we re-expand at $\xi=2\xi_\cO-c$, $c\in[0,2\xi_\cO)$, the coefficients remain positive. This suggests that in the bootstrap problem we may start from the positive-coefficient assumption, and different crossing solutions may correspond to the same GCFT, like a gauge structure with respect to $\xi$ on the space of GCFT.



\section*{Acknowledgments}
We would like to thank Luis Apolo, Jiaxin Qiao, Wei Song, Ning Su, Gang Yang, Yufan Zheng for valuable discussions.
The work is in part supported by NSFC Grant No. 11335012, No. 11325522 and No. 11735001.

\vspace{1cm}
\appendix
\renewcommand{\appendixname}{Appendix~\Alph{section}}

\section{Review of LCFTs} \label{LCFT}
In this section, we give a quick review on the multiplets in the logarithmic conformal field theories(LCFTs).
We will review the basic structures and results in the LCFTs, and also review the key points and tricks to access them. We will only give some short and simple examples. For more discussions, please see the references \cite{Nagi:2005sb,Flohr:2001tj,Flohr:1997wm,Cardy:2013rqg,Hogervorst:2016itc}. We can see that the multiplets in the GCFTs can be discussed in a parallel way.

Consider the primary states in the conformal field theories (due to the state-operator correspondence, we use the following notation),
\begin{equation}
L_n\cO=0,\ \ n>0,\ \ \ L_0\cO=\Delta \cO
\end{equation}
where $\Delta$ is a matrix. In the LCFTS, $\Delta$ becomes triangular, which can be written into the Jordan blocks. A rank-$2$ multiplet has the following properties,
\begin{equation}
L_0\cO_0=\Delta \cO_0+\cO_1,\ \ L_0\cO_1=\Delta \cO_1.
\end{equation}
In the following discussion, we use the same notation $\Delta$ for the matrix, as well as for its diagonal elements. In the LCFTs, the correlation functions of multiplets have logarithmic behaviour.  What's more, there are negative norm states in such theories, where the Hilbert space is not positive definite. Please see \cite{Cardy:2013rqg,Hogervorst:2016itc} for more detailed discussions on the norm.

Generally, we denote the primary operators in a rank-$r$ multiplet as $\cO_i,i=0,\cdots,r-1$.
From the BCH formula, one gets in general
\begin{equation}
[L_n,\cO(x)]=(x^{n+1}\partial_x+(n+1)x^n\Delta)\cO(x).
\end{equation}
Now note that $\Delta$ is a matrix. For example, for a rank-$2$ multiplet,
\be
[L_n,\cO_0(x)] =(x^{n+1}\partial_x+(n+1)x^n\Delta)\cO_0(x)+(n+1)x^n\cO_1,\ee
\be
[L_n,\cO_1(x)] =(x^{n+1}\partial_x+(n+1)x^n\Delta)\cO_1(x).
\ee
This in turn gives the infinitesimal transformation of the multiplets,
\begin{equation}
x\rightarrow x+\epsilon(x),\ \ \delta_\epsilon \cO_i(x)=\Delta_{ij}\partial_x\epsilon \cO_i+\epsilon\partial_x \cO_i
\end{equation}
and the OPE between the stress tensor and the multiplets,
\begin{equation}
T(x)\cO_i(0)\sim\frac{\Delta_{ij}}{x^2}\cO_i(0)+\frac{1}{x}\partial_{x'}\cO_i(x')|_{x'\rightarrow0}.
\end{equation}
Compared to the usual conformal field theories without multiplets,
\begin{equation}
\delta_\epsilon \cO(x)=\Delta\partial_x\epsilon \cO+\epsilon\partial_x \cO,
\end{equation}
\begin{equation}
T(x)\cO(0)\sim\frac{\Delta}{x^2}\cO(0)+\frac{1}{x}\partial_{x'}\cO(x')|_{x'\rightarrow0},
\end{equation}
there is a simple replacement rule in the expressions above, which  is linear in $\Delta$,
\begin{equation}\label{reprule1}
\cO\rightarrow \cO_i,\ \ \ \Delta\rightarrow \Delta_{ij}
\end{equation}
From the infinitesimal transformation, one can get the finite transformation behaviour of the multiplets: under $x\rightarrow F(x)$,
\begin{equation}
\cO_i(x)\rightarrow\sum_{k=0}^{r-i-1}\frac{1}{k!}\partial_\Delta^k(F'(x))^\Delta \cO_{i+k}.
\end{equation}
Note that here is another replacement rule,
\begin{equation}\label{reprule2}
f(\Delta,\cO)\rightarrow \sum_{k=0}^{r-i-1}\frac
{1}{k!}\partial_\Delta^k f(\Delta,\cO_{i+k})
\end{equation}
where $f$ is a general function containing the operators in discussion. Note that \eqref{reprule1}(where $f$ is linear of $\Delta$) is a special case of \eqref{reprule2}.

One can understand this replacement in the following heuristic way. Define
\begin{equation}\label{trick}
\tilde{\cO}(x,\alpha)=\sum_{k=0}^{r-1}\alpha^{r-1-k}\cO_k(x),\ \ \alpha^r=0,
\end{equation}
on which there is
\begin{equation}
L_0\tilde{\cO}(x,\alpha)=(\Delta+\partial_\alpha)\tilde{\cO}(x,\alpha).
\end{equation}
One can then expand the expression with respect to $\alpha$ to get the $\cO_i$ terms. In general,  the action of $f(L_0)$ on $\cO$ in usual CFTs without multiplet reads
\begin{equation}
f(L_0) \cO=f(\Delta,\cO).
\end{equation}
Expanding in both $L_0$ and $\alpha$, one can get the replacement rule \eqref{reprule2},
\begin{equation}
f(L_0)\cO_i=\sum_{k=0}^{r-1-i}\frac{1}{k!}\partial_\Delta^kf(\Delta)\cO_{i+k}.
\end{equation}
For more than one operators with weight $\Delta_i$ involved, the replacement rule becomes
\begin{equation}\label{reprule3}
f(\Delta_1,\Delta_2,\cdots,\cO_{1},\cO_{2},\cdots)\rightarrow\sum_{k_1=0}^{r_1-i_1-1}\sum_{k_2=0}^{r_2-i_2-1}\cdots\frac
{1}{k_1!}\partial_{\Delta_1}^{k_1}\frac
{1}{k_2!}\partial_{\Delta_2}^{k_2} \cdots f(\Delta_1,\Delta_2,\cdots,\cO_{1,i_1},\cO_{2,i_2},\cdots).
\end{equation}

We now review the two-point functions of a rank-$r$ multiplet. We denote
\begin{equation}
G_{i,j}(x_1,x_2)=\langle \cO_i(x_1)\cO_j(x_2)\rangle,\ \ i,j=0,\cdots r-1.
\end{equation}
We can also define the matrix
\begin{equation}\label{lbc}
G_{i,j}(x_1,x_2)=0,\hs{3ex}  \mbox{$i$ or $j \geq r$,}
\end{equation}
to write the following differential equations in a unified way.
The vacuum is invariant under $L_0,L_{\pm1}$, thus by the Ward identities we have
\begin{equation}
DG_{i,j}=0
\end{equation}
where $D$ are the differential operators being comprised of $L_0,L_{\pm1}$. For $L_{-1}$, there is
\begin{equation}
(\partial_{x_1}+\partial_{x_2})G_{i,j}=0,
\end{equation}
which means $G_{i,j}$ is translationally invariant,
\begin{equation}
G_{i,j}(x_1,x_2)=G_{i,j}(x)
\end{equation}
with
\begin{equation}
x=x_1-x_2.
\end{equation}
Considering $L_0$, we get
\begin{equation}
x\partial_x G_{i,j}(x)=-(\Delta_1+\Delta_2)G_{i,j}(x)-G_{i+1,j}(x)-G_{i,j+1}(x).
\end{equation}
Considering $L_1$, we find
\begin{equation}
x\partial_x G_{i,j}(x)+(\Delta_1-\Delta_2)=-(\Delta_1+\Delta_2)G_{i,j}(x)-2G_{i+1,j}(x).
\end{equation}
Now, from the action of $L_0$ and $L_1$, we  read
\begin{equation}
\Delta_1=\Delta_2,\ \ \ G_{i,j}=G_{j,i}.
\end{equation}
If there is no $G_{i+1,j}$ term in the differential equations, the solution is the usual one
\begin{equation}
G(x)=\frac{1}{x^{2\Delta}}
\end{equation}
Therefore, we can make the ansatz
\begin{equation}
G_{i,j}(x)=\frac{1}{x^{2\Delta}}\tilde{G}_{i,j}(x).
\end{equation}
From the properties of the two-point functions and the differential equations, we know that
\begin{equation}
G_{i,j}=G_{j,i}
\end{equation}
and  only the summation of $i$ and $j$ matters in the differential equation. Therefore we may introduce
\begin{equation}
\G_k=G_{i,j}, \hs{2ex}\tilde{\G}_k=\tilde G_{i,j},\hs{2ex} \mbox{with $k=i+j$.}
\end{equation}
From the differential relations above, we have
\begin{equation}
x\partial_x\tilde{\G}_k(x)=-2\tilde{\G}_{k+1}
\end{equation}
and
\begin{equation}
\tilde{\G}_{k}=0, \hs{2ex}\mbox{for $k\geq r$.}
\end{equation}
 Finally we  get the solutions
\begin{equation}\label{solution1}
\G_i=\sum_{k=0}^{r-i-1}\frac{1}{k!}N_k\partial_\Delta^k (\frac{1}{x^{2\Delta}})
\end{equation}
which can also seen as the replacement rule \eqref{reprule3}, with the condition
\begin{equation}
\G_{r}=0,\ \ \G_{r-1}=\frac{1}{x^{2\Delta}}.
\end{equation}
Note that there are $r$ undetermined constants $N_0,\cdots,N_{r-1}$ in \eqref{solution1}. But there are further degrees of freedom to change the basis. For example, in the rank-2 case, one has
\begin{equation}
\cO_0\rightarrow a\cO_0+b\cO_1,
\end{equation}
\begin{equation}
\cO_1\rightarrow a\cO_1
\end{equation}
to keep the multiplet relation invariant,
\begin{equation}
L_0\cO_0=\Delta \cO_0+\cO_1,\ \ L_0\cO_1=\Delta \cO_1.
\end{equation}
For the rank-$r$ case,  there are totally $r$ degrees of freedom so that one can make the two-point function into the canonical form
\begin{equation}
\G_{r-k-1}=\frac{1}{k!}\partial_\Delta^k(\frac{1}{x^{2\Delta}}),\ \ k\geq0.
\end{equation}
For the rank-2 case, we write them explicitly as follows,
\begin{equation}
G_{11}=0,\ \ G_{01}=G_{10}=\frac{1}{x^{2\Delta}},\ \ G_{00}=\frac{-2\log x}{x^{2\Delta}}.
\end{equation}

One can get the three-point function in the similar way (please see \cite{Nagi:2005sb,Flohr:2001tj,Hogervorst:2016itc} for details),
\begin{equation}
G_{ijk}=\langle \cO_{1,i}(x_1)\cO_{2,j}(x_2)\cO_{3,k}(x_3)\rangle.
\end{equation}
But now, one cannot rotate the basis anymore, so there are undetermined constants,
\begin{equation}
G_{r_1-1,r_2-1,r_3-1}=c_{r_1-1,r_2-1,r_3-1}x_{12}^{-\Delta_{123}}x_{23}^{-\Delta_{231}}x_{31}^{-\Delta_{312}},
\end{equation}
where
\begin{equation}
x_{ab}=x_a-x_b,
\end{equation}
\begin{equation}
\Delta_{abc}=\Delta_a+\Delta_b-\Delta_c,
\end{equation}
\begin{equation}
G_{r_1-1,r_2-1,r_3-2}=\partial_{\Delta_3}G_{r_1-1,r_2-1,r_3-2}+c_{001}x_{12}^{-\Delta_{123}}x_{23}^{-\Delta_{231}}x_{31}^{-\Delta_{312}}.
\end{equation}
Generically ,
\begin{equation}
G_{r_1-1-i,r_2-1-j,r_3-1-k}=\sum_{q=0}^{i}\sum_{w=0}^{j}\sum_{e=0}^{k} c_{qwe}\frac{1}{q!}\frac{1}{w!}\frac{1}{e!}\partial_{\Delta_1}^q\partial_{\Delta_2}^w\partial_{\Delta_3}^eG_{000}.
\end{equation}

From the finite transformation rule, one can define the conjugate states,
\begin{equation}
\langle \cO_i|=\lim_{x\rightarrow \infty}\sum_{k=0}^{r-i-1}\langle 0|\cO_{i+k}(x)\partial_{\Delta}^k x^{2\Delta}\frac{1}{k!}.
\end{equation}
And the inner products of the primary states in a rank-$r$ multiplet are
\begin{equation}
\langle \cO_i|\cO_j\rangle=\delta_{i+j,r-1}.
\end{equation}
This is helpful to the discussion of the global blocks. 

The differential equations could be read from another point of view. For example, in the rank-2 case, one has
\bea
L_0\cO_0=\Delta \cO_0+\cO_1 &\Rightarrow &
(L_0-\Delta)\cO_0=\cO_1, \\
L_0\cO_1=\Delta \cO_1
& \Rightarrow
&
(L_0-\Delta)^2\cO_0=0.
\eea
Such structure holds in generic case. For a rank-$r$ multiplet, there is
\be
(L_0-\Delta)^{r-i}\cO_i=0,\hs{3ex}i=0,\cdots r-1,
\ee
where $\cO_i$ is the $i$-th operator in the multiplet. This may lead to the logarithmic behavior in the correlation function. Consider the function in usual CFT obeying the differential equation
\be
Df=d(\Delta)f,
\ee
where $D$ is a differential operator and $f$ is a function of the operator, then in a LCFT, the corresponding differential equation becomes
   \begin{equation}
 (D-d(\Delta))^{r-i}f_i=0 \label{fieq}
\end{equation}
where  $f_i=f(\cO_i)$ is the function of the $i$-th operator $\cO_i$ in a multiplet.
One can see this from the replacement rule. Actually, from \eqref{reprule2}, there is
\begin{equation}
Df_i=\sum_{k=0}^{r-1-k}\frac{1}{k!}\partial_\Delta^kd(\Delta)f_{i+k}=\sum_{k=1}^{r-1-k}\frac{1}{k!}\partial_\Delta^kd(\Delta)f_{i+k}+d(\Delta)f_i,
\end{equation}
which gives
\begin{equation}
(D-d(\Delta))f_i=\sum_{k=1}^{r-1-k}\frac{1}{k!}\partial_\Delta^kd(\Delta)f_{i+k}
\end{equation}
with the condition
\begin{equation}
f_r=0.
\end{equation}
One gets \eqref{fieq}.
In order to  find the general solutions of \eqref{fieq}, one may consider the following differential equation
\begin{equation}
(D-d(\Delta))^{r}g=0, \label{rankrdiff}
\end{equation}
Using the trick \eqref{trick}, one can see that the linearly independent solutions of \eqref{rankrdiff} reads
\begin{equation}\label{solution}
\frac{1}{k!}\partial_\Delta^k f_{r-1},\hs{3ex} k=0, 1\cdots r-1,
\end{equation}
where $f_{r-1}$ is the eigenfunction obeying
\begin{equation}
(D-d(\Delta))f_{r-1}=0.
\end{equation}
To see why the solutions read \eqref{solution}, one can consider acting $\partial_\Delta$ on
\begin{equation}
(D-d(\Delta))f_{r-1}=0.
\end{equation}
So we have
\begin{equation}
\partial_\Delta[(D-d(\Delta))f_{r-1}]=0,
\end{equation}
and
\begin{equation}
D\partial_\Delta[f_{r-1}]-d(\Delta)\partial_\Delta f_{r-1}=\partial_\Delta d(\Delta) f_{r-1}.
\end{equation}
This is actually,
\begin{equation}
Df_{r-2}-d(\Delta) f_{r-2}=\partial_\Delta d(\Delta) f_{r-1}.
\end{equation}
So we have checked that $\partial_\Delta f_{r-1}$ is a solution of $f_{r-2}$. Actually, it is straightforward to check that $\p_\D f_i$ is a solution of $f_{i-1}$. Then starting from $f_{r-1}$, we can read $f_i$ by acting $\partial_\Delta$ $(r-i-1)$ times on it. In this way, we may read the generic solution to \eqref{rankrdiff}:
\begin{equation}
g=\sum_{k=0}^{r-1}\frac{1}{k!}\partial_\Delta^kf_{r-1}N_k
\end{equation}
with $r$ constants $N_0,\cdots,N_{r-1}$. The above discussion may be extended to the case that the function $f$ is not only a function of $\cO_i$, but also involves multiple operators. When it applies to the multi-point function, the logarithmic behavior appears.

The differential equation \eqref{fieq} shows the basic structures in the LCFTs where multiplets appear. Such structure appears in  the three-point functions and the global blocks as well. Consider the blocks expansion of the four-point functions with rank-1 quasi-primary operators. The propagating operator may be some multiplets. The Casmir equation reads (now we have the $L_0$ action on the multiplets, and we can consider the action of the Casmir operator on the propagating operators),
\begin{equation}
(C-c(\Delta))^rG_{\Delta}=0
\end{equation}
where $c(\Delta)$ is the eigenvalue in the usual CFTs without multiplet, and $G_{\Delta}$ is the stripped block. The solutions read,
\begin{equation}
\frac{1}{k!}\partial_\Delta^k G_{\Delta}P_k
\end{equation}
where $P_k$ is the undetermined coefficient, related to the three-point coefficients. The relation can be derived in the OPE limit. Here we provide a heuristic way to see the key structures in the LCFTs, where the multiplets appear. For more details of the blocks, please see \cite{Hogervorst:2016itc}. For more discussion on the trick, please see \cite{Flohr:1997wm} and related papers.

\section{Details on the multiplets in GCFTs} \label{MutiGCFT}
In GCFTs, we consider the $M_0$ multiplets, where $M_0$ is not diagonal. For example, the rank-$2$ case reads,
\begin{equation}
 M_0\cO=\xi \cO
\end{equation}
where $\xi$ is a matrix.
\begin{equation}
M_0\cO_0=\xi \cO_0+\cO_1,\ \ M_0\cO_1=\xi \cO_1.
\end{equation}
These multiplets have similar structures as the ones in LCFTs. The trick \eqref{trick} shows the reason why this is the case. But now, comparing to the LCFTs one should replace
\begin{equation}\label{reprule4}
\partial_\Delta\rightarrow\partial_\xi
\end{equation}
in the discussion. All the discussions are parallel to the ones in the section above.

From BCH formula, we find
\begin{equation}
[L_n,\cO(x,y)]=[(x^{n+1}\partial_x+(n+1)x^ny\partial_y)I+(n+1)(x^n \tilde{\Delta}-nx^{n-1}\tilde{\xi})]\cO(x,y),\ \ n\geq -1
\end{equation}
\begin{equation}
[M_n,\cO(x,y)]=[-(x^{n+1}\partial_y)I+(n+1)x^n\tilde{\xi}]\cO(x,y),\ \ n\geq -1
\end{equation}
where we denote the identical matrix as $I$ explicitly. Now we consider the two-point functions first. For a multiplet of rank $r$, we denote
\begin{equation}
G_{i,j}(x_1,y_1,x_2,y_2)=\langle \cO_i(x_1,y_1) \cO_j(x_2,y_2)\rangle,\ \ i,j=0,1,\cdots r-1.
\end{equation}
Since the vacuum is invariant undert $\{L_0,L_{\pm1},M_0,M_{\pm1}\}$, we can get the differential equations which the two-point functions obey. From $L_{-1}$ and $M_{-1}$, we have
\begin{equation}
L_{-1}:(\partial_{x_1}+\partial_{x_2})G_{i,j}=0,
\end{equation}
\begin{equation}
M_{-1}:-(\partial_{y_1}+\partial_{y_2})G_{i,j}=0,
\end{equation}
which means $G_{i,j}=G_{i,j}(x,y)$ with $x=x_1-x_2,y=y_1-y_2$. Moreover, we have
\begin{equation}
L_0:(x\partial_x+y\partial_y+\tilde{h}\times I+I\times \tilde{h})G_{i,j}=0,
\end{equation}
\begin{equation}
L_1:[(x_1^2\partial_{x_1}+2x_1y_1\partial_{y_1}+2(x_1\tilde{h}\times I-y_1\tilde{\xi}\times I)+(x_2^2\partial_{x_2}+2x_2y_2\partial_{y_2}+2(x_2I\times\tilde{h} -I\times y_2\tilde{\xi})]G_{i,j}=0,
\end{equation}
\begin{equation}
M_0:(-x\partial_y+\tilde{\xi}\times I+I\times\tilde{\xi})G_{i,j}=0,
\end{equation}
\begin{equation}
M_1:(-x_1^2\partial_{y_1}-x_2^2\partial_{y_2}+2x_1\tilde{\xi}\times I+2x_2I\times\tilde{\xi})G_{i,j}=0.
\end{equation}
Similar to the way in the LCFTs, from $M_0$ and $M_{1}$, one gets
\begin{equation}
x\partial_y G_{i,j}(x,y)=2\xi G_{i,j}(x,y)+G_{i+1,j}+G_{i,j+1},
\end{equation}
\begin{equation}
G_{i,j}=G_{j,i}.
\end{equation}
Introducing
\begin{equation}
\G_{k}=G_{i,j}, \ \ \ k=i+j,
\end{equation}
we have
\begin{equation}
x\partial_y \G_{k}(x,y)=2\xi \G_{k}(x,y)+2\G_{k+1}
\end{equation}
The $y$-dependent part reads,
\begin{equation}
\G_k(x,y)=\sum_{i=0}^{r-1-k}\frac
{1}{i!}N_if(x)\partial_\xi^ie^{2\xi\frac{y}{x}}.
\end{equation}
Using $L_0$ and $L_1$, we get
\begin{equation}
f(x)=\frac{1}{x^{2\Delta}}.
\end{equation}
Turning into the canonical form, we have
\begin{equation}
\G_k(x,y)=\frac{1}{k!}\partial_\xi^{r-1-k}(\frac
{1}{x^{2\Delta}}e^{2\xi\frac{y}{x}}).
\end{equation}

Let us give an explicit example: the two-point functions of a rank-$2$ multiplet in the GCFTs. We consider the following rank-2 multiplet $\cO=(\cO_0,\cO_1)^T$,
\begin{equation}
L_n\cO=0,\hs{3ex}
M_n\cO=0,\ \ n\geq 1
\end{equation}
\begin{equation}
L_0\cO=\tilde{\Delta}\cO,\ \ M_0\cO=\tilde{\xi}\cO,
\end{equation}
where
\begin{equation}
\tilde{\Delta}=
\begin{pmatrix}
 \Delta&0 \\
  0& \Delta\\
\end{pmatrix}_{2\times 2},\hs{3ex}
\tilde{\xi}=
\begin{pmatrix}
 \xi&1 \\
  0& \xi\\
\end{pmatrix}_{2\times 2}.
\end{equation}
The two-point functions now read
\begin{equation}
G_{1,1}=0,\ \ G_{0,1}=G_{1,0}=\frac{1}{x^{2\Delta}}e^{2\xi\frac{y}{x}},\ \ \ G_{0,0}=\partial_\xi G_{01}.
\end{equation}
This result matches with the replacement rule \eqref{reprule4}. 

Note that we use a different notation in the paper, compared to the LCFTs case,
\begin{equation}
\cO_0\leftrightarrow \cO_1
\end{equation}
Generally,
\begin{equation}
\cO_i\leftrightarrow \cO_{r-1-i}
\end{equation}

From the discussion of the above sections, one can confirm the more general results in the paper in a straightforward way.

\section{Number of the operators in GGFT}  \label{opernumber}
We want to find the local operators which appear in the $\cO_1\cO_2$ OPE. In other words, they have non-vanishing three-point functions with $\cO_1\cO_2$. Since we are discussing the GGFTs now, they are the composite operators comprising of $\cO_1,\cO_2$. Moreover, to have non-vanishing three-point functions, only one $\cO_1$ and one $\cO_2$ can appear  in the composite operator due to the normal ordering and the Wick theorem, since only the terms from the contraction of $\cO_1$(or $\cO_2$) in the composite operator with the operators outside the composite operator are left. $L_{-1}$ and $M_{-1}$ can appear in the local composite operator, since they correspond to $\partial_x$ and $\partial_y$ respectively. In short, we want to construct the operators like $\cO=L_{-1}^{a}M_{-1}^{b}\cO_1L_{-1}^{c}M_{-1}^{d}\cO_2$ and find their linear combinations which are quasi-primary operators.

More concretely, we would like to find the quasi-primary operators at level $N$. By level $N$, we mean that there are totally $N$ $L_{-1}$ and $M_{-1}$ in the composite operator, since both $L_{-1}$ and $M_{-1}$ are of weight $1$. The problem is  how many independent quasi-primary $\cO$s at level $N$. Firstly, there are two different operators ($L_{-1}$ or $M_{-1}$) inserted at two distinguished position (before $\cO_1$ or $\cO_2$) so that the partition function is
\begin{equation}
Z(q)=\frac{1}{(1-q)^4}
\end{equation}
where $4=2\times 2$. After doing expansion with respect to $q$, we can get the number of independent operators at level $N$ by reading the coefficient of $q^N$.
Secondly, starting with a quasi-primary operator, we can find its global descendent operators by acting $L_{-1}^{a}M_{-1}^{b}$ on it. At level $k=a+b$, the number of independent global descendent operators related to a specified quasi-primary operator is $f(k,2)$, where $f(k,2)$ is the number of different ways of the binary partition of the integer $k$. The number of independent quasi-primary operators at level $N$ is the total number of independent operators minus the number of global descendent ones at level $N$.

At level 0, the composite operator $\cO^{(1)}=\cO_1\cO_2$ is a quasi-primary operator, since $\cO_1$ and $\cO_2$ are quasi-primary operators. Consequently at level 1, there are $f(1,2)=2$ global descendent operators related to $\cO^{(1)}$. As totally there are 4 independent operators at level $1$, there remains two quasi-primary operators at level $1$. At level 2, there are $f(2,2)=3$ global descendent operators related to $\cO^{(1)}$. And there are $2\times f(1,2)=4$ global descendent operators related to quasi-primary operators at level 1. Totally there are 10 independent operators at level 2, so there remains 3($=10-4-3$) quasi-primary operators at level 2. In this way, we get the numbers of quasi-primary operators at each level.

Considering the general action of $M_0$ on the composite operator $\cO=L_{-1}^{a}M_{-1}^{b}\cO_1L_{-1}^{c}M_{-1}^{d}\cO_2$, it will give one term proportional to $\cO$ (the coefficient is $\xi_\cO=\xi_1+\xi_2$) and additional terms with one of the $L_{-1}$'s being replaced by $M_{-1}$ due to the commutation relation
\begin{equation}
[L_{-1},M_{0}]=-M_{-1}.
\end{equation}
By acting $(M_{0}-\xi_\cO)$ $k$ times on $\cO$, we get zero, where $k$ is the number of $L_{-1}$'s in $\cO$. In other words,  the composite operator $\cO$ belongs to a rank-$k$ multiplet if there are $k$ $L_{-1}$ in its construction. For a level-$N$ operator, there are at most $N$ $L_{-1}$s in $\cO$. Therefore a level-$N$ quasi-primary operator is at most of rank $N$.
On the other hand, we can show  that it is at least rank-$N$, which means $L_{-1}^{a}\cO_{1}L_{-1}^{b}\cO_2$ with $a+b=N$ must appear in the linear combination in at least one of the quasi-primary operators. Supposing not, we can count the number of the quasi-primary operators constructed without these terms. We denote the number of the quasi-primary operators with(/without) these terms as $Q(N)$(/$A(N)$).
\begin{equation}
A(N)=Q(N)-f(N,2)+1\times N=Q(N)-1
\end{equation}
where $f(N,2)$ comes from the terms of the form $L_{-1}^{a}\cO_{1}L_{-1}^{b}\cO_{2}$, and $1$ comes from that there are one less global descendant operators related to each quasi-primary operators at level $0,1,2,\cdots,N-1$. Thus, there must be one quasi-primary operator at level $N$ containing terms $L_{-1}^{a}\cO_{1}L_{-1}^{b}\cO_{2}$. To conclude, at level $N$, there is at least one rank-$N$ multiplet of the quasi-primary operators. From the counting of the number of the quasi-primary operators, we know that at level-$N$, there is one rank-$N$ multiplet.

\section{Proof of the cancelation in section \ref{LimitGFT}}\label{Cancelation}
In this section, we will show the cancellation of the superficially divergent terms in taking the non-relativistic limit of the global block expansion of a 2d GFT. The cancellation is due to some combinatorial identities.

We consider the general case $\Delta=2\Delta_\cO+n\equiv\Delta_n$, which is just the contribution of the rank-($n+1$) multiplet as  we will show very soon. At rank $n+1$,  we have:
\begin{equation}
\begin{split}
&a_n\bar{a}_0(G_{\Delta_n,2\xi_\cO}+n\epsilon\partial_\xi G_{\Delta_n,2\xi_\cO}+\frac{n^2\epsilon^2}{2!}\partial_\xi^2G_{\Delta_n,2\xi_\cO}+...)\\
+&a_{n-1}\bar{a}_1(G_{\Delta_n,2\xi_\cO}+(n-2)\epsilon\partial_\xi G_{\Delta_n,2\xi_\cO}+\frac{(n-2)^2\epsilon^2}{2!}\partial_\xi^2G_{\Delta_n,2\xi_\cO}+...)\\
+&a_{n-2}\bar{a}_2(G_{\Delta_n,2\xi_\cO}+(n-2)\epsilon\partial_\xi G_{\Delta_n,2\xi_\cO}+\frac{(n-2)^2\epsilon^2}{2!}\partial_\xi^2G_{\Delta_n,2\xi_\cO}+...)\\
+&...\\
+&a_0\bar{a}_n(G_{\Delta_n,2\xi_\cO}+(-n)\epsilon\partial_\xi G_{\Delta_n,2\xi_\cO}+\frac{(-n)^2\epsilon^2}{2!}\partial_\xi^2G_{\Delta_n,2\xi_\cO}+...)
\end{split}
\end{equation}
where the ellipses  denote the higher order terms in the Taylor expansions. The point is that all the $a\bar{a}$ terms have at most $(\frac{1}{\epsilon})^n$ singular behaviour,   so the Taylor expansions are
truncated to  the first $n+1$ terms as $\epsilon\to0$, corresponding to a multiplet of  rank $n+1$. Now we calculate the above summation, it can be written as:
\begin{equation}
\sum_{k=0}^\infty A_{n,k}\partial_\xi^kG_{\Delta_n,2\xi_\cO},
\end{equation}
with
\begin{equation}
A_{n,k}=\sum_{q=0}^{n}a_q\bar{a}_{n-q}\frac{(2q-n)^k}{k!}\epsilon^k.
\end{equation}
Using the relations
\begin{equation}
a_n=\frac{(2h_\cO)_n^2}{n!(4h_\cO+n-1)_n},  \qquad \bar{a}_n=\frac{(2\bar{h}_\cO)_n^2}{n!(4\bar{h}_\cO+n-1)_n},\nn
\end{equation}
and
\begin{equation}
h_\cO=\Delta_\cO+\frac{\xi_\cO}{\epsilon}, \qquad \bar{h}_\cO=\Delta_\cO-\frac{\xi_\cO}{\epsilon}\nn
\end{equation}
we have:
\begin{equation}
\begin{split}
A_{n,k}&=\frac{1}{k!}\sum_{q=0}^n\frac{\Gamma^2(2h_\cO+q)\Gamma(4h_\cO+q-1)}{q!\Gamma^2(2h_\cO)\Gamma(4h_\cO+2q-1)}
\frac{\Gamma^2(2\bar{h}_\cO+n-q)\Gamma(4\bar{h}_\cO+n-q-1)}{(n-q)!\Gamma^2(2\bar{h}_\cO)\Gamma(4\bar{h}_\cO+2n-2q-1)}(2q-n)^k\epsilon^k\\
&=\frac{1}{k!}\sum_{q=0}^n\frac{1}{q!(n-q)!}(2q-n)^k\frac{1}{\epsilon^n}\epsilon^k\frac{\prod_{i=0}^{q-1}[\xi_\cO+\epsilon(\Delta_\cO+i/2)]^2}{\prod_{j=q-1}^{2q-2}[\xi_\cO+\epsilon(\Delta_\cO+j/4)]}
\frac{\prod_{i=0}^{n-q-1}[-\xi_\cO+\epsilon(\Delta_\cO+i/2)]^2}{\prod_{j=n-q-1}^{2n-2q-2}[-\xi_\cO+\epsilon(\Delta_\cO+j/4)]}
\end{split}
\end{equation}
from which we easily get that when $k>n$, $A_{n,k}\to0$ as $\epsilon\to0$.  This truncation avoids the appearance of infinite-rank multiplet and in fact lead to a rank-($n+1$) multiplet.
For $k=n$, $A_{n,k}$ is finite; for $k<n$, the above expression  has superficial divergent terms of order $\epsilon^{k-n}$, $\epsilon^{k-n+1}$, ... $\epsilon^{-1}$.  Interestingly, all these divergent terms actually vanish. To see this fact, let us
rewrite $A_{n,k}$ as:
\begin{equation}
A_{n,k}=\frac{\epsilon^{k-n}}{k!}\sum_{q=0}^n\frac{1}{q!(n-q)!}(2q-n)^k\xi_\cO^n(-1)^{n-q}\frac{\prod_{i=0}^{q-1}(1+\epsilon\frac{\Delta_\cO+i/2}{\xi_\cO})^2}{\prod_{j=q-1}^{2q-2}(1+\epsilon\frac{\Delta_\cO+j/4}{\xi_\cO})}
\frac{\prod_{i=0}^{n-q-1}(1-\epsilon\frac{\Delta_\cO+i/2}{\xi_\cO})^2}{\prod_{j=n-q-1}^{2n-2q-2}(1-\epsilon\frac{\Delta_\cO+j/4}{\xi_\cO})}. \label{Ank}
\end{equation}
To calculate the order $\epsilon^{k-n+l}$ terms directly is not easy because the above expression is quite complicated. However, since we expect all the  terms of order $\epsilon^{k-n+l}$ vanish for $l<n-k$, we can use mathematical induction to prove the cancellation. Let us start from  the simplest case,
the most singular part involving the terms of order $\epsilon^{k-n}$ ($l=0$), which is of the form
\begin{equation}
\frac{\epsilon^{k-n}\xi_\cO^n}{k!}\sum_{q=0}^n\frac{1}{q!(n-q)!}(2q-n)^k(-1)^{n-q}.   \label{gk}
\end{equation}
For $k<n$, the above summation vanish. In fact, when $k=0$, the involved summation is:
\begin{equation}
\sum_{q=0}^n\frac{1}{q!(n-q)!}(-1)^{n-q}=\frac{1}{n!}\sum_{q=0}^nC_n^q(-1)^{n-q}=\frac{1}{n!}(1-1)^n=0  \label{0k}
\end{equation}
For general $k$, we can expand $(2q-n)^k$ in \eqref{gk} and  the summation for every individual $q^m$, $0\leq m\leq k < n$ term actually vanishes, namely:
\begin{equation}
\sum_{q=0}^n\frac{1}{q!(n-q)!}q^m(-1)^{n-q}=0,  \qquad m< n.    \label{gm}
\end{equation}
Now we use mathematical induction to prove this result. For $m=0$, we have already proven it in  \eqref{0k}. If \eqref{gm} holds for $m< n$, then for $m+1,n+1$, we have
\begin{equation}
\begin{split}
&\sum_{q=0}^{n+1}\frac{1}{q!(n+1-q)!}q^{m+1}(-1)^{n+1-q}\\
=&\sum_{q=1}^{n+1}\frac{1}{(q-1)!(n+1-q)!}q^{m}(-1)^{n+1-q}\\
=&\sum_{q=0}^{n}\frac{1}{q!(n-q)!}q^{m}(-1)^{n-q}\\
=&0.
\end{split}      \label{rec}
\end{equation}
Note that $m<n$ is necessary, and $m+1<n+1$ is automatically true when $m<n$ is given.

\par In the above discussion, using the general result \eqref{gm} we find that the  terms of order $\epsilon^{k-n}$ ($l=0$) always vanish. Following the same logic, we can show that the terms of  general order $\epsilon^{k-n+l}$ vanish as well. In fact, from \eqref{Ank}  we need only to show that in the $\epsilon$ expansion of the product
\begin{equation}
\frac{\prod_{i=0}^{q-1}(1+\epsilon\frac{\Delta_\cO+i/2}{\xi_\cO})^2}{\prod_{j=q-1}^{2q-2}(1+\epsilon\frac{\Delta_\cO+j/4}{\xi_\cO})}
\frac{\prod_{i=0}^{n-q-1}(1-\epsilon\frac{\Delta_\cO+i/2}{\xi_\cO})^2}{\prod_{j=n-q-1}^{2n-2q-2}(1-\epsilon\frac{\Delta_\cO+j/4}{\xi_\cO})},     \label{la}
\end{equation}
the order-$\epsilon^l$ terms, as polynomials in $q$, are at most of order $q^l$. Then as $m+l \leq k+l <n$, by \eqref{gm}, the summation vanishes. However, the terms of order $\epsilon^l$ in the expansion of \eqref{la} are superficially the polynomials of order $q^{2l}$. Next we would like to  show that those terms of order $q^m$ for $m>l$ always vanish. This can also be proved by using
mathematical induction. For $l=0$, \eqref{la} is $1$ and is of course of order $q^0$. Now, suppose that for $m=0,1,2...l$, the order-$\epsilon^m$ terms are polynomials in $q$ of at most order $q^m$, then for $m=l+1$, we denote
\eqref{la} as $a_{n,q}$, and rewrite $a_{n,q+1}$ as:
\begin{equation}
a_{n,q+1}\equiv a_{n,q}b_{n,q},
\ee
where
\be
\begin{split}
a_{n,q+1}&=a_{n,q}\frac{(1+\epsilon\frac{\Delta_\cO+\frac{q}{2}}{\xi_\cO})}{(1-\epsilon\frac{\Delta_\cO+\frac{n-q-1}{2}}{\xi_\cO})}
\frac{(1+\epsilon\frac{\Delta_\cO+\frac{q-1}{4}}{\xi_\cO})}{(1+\epsilon\frac{\Delta_\cO+\frac{2q-1}{4}}{\xi_\cO})}
\frac{(1-\epsilon\frac{\Delta_\cO+\frac{2n-2q-3}{4}}{\xi_\cO})}{(1-\epsilon\frac{\Delta_\cO+\frac{n-q-1}{4}}{\xi_\cO})}.\\
\end{split}
\end{equation}
Denote the order $\epsilon^m$ term of $a_{n,q}$ as $a_{n,q}^m$ and the order $\epsilon^m$ term of $b_{n,q}$ as $b_{n,q}^m$, then
\begin{equation}
a_{n,q+1}^{l+1}=\sum_{r=0}^{l+1}a_{n,q}^{r}b_{n,q}^{l+1-r},
\end{equation}
so that
\begin{equation}
a_{n,q+1}^{l+1}-a_{n,q}^{l+1}=\sum_{r=0}^{l}a_{n,q}^{r}b_{n,q}^{l+1-r} . \label{laa}
\end{equation}
If we can prove the RHS in \eqref{laa} is at most of order $q^l$, then $a_{n,q}^{l+1}$ is at most of order $q^{l+1}$ and we are done.
By  induction, we know that $a_{n,q}^{r}$ is at most of order $q^r$ for $0\leq r\leq l$, so what remains is to show $b_{n,q}^{l+1-r}$ is a polynomials in $q$ of  order $q^{l-r}$ at most. In other words, we need to show that
the highest order term, which is of order $q^{l+1-r}$, vanishes. In fact, for this term, we can effectively calculate the order $\epsilon^{l+1-r}$ ($r=0,1,2...l$) terms for the following expression:
\begin{equation}
\frac{(1+\epsilon\frac{\frac{q}{2}}{\xi_\cO})}{(1-\epsilon\frac{\frac{-q}{2}}{\xi_\cO})}
\frac{(1+\epsilon\frac{\frac{q}{4}}{\xi_\cO})}{(1+\epsilon\frac{\frac{2q}{4}}{\xi_\cO})}
\frac{(1-\epsilon\frac{\frac{-2q}{4}}{\xi_\cO})}{(1-\epsilon\frac{\frac{-q}{4}}{\xi_\cO})}.
\end{equation}
It easy to see that this gives exactly $1$. So all the order $\epsilon^{l+1-r}$ terms vanish. Then from the above analysis, we have finished the proof.

\par Furthermore, we can also show that the remaining finite pieces of the 2d GFT coincide with the GGFT result from other methods. From \eqref{rec}, if we set $m=n$, then:
\begin{equation}
\sum_{q=0}^{n}\frac{1}{(q)!(n-q)!}q^{n}(-1)^{n-q}=\frac{1}{0!0!}0^{0}(-1)^{0-0}=1.
\end{equation}
What remains to obtain $A_{n,k}$ is to calculate the coefficient of $q^{n-k}$ in $a_{n,q}^{n-k}$. One can try to calculate it directly or again use mathematical induction method, and will find our GGFT result can be reproduced by the limiting method.

\section{Check the spectral density of GGFTs} \label{check density}
In this section, we will show that GGFTs satisfy the requirement of using the tauberian theorem and
verify that the integrated spectral density function of GGFTs has the asymptotical behaviour predicted by the tauberian theorem in section \ref{Tauberian}. This gives a cross check of the calculation in both sides.

Firstly, we need to see whether the condition of the Hardy-Littlewood tauberian theorem \eqref{tauberian condition} hold in the GGFT case. Let's focus on the order $\kappa^0$ case, for which we have the relaxed condition \eqref{tauberian condition} for the spectral density function. It is easy to see that when $\Delta_\cO>1/4$\footnote{Note that in the case of order $\kappa^0$, the coefficients $P^t_{n,0}$ have no $\xi$ dependence, generally, we also need the condition $\xi_{\cO}> 0$ to ensure the positivity of $P^t_{n,k}$.}, the coefficients \eqref{coef1} is positive, so the requirement \eqref{tauberian condition} is satisfied. In fact, the condition $\Delta_\cO>1/4$ is just the one that ensure the order $\kappa^0$ part of the four point function divergent when $\beta\to0$ in \eqref{asymptotical behaviour}. Note that this divergent case is the standard situation for which tauberian theorems are applied. Nevertheless, as noted in \cite{2007Tauberian}, the Hardy-Littlewood tauberian theorem also holds for the convergent case (corresponding to $\Delta_\cO\leq1/4$ here). In general, the coefficients \eqref{coef1} can be negative\footnote{In the GFT case, unitarity bounds ensure the positivity of its coefficients, but as we emphasize in the main body of the paper, there is no unitarity for GCFTs.}, nevertheless, the condition \eqref{tauberian condition} still hold for GGFTs. In order to show this, we only need to check the case $\Delta_\cO\leq1/4$, calculate the asymptotical behaviour of $P^t_{n,0}$ as $n\to\infty$ (equivalently $\Delta=2\Delta_\cO+n\to\infty$) and then compare with the right hand side of \eqref{tauberian condition}. Note that both of these two quantities goes to $0$ and are negative when $\Delta\to\infty$. We write $P^t_{n,0}$ explicitly:
\begin{equation}
P^t_{n,0}=2^{4\Delta_\cO-2}(4\Delta_\cO+2n-2)\frac{\Gamma(4\Delta_\cO+n-2)}{\Gamma(4\Delta_\cO-2)\Gamma(n+1)}
\end{equation}
we want to know its behaviour when $n\to\infty$. Using Stirling's approximation:
\begin{equation}
\Gamma(x+1)\sim \sqrt{2\pi x}\left(\frac{x}{e}\right)^x, \qquad x\to\infty
\end{equation}
we find
\begin{equation}
\begin{split}
&P^t_{n,0}=2^{4\Delta_\cO-2}(4\Delta_\cO+2n-2)\frac{\Gamma(4\Delta_\cO+n-2)}{\Gamma(4\Delta_\cO-1)\Gamma(n+1)}\\
\sim&2^{4\Delta_\cO-2}\frac{2n}{\Gamma(4\Delta_\cO-1)}\frac{\sqrt{2\pi (4\Delta_\cO+n-3)}(\frac{4\Delta_\cO+n-3}{e})^{4\Delta_\cO+n-3}}{\sqrt{2\pi n}(\frac{n}{e})^n}\\
\sim&2^{4\Delta_\cO-2}\frac{2n}{\Gamma(4\Delta_\cO-1)}e^{3-4\Delta_\cO}(1+\frac{4\Delta_\cO-3}{n})^{n}(4\Delta_\cO+n-3)^{4\Delta_\cO-3}\\
\sim&2^{4\Delta_\cO-2}\frac{2n}{\Gamma(4\Delta_\cO-1)}e^{3-4\Delta_\cO}e^{4\Delta_\cO-3}n^{4\Delta_\cO-3} \\
\sim&2^{4\Delta_\cO-1}\frac{n^{4\Delta_\cO-2}}{\Gamma(4\Delta_\cO-1)} \\
\sim&2^{4\Delta_\cO-1}\frac{\Delta^{4\Delta_\cO-2}}{\Gamma(4\Delta_\cO-1)}    \label{check P}
\end{split}
\end{equation}
The $\Delta$ dependence of this behaviour is just the same as the right hand side of \eqref{tauberian condition} (so we say GGFTs marginally satisfy the condition in section \ref{Tauberian}). However, the condition \eqref{tauberian condition} only needs to hold for some constant $C$, and for any $\Delta_\cO\leq1/4$, we can always find such a $C$ which makes \eqref{tauberian condition} holds \footnote{In fact, \eqref{check P} is true for any $\Delta_\mathcal{O}$. So for $\Delta_\cO>1/4$, we can choose any $C\geq0$ to make \eqref{tauberian condition} holds. }. So we find the GGFTs satisfy the requirement of using the tauberian theorem.

Our next goal is to calculate the asymptotical behaviour of integrated spectral density function of GGFTs, it is expected to coincide with \eqref{integrated}. We want to use a quick way to check this result and do not calculate
the integrated spectral density explicitly. From the behaviour \eqref{integrated}, we know
\begin{equation}
F(n)\sim2^{4\Delta_\cO-1}\frac{(2\Delta_\cO+n)^{4\Delta_\cO-1}}{\Gamma(4\Delta_\cO)}
\end{equation}
so
\begin{equation}
\begin{split}
&P^t_{n,0}=F(n)-F(n-1)\\
\sim&[n-(n-1)]F'(n) \\
\sim&2^{4\Delta_\cO-1}(4\Delta_\cO-1)\frac{(2\Delta_\cO+n)^{4\Delta_\cO-2}}{\Gamma(4\Delta_\cO)} \\
\sim&2^{4\Delta_\cO-1}\frac{\Delta^{4\Delta_\cO-2}}{\Gamma(4\Delta_\cO-1)}
\end{split}
\end{equation}
which is precisely \eqref{check P}. So we have finished the cross check.


\bibliographystyle{JHEP}
\bibliography{Refs}

\end{document}